\pdfoutput=1
\documentclass[structabstract]{aa}  
\usepackage{float}
\usepackage{natbib}
\usepackage{graphicx}
\usepackage{txfonts}

\newcommand{\ntot}{n_{\left<{\rm H}\right>}}

\newcommand{\rpr}{R_{i\leftarrow jl}^{\rm Xpr}}
\newcommand{\rse}{R_{i\leftarrow j}^{\rm Xsec}}

\begin{document}
   \title{FUV and X-ray irradiated protoplanetary disks: a grid of models}

   \subtitle{I. The disk structure}

   \author{R. Meijerink\inst{1,2} \and G. Aresu\inst{1} \and
     I. Kamp\inst{1} \and M. Spaans\inst{1} \and W.-F. Thi\inst{3} \and P. Woitke\inst{4}}

   \institute{ Kapteyn Astronomical Institute, PO Box 800, 9700 AV Groningen, The Netherlands \\ 
     \email{meijerink@astro.rug.nl} \and
     Leiden Observatory, Leiden University, P.O. Box 9513, NL-2300 RA
     Leiden, Netherlands \and
     UJF-Grenoble 1/CNRS-INSU, Institut de Plan\'etologie
     d'Astrophysique (IPAG) UMR5274, Grenoble, F-38041, France \and
     SUPA, School of Physics \& Astronomy, University of St. Andrews,
     North Haugh, St. Andrews KY16 9SS, UK     \\
    }

   \date{Received ??; accepted ??}

   \abstract{Planets are thought to eventually form from the mostly
     gaseous ($\sim 99\%$ of the mass) disks around young stars. The
     density structure and chemical composition of protoplanetary
     disks are affected by the incident radiation field at optical,
     far-ultraviolet (FUV), and X-ray wavelengths, as well as by the
     dust properties.}{The effect of FUV and X-rays on the disk
     structure and the gas chemical composition are investigated. This
     work forms the basis of a second paper, which discusses the
     impact on diagnostic lines of, e.g., C$^+$, O, H$_2$O, and Ne$^+$
     observed with facilities such as Spitzer and Herschel.}{A grid of
     240 models is computed in which the X-ray and FUV luminosity,
     minimum grain size, dust size distribution, and surface density
     distribution are varied in a systematic way. The hydrostatic
     structure and the thermo-chemical structure are calculated using
     \underline{Pro}toplanetary \underline{Di}sk \underline{Mo}del
     (ProDiMo), with the recent addition of X-rays.}{The abundance
     structure of neutral oxygen is very stable to changes in the
     X-ray and FUV luminosity, and the emission lines will thus be
     useful tracers of the disk mass and temperature. The C$^+$
     abundance distribution is sensitive to both X-rays and FUV. The
     radial column density profile shows two peaks, one at the inner
     rim and a second one at a radius $r=5-10$~AU. Ne$^+$ and other
     heavy elements have a very strong response to X-rays, and the
     column density in the inner disk increases by two orders of
     magnitude from the lowest ($L_{\rm X} = 10^{29}$~erg\,s$^{-1}$)
     to the highest considered X-ray flux ($L_{\rm X} =
     10^{32}$~erg\,s$^{-1}$). FUV confines the Ne$^+$ ionized region
     to areas closer to the star at low X-ray luminosities ($L_X =
     10^{29}$~erg\,s$^{-1}$). H$_2$O abundances are enhanced by
     X-rays due to higher temperatures in the inner disk than in the
     FUV only case, thus leading to a more efficient neutral-neutral
     formation channel. Also, the higher ionization fraction provides
     an ion-molecule route in the outer disk. The line fluxes and
     profiles are affected by the effects on these species, thus
     providing diagnostic value in the study of FUV and X-ray
     irradiated disks around T Tauri stars.}{}


%
%________________________________________________________________

\maketitle

\section{Introduction}

New observing facilities in the past decade pushed our understanding
of protoplanetary disks from a rough picture of a vertically layered
structure to a wealth of details on the composition and
two-dimensional structure of the gas and dust of these disks. In the
infrared, the Spitzer Space Telescope performed systematic studies of
nearby star-forming regions. The spectral energy distributions (SEDs)
revealed the physical structure of disks, e.g., the presence of gaps,
source-to-source variations and important statistics on SED types,
which allow evolutionary scenarios to be built
\citep[cf.][]{Merin2010}. Extensive gas-phase emission line studies
with Spitzer provide a first indication of chemical diversity across
stellar spectral types
\citep[cf.][]{Najita2009,Pontoppidan2010,Lahuis2007}. In the
near-infared, there have been large ground-based studies with, e.g.,\
the VLT and the Keck telescope resolving the lines to study the
kinematics (spatial origin of the lines in hot disk surfaces and
winds) and their excitation mechanisms \citep[thermal and
fluorescence; ][]{Brittain2010, Pontoppidan2011, Fedele2011}. In the
past two years, the Herschel Space Observatory extended the spectral
window to the far-IR and submm, with spectral line scans within the
Dust, Ice, and Gas in Time (DIGIT) key program (PI: N. Evans) for many
Herbig disks, e.g., HD~100549 \citep{Sturm2010}, water studies of a
few selected targets, such as TW Hya \citep{Hogerheijde2011} from the
Water in Star-forming Regions with Herschel (WISH) key program (PI:
E.van Dishoeck), and large statistical gas surveys targeting the
dominant cooling lines \citep[e.g.,][]{Mathews2010,Meeus2012,Dent2012,
  Riviere-Marichalar2012} in the Gas in Protoplanetary Systems (GASPS)
key program (PI: B. Dent). The HIFI and some PACS line detections deal
with the cold water, while the 63.3~$\mu$m H$_2$O line possibly probes
the inner water reservoir.

Observational studies report trends in emission line strength with the
irradiation of the central star. \citet{Guedel2010} find that the
[Ne\,{\sc ii}] emission at 12.81\,$\mu$m correlates with the X-ray
luminosity of the central star; the slope of the correlation for
nonjet sources is $0.44\pm 0.18$. \citet{Pinte2010} and
\citet{Meeus2012} show that the [O\,{\sc i}]\,63\,$\mu$m line flux
increases with stellar luminosity and that this is largely driven by
the strength of the far-ultraviolet (FUV)
luminosity. \citet{Riviere-Marichalar2012} report a tentative trend of
the 63.3\,$\mu$m water line flux with X-ray luminosity.

It is obvious from previous works that the most relevant parameter for
the SED is the total stellar luminosity; in most cases, $L_{\rm X}$
and $L_{\rm FUV}$ present only a small fraction of this and hence by
themselves do not cause significant SED changes. However, recent
near-IR high-contrast imaging with HiCIAO on the Subaru 8.2\,m
telescope allowed probing of the inner disk structure on scales below
0.1\arcsec\ \citep{Thalmann2010,Hashimoto2011}. This offers the
possibility of a direct probe, e.g., the height of the inner rim as
predicted by protoplanetary disk models. \citet{Thi2011} predicted
that the height of the inner rim could be a factor two higher when the
vertical hydrostatic structure takes into account the gas temperature
at the inner rim. \citet{Aresu2011} show that the height of the inner
rim increases in disk models with the impinging stellar $L_{\rm X}$.

In the last decade, much theoretical progress has taken place to
support the interpretation of the wealth of new observations,
specifically the gas observations. The foundation was laid by
\citet{Chiang1997}, \citet{D'Alessio1998}, and \citet{Dullemond2001},
whose layered dust disk models have helped to interpret SEDs for
irradiated disks (see Dullemond et al. \citeyear{Dullemond2007} for a
review of disk structure models). Based on this, several groups
focussed on the effects of FUV and X-rays on the thermal and chemical
structure of the gas in a pre-prescribed protoplanetary disk model,
e.g., \citet{vanZadelhoff2003}, \citet{Kamp2004},
\citet{Jonkheid2004}, and \citet{Glassgold2004}. In a subsequent step,
\citet{Nomura2005}, \citet{Gorti2008}, and \citet{Woitke2009} studied
the chemical structure of disks, while self-consistently solving for
the vertical hydrostatic equilibrium using the gas temperature. Most
recently, \citet{Aresu2011} performed an exploratory study to assess
the relative importance of FUV and X-rays by expanding the existing
\underline{Pro}toplanetary \underline{Di}sk \underline{Mo}del
(ProDiMo) code to include X-ray processes. The modeling efforts that
solve for the vertical disk structure are computationally expensive
because the chemical networks, heating/cooling balance, 2D continuum
radiative transfer, and hydrostatic equilibrium have to be solved
iteratively. Hence, these studies have largely focussed on a single
representative disk model, or at most a handful of models varying one
specific parameter. However, large-model grids are in principle
required to understand the potential diagnostic power of certain
observables. So far examples for this are large grids (of the order of
200\,000-300\,000 models) of parametrized dust disk models to study
SED diagnostics \citep{Robitaille2006} and of parametrized gas/dust
disks to study the gas emission line diagnostics
\citep{Woitke2011,Kamp2011}.

In this paper, we perform for the first time an extensive analysis of
a grid of 240 self-consistent disk models (including the vertical disk
structure) to study the effects of X-rays, FUV, and the relative
importance of grain size and gas surface density distribution on the
thermal, chemical, and physical structure of disks around T Tauri
stars. This paper builds on the implementation of X-rays into the
ProDiMo code as described in \citet{Aresu2011}. While we discuss here
foremost the physical and chemical structure of the models and how
they change with irradiation, a companion paper \citep[][paper
II]{Aresu2012}, discusses the power of line diagnostic in
disentangling these effects. The paper is structured in the following
way: Sect.~\ref{updates} describes the updates on the code and the
range of parameters used in the grid. The effects on disk temperature
and density structure will be discussed in Sect.~\ref{dens_temp}. The
resulting distribution of various key species and its key reactions
are extensively described in
Sect.~\ref{chemistry}. Sect.~\ref{rad_col} shows the radial column
density distributions. Those are key in understanding the line
profiles, which is the topic of the accompanying paper
II. Sect. \ref{conclusions} summarizes the conclusions and
implications of the paper.

\section{Updates on ProDiMo and the calculated grid}\label{updates}

The original ProDiMo code \citep{Woitke2009} is based on the code as
discussed by \citet{Kamp2004}. The original model includes (1)
frequency-dependent two-dimensional dust continuum radiative transfer,
(2) kinetic gas-phase and FUV photo-chemistry, (3) ice formation, and
(4) detailed non-local thermodynamic equilibrium (non-LTE) heating and
cooling with (5) a consistent calculation of the hydrostatic disk
structure. The models are characterized by a high degree of
consistency among the various physical, chemical, and radiative
processes, since the mutual feedbacks are solved
iteratively. \citet{Aresu2011} included X-ray heating and chemical
processes, guided by work from \citet{Maloney1996},
\citet{Glassgold2004}, and \citet{Meijerink2005}, while most recently
the X-ray chemistry processes are updated following
\citet{Adamkovics2011}. This includes an extension of our chemical
network with species such as Ne, Ar, and their singly and doubly
ionized states, as well as other heavy elements. We added Ne$^+$,
Ne$^{2+}$, Ar$^+$ and Ar$^{2+}$ in the heating/cooling balance and
also implemented an extended sulfur chemistry \citep[following][e.g.,
SO, SO$_2$, HS, and their related reactions were added]{Leen1988,
  Meijerink2008} to achieve a proper calculation of the sulfur-based
species abundances. This allowed us to make more realistic predictions
of, e.g., the sulfur fine-structure lines. A detailed description on
the treatment of X-ray chemistry in the code is given in Appendix
\ref{xray_chem}.

\begin{figure}
  \centering
  \includegraphics[width=9cm,clip]{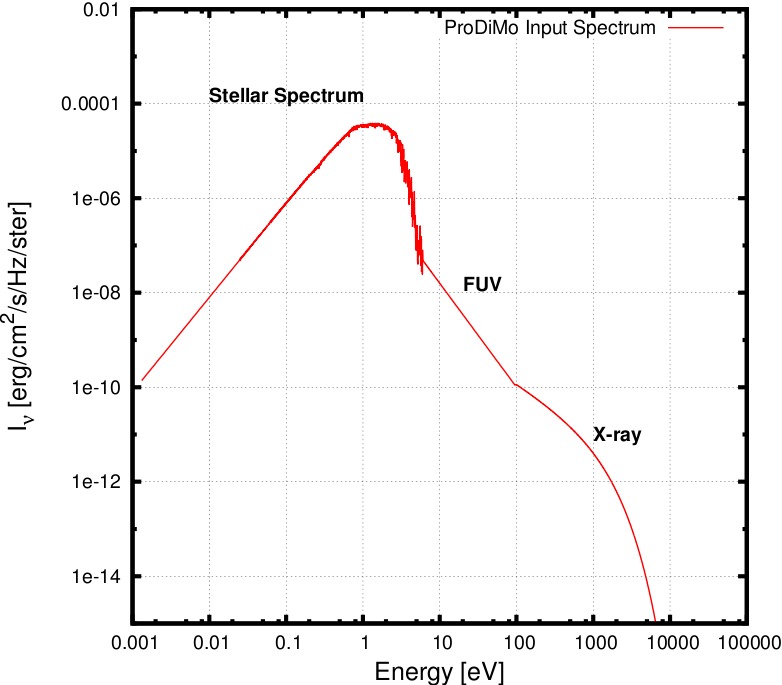}
  \caption{Input spectrum for model with $L_{\rm
      X}=10^{30}$~erg~s$^{-1}$ and $L_{\rm FUV}=10^{31}$~erg~s$^{-1}$.}
  \label{spec}
\end{figure}

Our input spectrum is composed of the stellar spectrum and a FUV
excess as described in \citet{Woitke2009}, along with an X-ray
component as outlined in \citet{Aresu2011}. The FUV excess (from 92.5
to 205 nm) spectrum is a power law with $I_\nu(\lambda) \propto
\lambda^{1.2}$. The X-ray spectrum (0.1 - 10~keV) has been chosen
following \citet{Glassgold2007} and is the same we used in
\citet{Aresu2011}. It is a bremsstrahlung spectrum with
$I_{\nu}(E)\propto 1/E \times \exp(-E/kT)$ (see Fig. \ref{spec}). We
varied those quantities that could potentially affect the penetration
of X-ray and FUV radiation and change the energy deposition through
the disk. For example, we use two different values for the dust size
distribution power law. The first value, generally used in the
literature for the ISM, is a power law index of $a_{\rm pow}=3.5$. It
is well known, though, that dust coagulation is the seed process that
in timescales of a few million years leads to the formation of
planetesimals. To model a disk that could have undergone dust
coagulation, we also adopt a shallower power law index of $a_{\rm
  pow}=2.5$. For the same reason, we vary the minimum dust size: 0.1,
0.3, and 1 $\mu m$. These two parameters ($a_{\rm min}$ and $a_{\rm
  pow}$) determine the FUV opacity and, as a result, impact the amount
of energy reprocessed by the photoelectric heating. Ultimately, the
parameters affect the gas thermal balance in the disk. The changing
opacity and optical depth also impact the dust temperature $T_{\rm
  dust}$.

\citet{Aresu2011} performed an exploratory study on the influence of
X-rays and FUV irradiation on the structure of protoplanetary disks,
for X-ray luminosities ranging from $L_x = 10^{29} -
10^{32}$~erg\,s$^{-1}$. The range of values for this particular
parameter is based on the outcome of the Taurus survey
\citep{Guedel2007}, which showed that X-ray-active T Tauri stars emit
in the range $L_{\rm X} = 10^{29} - 10^{31}$~erg\,s$^{-1}$. The model
with $L_{X}=10^{32}$~erg\,s$^{-1}$, on the other hand, is an extreme
case with interesting implications in terms of modeling. FUV excess
emission from classic T Tauri stars between 6 and 13.6 eV is believed
to arise from accretion spots on the surface of the star and stellar
activity. The disk is truncated to several stellar radii from the star
magnetic field, which channels matter toward the stellar
photosphere. Shocks in the outer layers of the star cause the
temperature to rise and emit FUV photons, which results in a FUV
excess with respect to the stellar luminosity in the same band
\citep{Gullbring2000}. \citet{Gorti2008} infer the FUV excess
luminosity (91.2 nm $< h\nu <$ 205 nm) to be between $L_{\rm
  FUV}/L_{*}\sim 10^{-2} - 10^{-3}$ . In our grid, we decided to scale
the FUV flux so that we obtain the same range of luminosities as we
use for the X-rays. The adopted values of $L_{\rm FUV}/L_{\rm *}$ are
then 2.6$\times 10^{-2}$, 2.6$\times 10^{-3}$, 2.6$\times 10^{-4}$,
and 2.6$\times 10^{-5}$. This way we can directly compare the input
energy radiation in the FUV and X-ray band as well as its effects on
the disk. We also explore two different values for the surface density
power law distribution $\epsilon$, which is defined as $\Sigma(r) =
\Sigma(r_0)\times (r/r_0)^{-\epsilon}$: \citet{Hayashi1981} derived
the value $\epsilon=1.5$ in his model for the minimum mass solar
nebular (MMSN) and their diagnostic value for our own solar system,
while \citet{Hartmann1998} suggested $\epsilon = 1$ for objects older
than 1 Myr.

In summary, the current paper discusses the effects of variations in
the following parameters: X-ray luminosity (five values), FUV luminosity
(four), surface density profile (two), and dust size distribution, varying
both the minimum grain size $a_{min}$ (three) as well as power law indices
(two). This yields a total of 240 models; the summary of the model
parameters is given in Table \ref{ParameterTable}.

\begin{table}[t]
\centering
\caption{Parameters used in the models.}
\begin{tabular}{l|c|c}
\hline
Quantity & Symbol & Value \\
\hline
\hline
Stellar mass       & $M_*$      & 1 $M_{\odot}$ \\
Effective temperature & $T_{\rm eff}$ & 5770~K \\
Stellar luminosity & $L_*$      & 1 $L_{\odot}$ \\
Disk mass          & $M_{\rm disk}$ & 0.01 $M_{\odot}$ \\
X-ray luminosity (0.1-50 keV)& $L_{\rm X}$   & $0,10^{29},10^{30}$\\
&    & $10^{31},10^{32}$~erg~s$^{-1}$ \\
FUV luminosity & $L_{\rm FUV}$ & $10^{29}$, $10^{30}$,\\
&    &  $10^{31}$, $10^{32}$~erg~s$^{-1}$ \\
Inner disk radius  & $r_{\rm in}$   & 0.5 AU \\
Outer disk radius  & $r_{\rm out}$   & 500 AU \\
Surface density power law index& $\epsilon$ & 1.0, 1.5 \\
Dust-to-gas mass ratio & $\rho_d / \rho$ & 0.01 \\
Min. dust particle size & $a_{\rm min}$ & 0.1, 0.3, 1.0 $\rm{ \mu m}$ \\
Max. dust particle size & $a_{\rm max}$ & 10 $\rm{ \mu m}$ \\
Dust size distribution power index& $a_{\rm pow}$ & 2.5, 3.5 \\
Dust material mass density & $\rho_{\rm gr}$ & $2.5$~g~cm$^{-3}$ \\
Strength of incident ISM FUV & $\chi^{\rm ISM}$ & 1 \\
Cosmic ray ionization rate of H$_2$ & $\zeta_{\rm CR}$ & $5\times 10^{-17}$~s$^{-1}$ \\
Abundance of PAHs relative to ISM & $f_{\rm PAH}$ & 0.12 \\
Viscosity parameter & $\alpha$ & 0 \\ 
%X-ray spectral shape & $F(E)$ & $ 1/E\cdot\mathrm{exp}(-E/kT_{\rm X})$ \\[1mm]
\hline
\end{tabular}
\label{ParameterTable}
\end{table}

\begin{figure}
  \centering
  \includegraphics[width=9cm,clip]{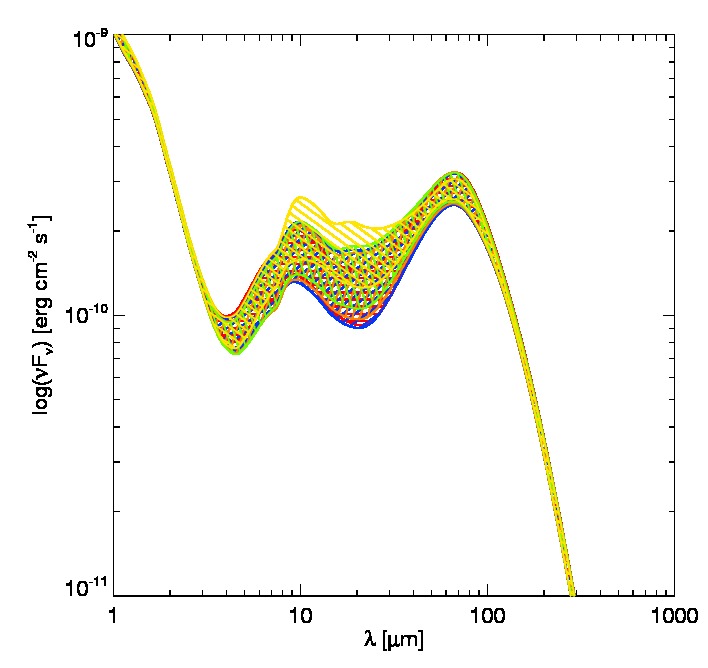}
  \caption{Average spectral energy distributions of models with
    $L_{\rm X}=0$ (red), $10^{29}$ (blue), $10^{30}$ (orange),
    $10^{31}$ (green), and $10^{32}$~erg~s$^{-1}$ (yellow).}
  \label{SED_models}
\end{figure}

\begin{figure*}
  \centering
  \includegraphics[width=4cm,clip]{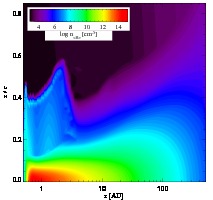}
  \includegraphics[width=4cm,clip]{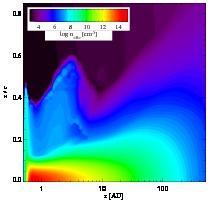}
  \includegraphics[width=4cm,clip]{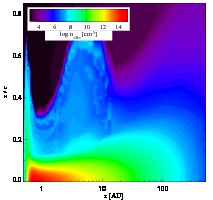}
  \includegraphics[width=4cm,clip]{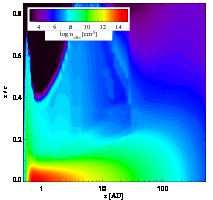}
  \includegraphics[width=4cm,clip]{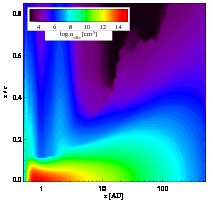}
  \includegraphics[width=4cm,clip]{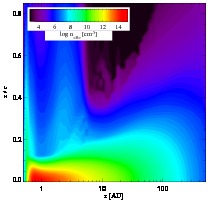}
  \includegraphics[width=4cm,clip]{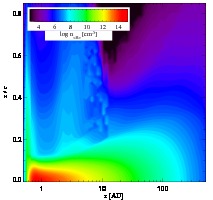}
  \includegraphics[width=4cm,clip]{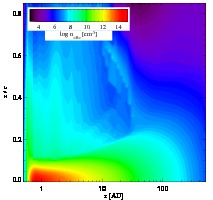}
  \includegraphics[width=4cm,clip]{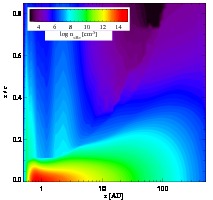}
  \includegraphics[width=4cm,clip]{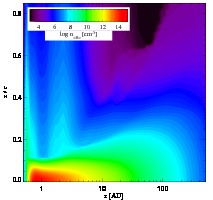}
  \includegraphics[width=4cm,clip]{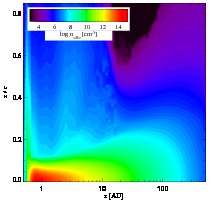}
  \includegraphics[width=4cm,clip]{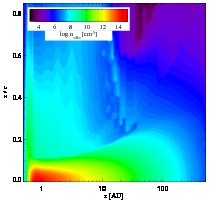}
  \includegraphics[width=4cm,clip]{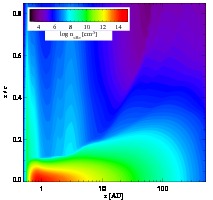}
  \includegraphics[width=4cm,clip]{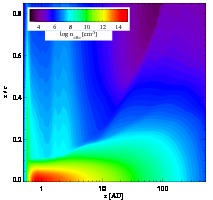}
  \includegraphics[width=4cm,clip]{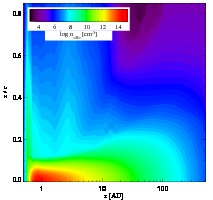}
  \includegraphics[width=4cm,clip]{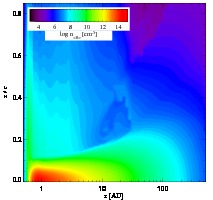}
  \includegraphics[width=4cm,clip]{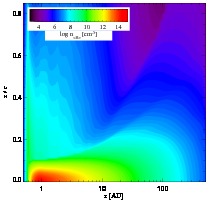}
  \includegraphics[width=4cm,clip]{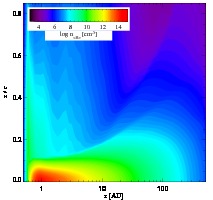}
  \includegraphics[width=4cm,clip]{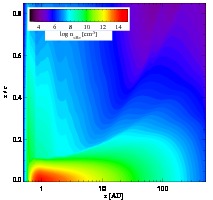}
  \includegraphics[width=4cm,clip]{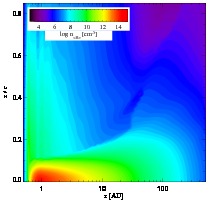}
  \caption{Total density: FUV luminosity increasing from $L_{\rm
      FUV}=10^{29}$ (left) to $10^{32}$~erg/s (right). X-ray luminosity
    increasing from $L_x=0$ (top) to $10^{32}$~erg/s (bottom).}
  \label{model_dens_struct}
\end{figure*}

\begin{figure*}
  \centering
  \includegraphics[width=4cm,clip]{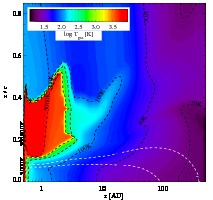}
  \includegraphics[width=4cm,clip]{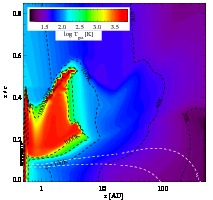}
  \includegraphics[width=4cm,clip]{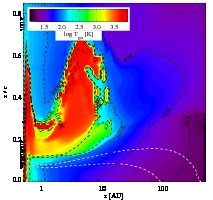}
  \includegraphics[width=4cm,clip]{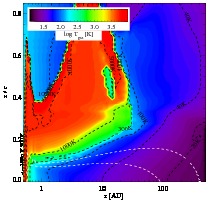}
  \includegraphics[width=4cm,clip]{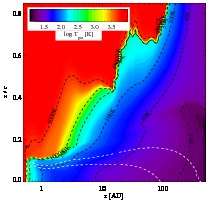}
  \includegraphics[width=4cm,clip]{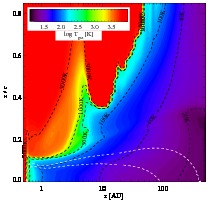}
  \includegraphics[width=4cm,clip]{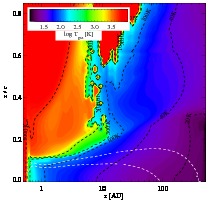}
  \includegraphics[width=4cm,clip]{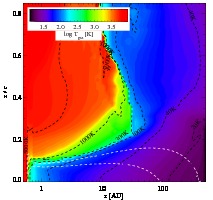}
  \includegraphics[width=4cm,clip]{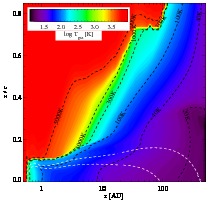}
  \includegraphics[width=4cm,clip]{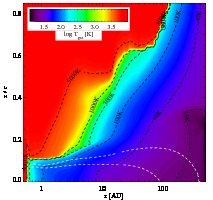}
  \includegraphics[width=4cm,clip]{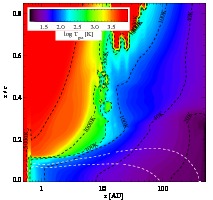}
  \includegraphics[width=4cm,clip]{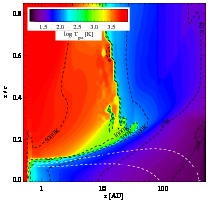}
  \includegraphics[width=4cm,clip]{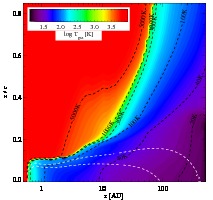}
  \includegraphics[width=4cm,clip]{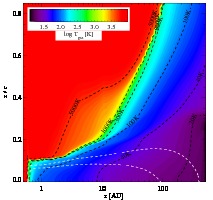}
  \includegraphics[width=4cm,clip]{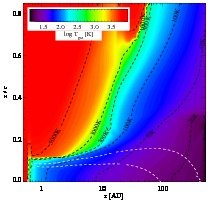}
  \includegraphics[width=4cm,clip]{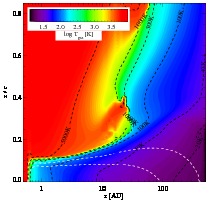}
  \includegraphics[width=4cm,clip]{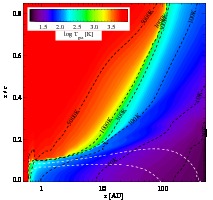}
  \includegraphics[width=4cm,clip]{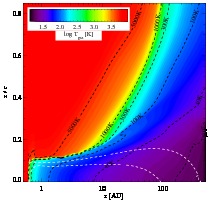}
  \includegraphics[width=4cm,clip]{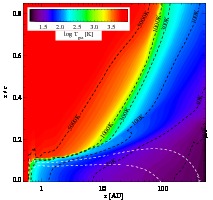}
  \includegraphics[width=4cm,clip]{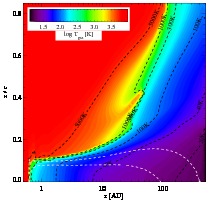}
  \caption{Gas temperature distribution: FUV luminosity increasing
    from $L_{\rm FUV}=10^{29}$ (left) to $10^{32}$~erg/s
    (right). X-ray luminosity increasing from $L_x=0$ (top) to
    $10^{32}$~erg/s (bottom). White contours represent the $A_{\rm
      v}=1$ and 10. The black contours represent the gas temperatures.}
  \label{Gas_temperature}
\end{figure*}

\begin{figure*}
  \centering
  \includegraphics[width=16cm,clip]{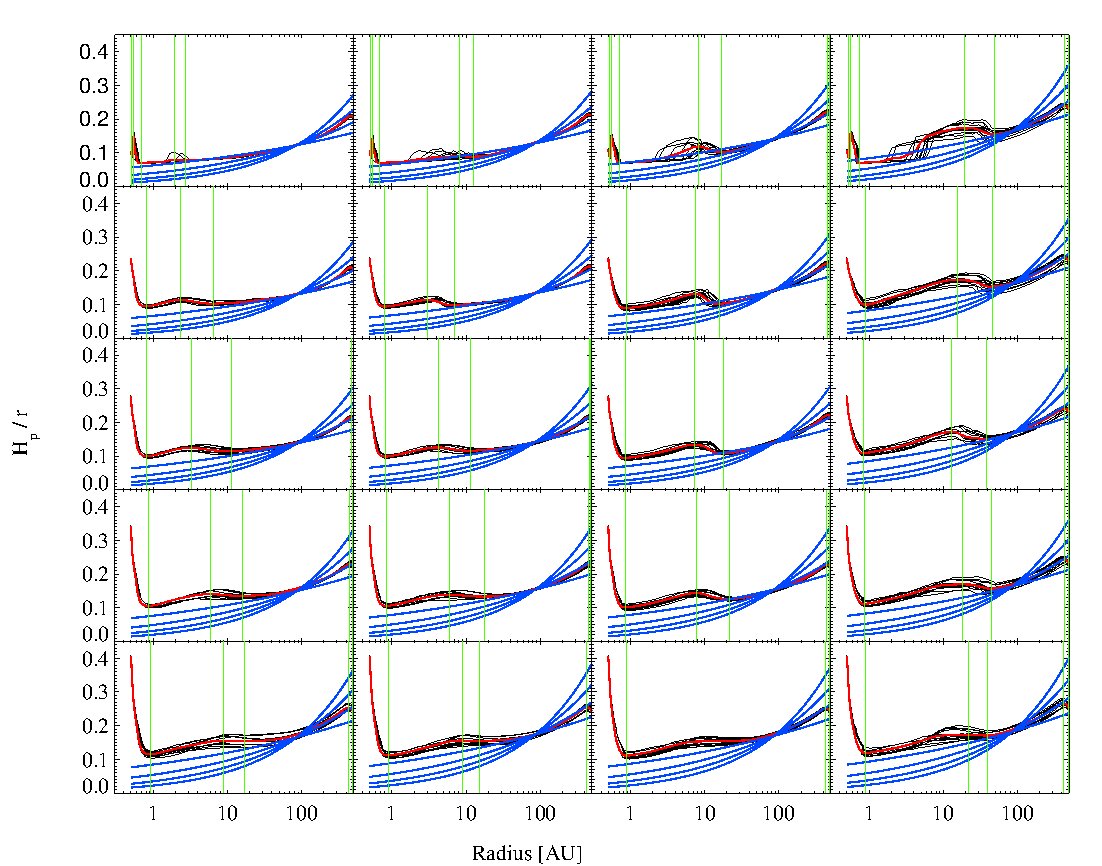}
  \caption{Scale height of the disk at $z/r=0.5$. The black lines show
    the 12 different models ($a_{\rm min}$, $a_{\rm max}$, and
    $\epsilon$), at a particular $L_{FUV}$ and $L_{\rm X}$. The
    vertical green lines indicate a maximum or minimum in the scale
    height. The red line is the average of the 12 models.}
  \label{Scale_height}
\end{figure*}

\section{Disk thermal and density structure}\label{dens_temp}

The coupling of gas and dust to the radiation field determines the
heating and cooling rates, which in return determine the pressure
balance and structure of a disk. FUV and X-rays couple differently to
the gas: FUV is absorbed by dust grains and ejects an electron with a
small surplus of energy in the form of kinetic energy, which then
consequently heats the gas. X-rays, on the other hand, are absorbed by
the K-, L-, or M-shell of atoms, where no distinction is made as to
whether this atom is part of a molecule, dust grain, or PAH. This
assumption might overestimate the X-ray absorption cross section by a
factor of approximately $\sim 2$ at energies $E < 1$~keV, since we do
not include self-shielding effects by large grains \citep[see, e.g.,
Fig. 1 of][]{Fireman1974}. On the other hand, not much is known about
gas and dust phase elemental abundances in disks. The observed
abundance of, e.g., neon (dominating the X-ray absorption cross
section between $\sim 0.9$ to $1.2$~keV, and, as such, the total
absorption of our 1 keV thermal source), already varies over a larger
range \citep[see][for a discussion on this topic]{Glassgold2004}. The
electron that is ejected has a large energy capable of exciting,
ionizing and heating the gas. The efficiency for X-ray heating is much
larger, about 10 to 40 percent, depending on the ionization fraction
of the gas, while that for FUV heating does not exceed one to three
percent. The coupling of X-rays to gas is weaker than for FUV, due to
the smaller cross section for the absorption, but the larger
efficiency makes sure that it is an important heating source, even
when the FUV excess is large. Their relative contribution to the
energy budget will result in a different structure and emitted line
fluxes. X-rays do not much affect the continuum fluxes, certainly less
than $a_{\rm min}$, $a_{\rm pow}$, and $\epsilon$ (see
Fig. \ref{SED_models}). It shows that the fluxes change at most by a
factor of three between $\lambda = 5 - 100$~$\mu$m due to X-rays. The
models with $L_{\rm X}=10^{32}$~erg~s$^{-1}$ are affected most, and
these high X-ray fluxes are rarely observed for T Tauri stars.

\citet{Woitke2009} have already pointed out that the cooling and
heating balance is very important in determining the physical
structure of the disk. They present in their Fig. 9 the density
structures for (1) a model where the gas temperature is decoupled from
the dust and (2) where the gas and dust temperature are coupled. Model
1 shows a very complex structure, where the gas is puffed up at the
inner rim and a second bump in the density is seen at a radius $r\sim
10$~AU. \citet{Aresu2011} extended this model by including X-rays and
showed for the particular model of \citet{Woitke2009} that the
vertical extension inward of 10~AU becomes smoothed out for increasing
X-ray fluxes and merges with the inner rim for the largest X-ray
fluxes. A minimum in the vertical extension remains visible, though.

In Figs. \ref{model_dens_struct} and \ref{Gas_temperature}, the
density and temperature of the disk are shown for FUV luminosities
ranging from $L_{\rm FUV}=10^{29}$, $10^{30}$, $10^{31}$, and
$10^{32}$~erg/s (left to right) and for X-ray luminosities $L_{\rm X}=0$,
$10^{29}$, $10^{30}$, $10^{31}$, and $10^{32}$~erg/s. There are 12
different possible variations with these X-ray and FUV luminosities
(due to the variations in $a_{\rm min}$, $a_{\rm max}$, and
$\epsilon$); as a baseline model, we always show the models with
$a_{\rm min}=0.1$~$\mu$m, $\epsilon=1.5$, and $a_{\rm pow}=3.5$.

\subsection{Density distribution} 

\noindent Because density and temperature structure are directly
related and we assume hydrostatic equilibrium, many of the effects
that are seen in the density structure will show up in the temperature
structure as well. The FUV-only models show that the inner rim becomes
higher and broader for increasing luminosity. The disk right behind
the inner wall is shielded, and a second, vertically more
extended density structure thus appears at radii between $r = 2$ and 6 to
30~AU (depending on the FUV luminosity).

The X-rays have a different effect on the density structure. While the
peak of the second bump in the density structure appears at larger
radii for higher FUV luminosities, the X-rays mostly affect the
region within 5 AU. The second bump behind the inner rim expands to
smaller and smaller radii for increasing X-ray luminosities (while
keeping FUV the same), and the absolute densities become higher (from
$n\sim 10^6$ to $10^8$~cm$^{-3}$). Eventually, the puffed-up inner rim
and the second bump merge. At the highest FUV luminosity, the merging
of the two density structures occurs at the smallest X-ray
luminosity.

\subsection{Temperature structure} 

\noindent When only FUV is included in the models, we find that the
temperature structure of the disk shows an inversion in the vertical
direction: The temperature reaches a maximum at a certain relative
height $z/r$ and shows a fast drop above it. For example, the models
with $L_{\rm FUV} = 10^{29}$~erg/s show a region with temperatures $T
> 1000$~K out to a radius $r\sim 3$~AU and between relative heights
$z/r \sim 0.1$ to 0.4. Noticeable is also that the hot region extends
to a slightly higher relative height at the inner rim (up to
$z/r=0.45$) and at a radius $r=3$~AU (out to $z/r=0.55$). This is the
result of shielding; the disk surface is less heated right behind the
inner rim. For larger FUV luminosities, shielding effects are even
more prominent and the drop in temperature right behind the inner rim
is increasingly pronounced.

Adding X-ray heating changes this picture. These models show a much
more extended region with temperatures higher than $T > 1000$~K, even
when the X-ray luminosity is much smaller than the FUV luminosity. At
radii $r < 5 - 10$~AU, the inversion layer disappears and the
temperature just smoothly increases toward larger $z/r$. At larger
radii ($r > 10 - 20$~AU), the inversion layer only disappears when the
X-ray luminosity is at least ten percent of the FUV luminosity. In
models where the X-ray luminosity is much smaller, temperature
distribution is also affected (or even dominated) by the FUV. An
example is $L_{\rm FUV} = 10^{32}$ erg/s in combination with $L_{\rm
  X}=10^{29}$ erg/s. Although the inversion layer disappears at small
radii ($r < 5$~AU), it is still present at larger radii.

\subsection{Heating and cooling processes}

Temperature and density structure are in the end determined by the
balance between the heating and cooling processes. Line cooling
processes (except Lyman $\alpha$ and [OI] 6300\AA\ line cooling) are
treated in an escape probability approximation. Before solving the
equations for statistical equilibrium, a global continuum radiation
transfer calculation is done to estimate the background mean
intensities for the radiative excitation and de-excitation
rates. Other heating and cooling processes are approximated by an
analytical expression. Over 50 different heating and 50 different
cooling processes are included in the code. Except for the treatment of
X-ray related heating processes, which are described in
\citet{Aresu2011}, we refer to \citet[][]{Woitke2009,Woitke2011} for a
more detailed description. The reason that so many processes are
included is that they each play a significant role, depending on the
ambient density and radiation field (FUV, X-rays, cosmic rays), which
vary over many orders of magnitude. It is beyond the scope of this
paper to describe all these processes in detail for the
models. However, we do present the dominant heating and cooling processes
in Figs. \ref{heating} and \ref{cooling} for the models for which we
are showing the temperature and density structure.

First, we consider the FUV-only models. At the lowest FUV luminosity
$L_{FUV}=10^{29}$~erg~s$^{-1}$, there are three main heating process
in the unattenuated part of the disk, namely, background heating by
C$^{+}$ at the lowest densities, $n_{H}=10^{5}$~cm$^{-3}$ (dark blue),
PAH heating within the inner rim and the second extension of the disk
(orange), and heating by carbon ionization in the outer disk
(blue-green). Increasing the FUV field makes the picture more
complicated. Heating by collisional de-excitation of H$_2$ becomes
important in the second bump (light blue), while background heating by
FeII is dominant at the inner rim. Locally, other processes, such as
photo-electric heating, are important. Going to the more shielded
regions of the disk, there is roughly a three-layered structure of
dominating heating processes: (1) infrared background heating CO
ro-vibrational transitions (black), (2) slightly deeper in the disk
heating by thermal accomodation on grains (white), and (3) in the
mid-plane cosmic ray heating (red). This three-layered structure
becomes more confined toward the mid-plane for higher FUV
luminosities.

X-ray Coulomb heating dominates in the upper part of the disk, when
the FUV to X-ray luminosity $L_{\rm FUV} / L_{\rm X} \leq 1$. When it
is higher than one, X-ray Coulomb heating is restricted to the higher
regions of the disk. Adding X-rays increases the temperature
significantly in the disk, making [FeII] also dominant in large parts
of the second-density extension at the lowest X-ray luminosity,
$L_X=10^{29}$~erg~s$^{-1}$. For large X-ray luminosities, the
structure of heating sources becomes less complicated. The entire
unshielded part of the disk is dominated by X-ray Coulomb heating,
followed by the three-layered structure described above (CO ro-vib,
thermal accommodiation on grains, cosmic ray). For the highest two
X-ray fluxes, an additional layer with X-ray H$_2$ dissociation
heating as the main heating source is located on top of this
three-layered structure.

In the FUV-only case, there are three main coolants in the
unattenuated part of the disk: C$^+$ line cooling (yellow), [FeII]
line cooling (blue-green), and [OI] line cooling. The size of the
region, where [FeII] line cooling dominates expands for higher FUV
luminosities, while reducing the size of region where [OI] line
cooling dominates. When X-rays are added, the region where C$^+$ line
cooling dominates is pushed to the outer part of the
disk. Temperatures are much higher in the upper part of the disk, and
as a result Lyman $\alpha$ cooling (black) dominates in increasingly
larger regions of disk, when X-ray luminosities become larger. At the
highest X-ray luminosities, we find a three-layered structure of Lyman
$\alpha$ cooling, [FeII] line cooling, and [OI] line cooling. The
shielded region of the disk shows a layered structure. CO
ro-vibrational (red) and H$_2$O rotational cooling (green-blue) are on
top. Closer to mid-plane, several smaller regions have their own major
coolant, such as HCN (purple), and HNC (orange).

\subsection{Vertical scale height} 

\noindent Another way to show the combined effects of X-rays and FUV on
the disk structure is by comparing the scale vertical height. Our
definition is the same as the one used by \citet{Woitke2009}, and is
approximately (i.e., assuming $z \ll r$) given by

\begin{equation}
\left(\frac{H}{r}\right)^2 \simeq \frac{2rc_T}{ G M_\star},
\end{equation}

\noindent where $H$ is defined as $\rho(z)\approx\rho(0)
\exp(−z^2/H^2)$ and $c_T$ is the isothermal sound speed. As mostly the
disk atmosphere is affected by the FUV and X-ray irradiation, we plot
$H/r$ at the relative height $z/r=0.5$ in Fig. \ref{Scale_height}. In
each panel, we overplot the scale height for all the different
parameters in black (twelve models in total), and the red line is the
average. Four blue lines are overplotted, indicating flaring index
$H/r\propto r^p$ with $p = 0.15 - 0.45$, where the lines are
normalized to the vertical scale height at $r=100$~AU.

A number of things stand out in these plots: (i) The vertical scale
height of the inner rim is mostly affected by the X-rays, and almost
no response is seen for different FUV luminosities. The scale height
is $H/r=0.15$ without X-rays at the inner rim, and it increases to
$\sim0.4$ at $L_X = 10^{32}$~erg/s. (ii) The second bump also shows up
when plotting $H/r$ and slowly moves outward for increasing X-ray
luminosities. (iii) The flaring index $p$ is 0.25 in the outer disk
($r > 100$~AU), similar to the $2/7$ exponent found by
\citet{Chiang1997}. The vertical scale height in the inner disk is
elevated, compared to the outer disk. (iv) The $H_0 / r \approx 0.1$
and $0.18$ at $r_0 =100$~AU and $z/r=0.1$ and 0.5, respectively. For a
direct comparison to \citet{Chiang1997}, we measure $H_0/r$ at
$r_0=10$~AU, because the approximation of the disk dust temperature
being vertically isothermal does not hold in our models at the larger
distances. We find a value $H_0 / r \approx 0.04$, which is very
similar to \citet{Chiang1997}, after correcting for the stellar mass
and their definition of the scale height. (v) The merging of the inner
rim with the second bump is clearly visible, as changes in the
vertical scale height in the inner disk ($r=1-10$~AU) become more
gradual for larger X-ray luminosities.

\section{The chemical balance in the disk}\label{chemistry}

The combined effects of X-rays and FUV irradiation on the disk are
discussed in this section. The density structure is in pressure
equilibrium as discussed in the previous section and is altered due to
irradiation effects. The chemical rates are density dependent, and as
such the pressure balance also influences the abundances of the
species.

\begin{figure*}
  \centering
  \includegraphics[width=4cm,clip]{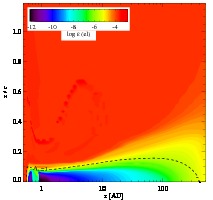}
  \includegraphics[width=4cm,clip]{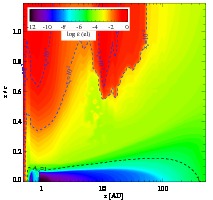}
  \includegraphics[width=4cm,clip]{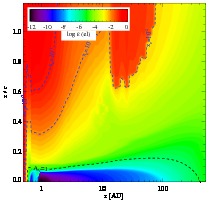}
  \includegraphics[width=4cm,clip]{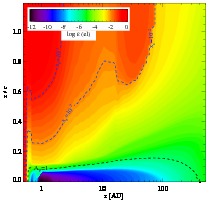}
  \caption{Electron abundance: FUV luminosity is fixed at
    $10^{31}$~erg/s. X-ray luminosity increases from $L_x=0$ (left) to
    $10^{32}$~erg/s (right). The black dashed line indicates $A_{\rm
      V} = 1$. The blue lines mark the ionization fractions
    $x_e=10^{-2}$ and 10$^{-1}$.}
  \label{model_electron_abundance}
\end{figure*}

\begin{figure*}
  \centering
  \includegraphics[width=4cm,clip]{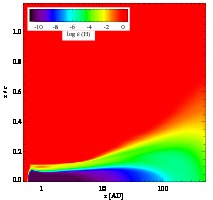}
  \includegraphics[width=4cm,clip]{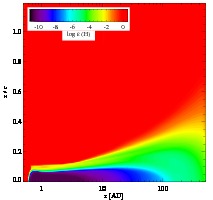}
  \includegraphics[width=4cm,clip]{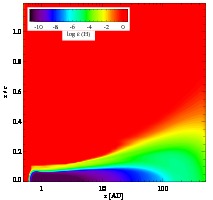}
  \includegraphics[width=4cm,clip]{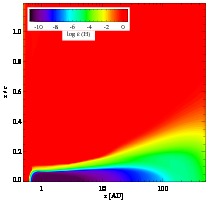}
  \includegraphics[width=4cm,clip]{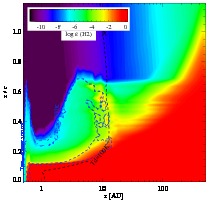}
  \includegraphics[width=4cm,clip]{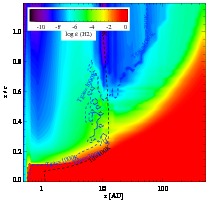}
  \includegraphics[width=4cm,clip]{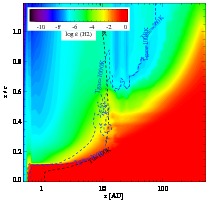}
  \includegraphics[width=4cm,clip]{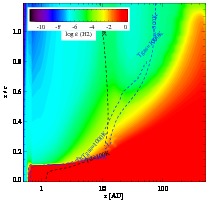}
  \caption{Abundance structure of atomic (top) and molecular (bottom)
    hydrogen. FUV and X-ray fluxes are the same as in
    Fig. \ref{model_electron_abundance}. The blue contours indicate
    the gas temperature $T_{\rm gas} = 300$ and 1000~K. The black line
    indicates $T_{\rm dust}=100$~K.}
  \label{hydrogen_abundance_struct}
\end{figure*}

\subsection{Electron abundances}\label{electron_abundance_section} 

\noindent The main electron donor in a FUV-only chemistry is atomic
carbon, because the incident FUV radiation field is not capable of
ionizing atomic hydrogen, leaving cosmic rays as the only ionization
source of atomic hydrogen. As a result, the maximum electron abundance
will not be higher than $x_{\rm e} \sim 10^{-4} - 10^{-3}$, as can be
seen in Figs. \ref{model_electron_abundance} and
Fig. \ref{model_electron_abundance_appendix}. The maximum electron
abundances occur where the disk is unshielded to the radiation
source. The inner rim has significant ionization fractions ($x_{\rm
  e}\sim 10^{-4} - 10^{-3}$) all the way down to the midplane of the
disk, while further out the electron abundances are only this high at
densities $n < 10^7$~cm$^{-3}$. The drop of the electron abundances
below $x_{\rm e} < 10^{-6}-10^{-5}$ for higher densities nicely
coincides with the $A_V=1$. Closer to the midplane, when the radiation
becomes more shielded and densities are higher, the electron abundance
exhibits a fast drop. This is the result of a combination of a
decreasing ionization rate, $\zeta \propto n$, and an increasing
recombination rate, $k_{\rm rec} \propto n^2$. The abundance drop
becomes more gradual at larger radii, which is a result of lower
ambient densities and thus lower recombination rates. The region where
the transition occurs is moving slightly toward smaller relative
heights, when FUV fluxes are higher. The change in the transition
location is not much due to the sharp increase of the density toward
the midplane ($k_{rec} \propto n^2$): A relatively small increase in
the density already compensates for larger FUV fluxes.

This picture changes quite significantly when the radiation fields
also contain X-rays. The absorption of an X-ray photon by an arbitrary
species creates a fast electron that can produce many ionizations,
e.g., a 1 keV electron is able to produce $\sim 27$ hydrogen
ionizations. As a result, the electron abundances are much higher when
X-rays are included. In the unattenuated part of the disk, ionization
fractions are of the order $x_e\sim 10^{-2}-10^{-1}$ (contours for
$x_e=10^{-2}$ and $10^{-1}$ are indicated). They can even be higher
than $x_{\rm e} > 10^{-1}$ at very small radii ($r < 0.5$~AU) and high
relative heights ($z/r \gtrsim 0.5$). Here, low densities ($n_{\rm H}
< 10^7$~cm$^{-3}$) and high temperatures ($T \gg 5000$~K) reduce the
recombination rates. Recombination rates decrease with temperature up
to $T\sim 10000$~K because they are dominated by radiative
recombination, which scales as $k_{\rm rec}\propto n_e n_i
(T/10^4)^{-X_{\rm rad}}$, with $X_{\rm rad}\sim 0.6-0.9$. We find
slightly higher ionization fractions than, e.g.,
\citet{Glassgold2004}. Their calculation at $r=1$~AU (see their
Fig. 4) is truncated at vertical column density $N_{\rm H} \approx
2\times 10^{18}$~cm$^{-2}$ and density $n_{\rm H} \approx
10^{7}$~cm$^{-3}$ . \citet{Ercolano2008} (their Figs. 1 and 2) show
calculations for lower column densities and densities, and they find
ionization fractions slightly higher than $x_e > 1$ at 0.07~AU at
vertical column densities $N_{\rm H} < 10^{16}$~cm$^{-2}$. Both
calculations are consistent with those presented here. The electron
abundance structure looks a little counterintuitive, especially for
the lowest two X-ray fluxes $L_{\rm X} = 10^{29}$ and
$10^{30}$~erg/s. The region with very high ($x_{\rm e} \approx 10^{-2}
- 10^{-1}$) becomes smaller for increasing FUV luminosities
(Fig. \ref{model_electron_abundance_appendix}). This is because the
inner rim is puffed up more and even merged with the second bump for
the highest FUV fluxes. The densities are elevated to $n_{\rm
  H}=10^8$~cm$^{-3}$ in these regions, and the recombination rates are
orders of magnitude higher as a result. The outer disk is shielded
from radiation. The regions with very high ionization fractions
($x_{\rm e} > 10^{-2}$) are of similar size for all FUV luminosities only
for the highest X-ray luminosity (see
Fig. \ref{model_electron_abundance_appendix}).

When the ionization fraction is increased by orders of magnitude, the
gas shows a radically different chemistry, as it will be dominated by
ion-molecule reactions. This will be discussed for some of the key
species, and water in particular.

\begin{figure*}
  \centering
  \includegraphics[width=4cm,clip]{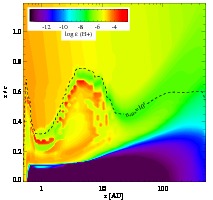}
  \includegraphics[width=4cm,clip]{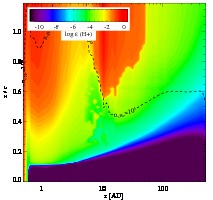}
  \includegraphics[width=4cm,clip]{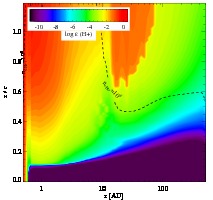}
  \includegraphics[width=4cm,clip]{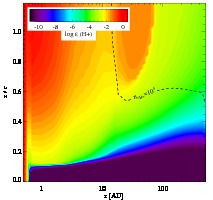}
  \includegraphics[width=4cm,clip]{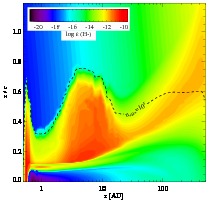}
  \includegraphics[width=4cm,clip]{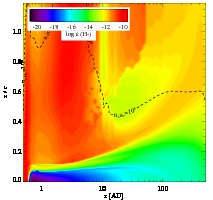}
  \includegraphics[width=4cm,clip]{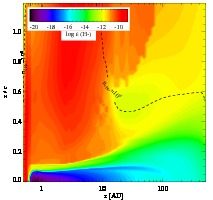}
  \includegraphics[width=4cm,clip]{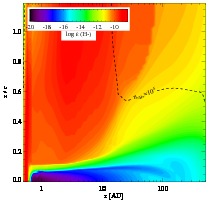}
  \includegraphics[width=4cm,clip]{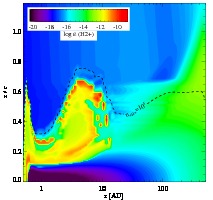}
  \includegraphics[width=4cm,clip]{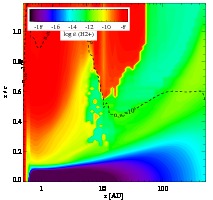}
  \includegraphics[width=4cm,clip]{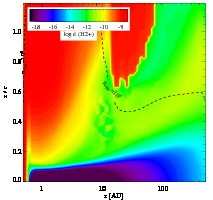}
  \includegraphics[width=4cm,clip]{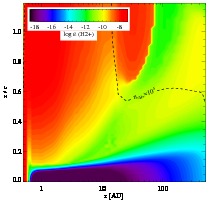}
  \includegraphics[width=4cm,clip]{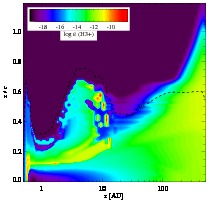}
  \includegraphics[width=4cm,clip]{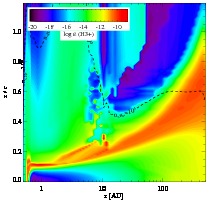}
  \includegraphics[width=4cm,clip]{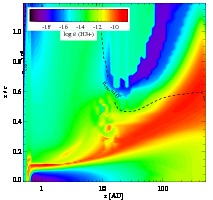}
  \includegraphics[width=4cm,clip]{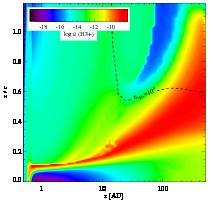}
  \caption{Abundance structure of H$^+$, H$^-$, H$_2^+$, and
    H$_3^{+}$. FUV and X-ray fluxes are the same as in
    Fig. \ref{model_electron_abundance}. Black contour indicates $n_{\rm
      H}=10^5$~cm$^{-3}$.}
  \label{ionic_hydrogen_abundance_struct}
\end{figure*}

\subsection{H and H$_2$ abundances} 

\noindent The abundance structure of atomic hydrogen is shown in the
top panel of Fig. \ref{hydrogen_abundance_struct} (and also in
Figs. \ref{Habundance_struct_appendix} and
\ref{H2abundance_struct_appendix}). The highest abundances ($x_{\rm
  H}\gtrsim 10^{-2}$) are obtained in the unshielded region of the
disk, where FUV and X-rays have high fluxes. At radii $r < 0.6$~AU,
the disk is atomic all the way down to the mid-plane of the disk, as
it is directly exposed to the central source. At larger radii, the
atomic fraction drops below $x_{\rm H} \sim 10^{-2}$ and makes a
transition to molecular hydrogen at a relative height of $z/r \sim
0.1$ between $r \sim 0.6 - 5$~AU. The transition occurs at
increasingly larger relative heights at larger distances to the
central source, and it is at approximately $z/r \sim 0.4 - 0.5$
(depending on the X-ray flux) at a distance $r \sim 100$~AU. The
transition occurs at slightly smaller relative heights at larger X-ray
fluxes.

The formation of H$_2$ on grains is extremely efficient ($\sim 1$)
when the dust temperatures are not high ($T_{\rm dust} \sim 10 - 50$~K
\citep{Cazaux2004}. At dust temperatures higher than $T_{\rm dust} >
100$~K, the H$_2$ efficiency drops rapidly and H$_2$ formation on dust
does not occur at $T_{\rm dust} > 1000$~K. H$_2$ is also formed
through the H$^-$ route, $\rm H^- + H \rightarrow H_2 + e^-$,
especially when X-rays are present, and the fractional abundance of
electrons is relatively high (and thus also H$^-$). It is possible to
maintain an efficient route to form molecular hydrogen in the gas
phase at high temperatures, $T_{\rm gas} > 300$~K and $T_{\rm dust} >
100$~K (see contours in Fig. \ref{hydrogen_abundance_struct}). Also
FUV is much more efficient in destroying H$_2$, than X-rays. The FUV
photo-dissociates H$_2$ ($\rm H_2 +$ FUV photon $\rightarrow \rm H_2^*
\rightarrow H + H$). X-rays predominantly ionize molecular hydrogen
indirectly by collisions with fast electrons produced after an X-ray
absorption ($\rm H_2 + e^{-*} \rightarrow H_2^+ + e^- + e^{-*}$),
while only a small fraction of H$_2$ is dissociated in this
process. After ionization, H$_2$ is able to reform H$_2$ through $\rm
H_2^+ + H \rightarrow H_2 + H^+$, but it will also be able to form
H$_3^+$ through the reaction $\rm H_2^+ + H_2 \rightarrow H_3^+ +
H$. This particular species is key in forming molecules through
ion-molecule reactions. Because X-rays are destroying H$_2$ less
efficiently and can provide an environment to form H$_2$ in the
gas-phase, a fast ion-neutral chemistry results, which makes it
possible to form molecules (and maintain significant abundances) at
much higher temperatures than when only FUV is present.

The bottom panel of Fig. \ref{hydrogen_abundance_struct} presents the
abundances of H$_2$. It shows that the H$_2$ abundance in the warm
atmosphere of the disk is largest for the highest X-ray fluxes (as
high as $x_{\rm H_2} \sim 10^{-6}$ in the region directly exposed to
the central source at $r = 0.6$~AU). This is the result not only of
the processes described above, but also of the scale height
of the inner rim, which increases with X-ray luminosities, an effect
described in the previous section.

\begin{figure*}
  \centering
  \includegraphics[width=4cm,clip]{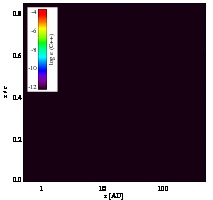}
  \includegraphics[width=4cm,clip]{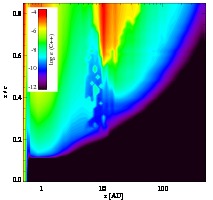}
  \includegraphics[width=4cm,clip]{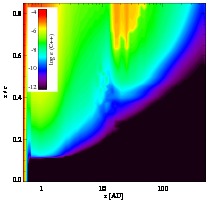}
  \includegraphics[width=4cm,clip]{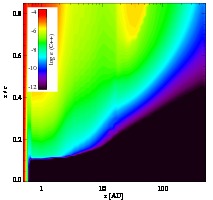}
  \includegraphics[width=4cm,clip]{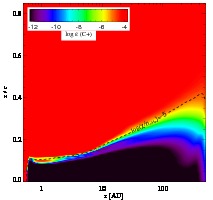}
  \includegraphics[width=4cm,clip]{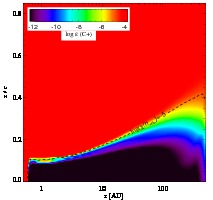}
  \includegraphics[width=4cm,clip]{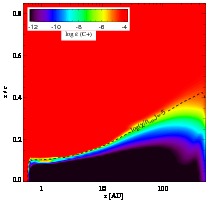}
  \includegraphics[width=4cm,clip]{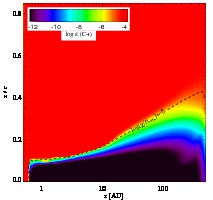}
  \includegraphics[width=4cm,clip]{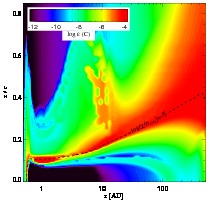}
  \includegraphics[width=4cm,clip]{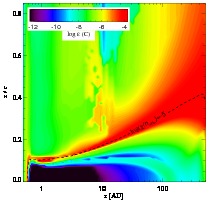}
  \includegraphics[width=4cm,clip]{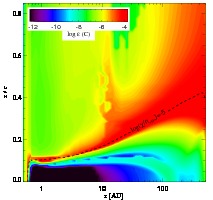}
  \includegraphics[width=4cm,clip]{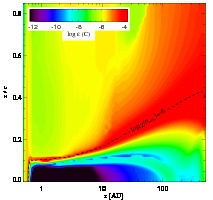}
  \includegraphics[width=4cm,clip]{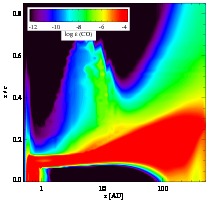}
  \includegraphics[width=4cm,clip]{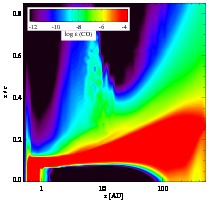}
  \includegraphics[width=4cm,clip]{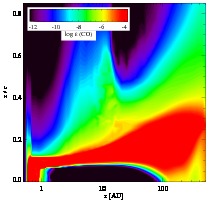}
  \includegraphics[width=4cm,clip]{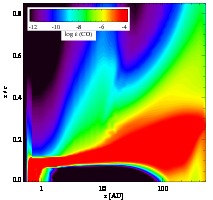}
  \caption{C$^{2+}$, C$^{+}$, C, CO abundances. The black dashed line
    indicates $\log(\chi/n_{\rm H})=-5$. Fluxes are the same as
    Fig. \ref{model_electron_abundance}.}
  \label{carbon_abundance_struct}
\end{figure*}

\begin{figure*}
  \centering
  \includegraphics[width=4cm,clip]{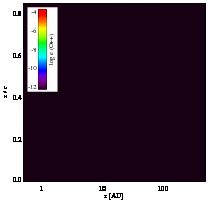}
  \includegraphics[width=4cm,clip]{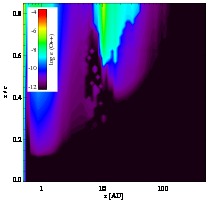}
  \includegraphics[width=4cm,clip]{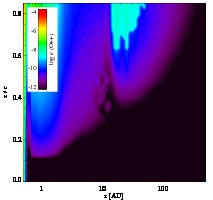}
  \includegraphics[width=4cm,clip]{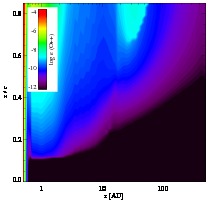}
  \includegraphics[width=4cm,clip]{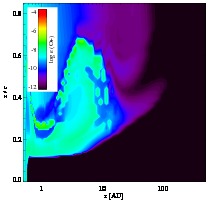}
  \includegraphics[width=4cm,clip]{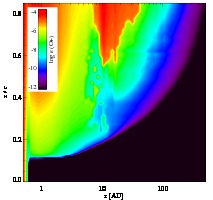}
  \includegraphics[width=4cm,clip]{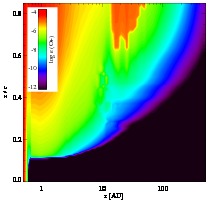}
  \includegraphics[width=4cm,clip]{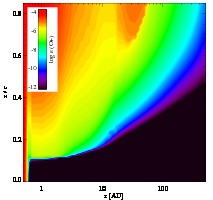}
  \includegraphics[width=4cm,clip]{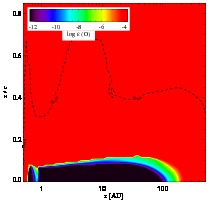}
  \includegraphics[width=4cm,clip]{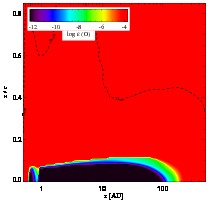}
  \includegraphics[width=4cm,clip]{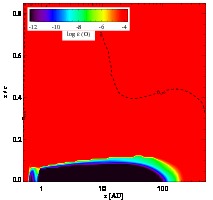}
  \includegraphics[width=4cm,clip]{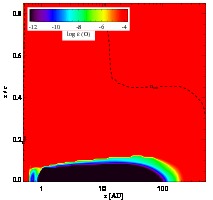}
  \caption{O$^{2+}$, O$^+$, and O abundances. Fluxes are the same as
    in Fig. \ref{model_electron_abundance}. The black contour in the
    lower panel indicates the critical density of the [OI] 63~$\mu$m
    line.}
  \label{oxygen_abundance_struct}
\end{figure*}

\subsection{H$^+$, H$^-$, H$_2^+$, and H$_3^+$ abundances} 

\noindent The main drivers of the chemistry in the atmospheres of
disks exposed to irradiation are ionic species, starting with H$^+$,
H$_2^+$, and H$_3^+$ (see also the discussion of the water
chemistry). The H$^-$ abundance structure is shown for completeness,
since it can be important in the production of H$_2$ (see
Fig. \ref{ionic_hydrogen_abundance_struct} and
Figs. \ref{H+abundance_struct_appendix} to
\ref{H3+abundance_struct_appendix}).

The abundance structure of ionized hydrogen shows abundances as high
as $x_{\rm H^+} \sim 10^{-3} - 10^{-2}$, even if there is only FUV
irradiation. FUV is not able to photo-ionize atomic hydrogen, and
hydrogen ionization is solely done by cosmic rays. This only happens
at very low densities ($n < 10^{4}$~cm$^{-3}$), though. Once densities
of the order of $n \sim 10^5$~cm$^{-3}$ are reached, the ionized
hydrogen fractions are closer to $x_{\rm e} \sim 10^{-5} - 10^{-4}$ or
smaller. The ionization balance is regulated by the charge exchange
reaction $\rm H + O^+ \leftrightarrow H^+ + O$. The backward reaction
has an energy barrier of $T = 227$~K, so the ratio of O$^+$ with
respect to H$^+$ decreases with temperature. The charge exchange rates
are much faster than the photo-ionization reaction or the
recombination rates of the species. The amount of H$^+$ is thus
directly coupled to the production rate of O$^+$, which is also
produced by cosmic ray ionization. Even when X-ray ionization
determines the ionization balance of the gas, the charge exchange
rates are much faster than the primary and secondary X-ray ionization
rates of the species, and the ion abundances are directly coupled.
X-rays produce much higher fractions of H$^+$ (as high as $x_{\rm H+}
> 0.1$ at the lowest densities). The abundance structure strongly
resembles that of the electrons, although the H$^+$ abundance drops
faster closer to the mid-plane, and other species such as Na$^+$ take
over as electron donor \citep[see also][]{Adamkovics2011}.
 
Once the gas gets more shielded to the radiation, it has higher
abundances of molecular species (see, e.g.,
Fig. \ref{H2abundance_struct_appendix}). In those regions, H$^+$ can
also be produced by other reaction paths, such as $\rm CO^+ + H
\rightarrow CO + H^+$ when only FUV is present and secondary X-ray
ionization of H$_2$ when X-rays are present as well. Once cosmic ray
ionization is the dominant source of ionization, H$^+$ is produced by
cosmic ray ionization of H$_2$ or charge exchange with He$^+$.

The H$^-$ ion is produced by cosmic ray ionization ($\rm H_2 + CR
\rightarrow H^+ + H^-$) or by radiative recombination (${\rm H} + {\rm
  e^{-}} \rightarrow {\rm H}^- + h\nu$). This last reaction is fairly
slow at low temperatures due to temperature barriers, but it becomes
important in the warm part of the atmosphere of the disk. When only
FUV is irradiating the disk, the H$^-$ abundance is high ($x_{\rm
  H^-}\sim 10^{-11} - 10^{-10}$) at the inner rim and in the second
bump extending out of the disk, and to a lesser extent in the
transition zone from atomic to molecular gas. There are two reasons
that there is not as much H$^-$ at high altitudes in the disk. The
first one is that the electron densities are two to three orders of
magnitude lower. The second one is that the ambient temperatures are
between 50 and 100~K, which is not very favorable as the reaction rate
contains an energy barrier. At the inner rim and in the second bump,
both temperatures ($T > 1000$~K) and densities ($n > 10^7$~cm$^{-3}$,
and thus also electron densities) are much higher. When X-ray
irradiation plays a role, the abundances structure changes a lot. The
temperatures are in excess of $T > 1000$~K to large relative heights
(see Fig. \ref{Gas_temperature}), and electron abundances are much
higher (due to higher ionization rates and slower recombination
rates). Especially when X-ray luminosities exceed $L_X >
10^{30}$~erg/s, the relative abundances of H$^{-}$ are one to three
orders of magnitude higher (close to $x_{\rm H^-} \sim 10^{-8}$). In
the FUV case, the H$^{-}$ route to form H$_2$ is at least three orders
of magnitude lower than the formation route on dust grains. It becomes
a significant contributor ($\sim 30\%$) to the total H$_2$ formation
rate at the highest X-ray luminosities when these extreme ($x_{\rm
  H^-} \sim 10^{-8}$) abundances are reached, and at the same time,
H$_2$ formation on dust is quenched by high dust temperatures.

H$_2^+$ is formed by cosmic ray ionization ($\rm H_2 + CR \rightarrow
H_2^+ + e^-$) in the regions where the gas is molecular and shielded
from FUV and X-ray radiation. When the H$_2$ abundance is high
($x_{\rm H_2}\sim 0.5$), however, the H$_2^+$ is quickly transformed
to H$_3^+$ ($\rm H_2^+ + H_2 \rightarrow H_3^+ + H$). It is above this
layer, that the H$_2^+$ abundance is increasing. Here, the gas is more
exposed to radiation (FUV or X-rays). If X-rays are present, H$_2$ is
ionized through secondary ionizations. In the FUV-only case, where we
do not have this last reaction, other routes contribute to the
production of H$_2^+$. One example is $\rm S^+ + H_2 \rightarrow H_2^+
+ S$. In the unshielded regions of the disk, where H$_2$ abundances
are low ($x_{\rm H_2} < 10^{-6}$), the only efficient way to form
H$_2^+$ is through radiative association ${\rm H^+} + {\rm H}
\rightarrow {\rm H_2^+} + h\nu$. However, this is only efficient at
relatively high temperatures ($T > 5000$~K) and significant abundance
levels of H$^+$. In Fig. \ref{ionic_hydrogen_abundance_struct}, the
cation H$^+$ is shown to be more abundant by orders of magnitude when
X-rays are present. It is obvious that the production of H$_2^+ $ is
much more efficient as well, and hence abundances $x_{\rm H_2^+} \sim
10^{-8} - 10^{-7}$ occur in the high warm atmosphere of the disk. In
the FUV-only case, there is only a significant abundance ($x_{\rm
  H_2^+} > 10^{-12}$) in the inner rim and the second bump.

The dominant way to form H$_3^+$ is through the ion-molecule reaction
$\rm H_2^+ + H_2 \rightarrow H_3^+ + H$. As a result, the H/H$_2$
transition layer and the region below is very suitable to produce
large abundances: H$_2$ is present in reasonable amounts ($x_{\rm H_2}
> 10^{-5} - 10^{-3}$, see Fig. \ref{hydrogen_abundance_struct}), and
radiation is available to produce H$_2^+$ (see previous paragraph, and
Fig. \ref{ionic_hydrogen_abundance_struct}). Once the H$_2^+$ drops
below an abundance $x_{\rm H_2^+} < 10^{-12}$, the H$_3^+$ abundance
also drops to values below $x_{\rm H_3^+} < 10^{-11}$. Consequently,
there is a distinct layer where H$_3^+$ is formed in the most optimal
way. We also saw that the H$_2^+$ is more efficiently formed when
X-rays are abundantly present in the disk. The H$_3^+$ abundances thus
also significantly increase for larger X-ray luminosities and can be
as high as $x_{\rm H_3^+} \gtrsim 10^{-8}$. In the FUV-only case, the
size of the region where H$_3^+$ is present in large abundances is
smaller, and the maximum abundance is at least an order of magnitude
lower compared to the models that include X-rays. It is a key species
in driving ion-molecule chemistry, and so the chemistry with and
without X-rays will obviously yield significantly different abundance
structures.

\subsection{C$^{2+}$, C$^+$, C and CO abundances}\label{carbon_sub}

\noindent The ionization potentials ($IP$) of the species C and C$^+$
are $IP=11.26$ and 24.38 eV, respectively. While FUV photons can
ionize neutral carbon, X-rays (or collisions with the resulting fast
electrons) and/or cosmic rays are needed to reach a higher degree of
ionization. As a result, the abundance patterns of the two species
C$^+$ and C$^{2+}$ respond radically differently to the incident
radiation field on the disk. The C$^{2+}$ abundance pattern is, of
course, strongly correlated with the X-ray luminosity, and C$^{2+}$ is
absent in models without X-rays. The region with significant C$^{2+}$
abundances ($x_{\rm C^{2+}}>10^{-9} - 10^{-8}$) extends to larger
radii and smaller relative heights for higher X-ray luminosities,
simply because there is more ionizing radiation available (and not
inhibited by dust, since the dominant opacity is caused by the
gas). However, the FUV plays an important role in shaping the disk and
therefore indirectly affects the abundance distribution as well. The
bump in the total H number density (see Fig. \ref{model_dens_struct})
extends to larger relative heights for larger FUV luminosities,
thereby increasing the density at larger relative heights. This causes
a larger attenuating column between the central source and the outer
part of the disk, thus reducing the amount of ionizing
photons. Furthermore, it allows H$_2$ to form at smaller radii. H$_2$
is very efficient in reducing C$^{2+}$ by charge exchange reactions,
such as $\rm C^{2+} + H_2 \rightarrow C^+ + H$. As a result, the FUV
luminosity confines the C$^{2+}$ to smaller radii (see
Fig. \ref{C2+abundance_struct_appendix}). C$^+$ is present throughout
the unattenuated part of the disk. It has an abundance $x_{\rm C^+}
\sim 10^{-4}$ down to the mid-plane in the inner rim ($r <
0.6$~AU). Beyond this radius, carbon is only significantly ionized
above a relative height $z/r\sim 0.1$ (at $r \sim 0.6$~AU), increasing
to $z/r\sim 0.4$ at $r\sim 200$~AU. The width $\Delta z/r$, over which
the transition from C$^+$ to C and CO occurs, is larger toward the
outer regions of the disk. Absolute densities of C$^+$ and e$^-$ in
the inner disk are larger, and recombination rates scale with
$n^2$. As a result, the transition from ionized to neutral carbon
occurs abruptly, similar to the abundance drop of electrons (at
approximately $z/r \sim 0.01$ at $r=1$~AU). However, in the outer disk
($r = 200$~AU), the transition stretches out from $z/r \sim 0.04$ to
0.01. Although there are variations of a factor a few in the absolute
abundance of C$^+$, the overall appearance of the abundance structure
is not significantly affected for different values of the X-ray and
FUV luminosity (see Fig. \ref{C+abundance_struct_appendix}).

The neutral carbon abundance pattern is very much affected by
variations in the X-ray and FUV luminosities. In the case where only
FUV is irradiating the disk, one finds a clear transition from C$^+$
to C to CO (see Fig. \ref{carbon_abundance_struct}), which is expected
in a FUV-dominated PDR \citep[cf. ][]{Hollenbach1999}. To guide the
eye, we added a contour with value $\log(\chi / n_{\rm H})=-5$ (with
$\chi$ the Draine field). This contour follows the C$^+$/C/CO
transition very well, indicating PDR physics. Neutral carbon is
confined in a layer between the C$^+$ and CO. This picture is more
complicated when X-rays are added. Perhaps not entirely intuitive, the
overall trend with increasing X-rays is that neutral carbon is more
and more abundant in the unattenuated part of the atmosphere (as high
as $x_{\rm C}\sim 10^{-5}-10^{-4}$). When X-rays are present, neutral
carbon is higher by at least one to two orders of magnitude in
abundance compared to models that exclude X-rays. The C and C$^+$
ionization rates are of course higher, but the recombination rates are
also higher owing to the increased electron abundances (although the
higher temperatures reduce the recombination rates, as already pointed
out in Sect. \ref{electron_abundance_section}). This is because
electrons are now also produced by, e.g., ionization of hydrogen and
helium. The fact that C$^+$ and C coexist in more or less equal
amounts, when X-rays are dominating the ionization fraction of gas
clouds, was already pointed out in papers by \citet{Maloney1996} (see
their Figs. 3a and 4a) and \citet{Meijerink2005} (see their Figs. 3
and 4).

The bulk of the CO is situated in the shielded, optically thick part
of the disk, and the total CO gas mass does not change when varying
the FUV and X-ray radiation field. This molecule could thus serve as a
tracer of the total mass of a protoplanetary disk, keeping in mind
that the conversion factor depends on the amount of ice formation. A
few details should be noted. There is a significant correlation
between CO and OH in the unattenuated part of the disk (compare
Fig. \ref{COabundance_struct_appendix} with
\ref{OHabundance_struct_appendix}). A route to form CO is through the
reaction $\rm C + OH \rightarrow CO + H$. While it is possible to
obtain an abundance level of $x_{\rm CO} \sim 10^{-7}$ through this
channel, it will never go up to $x_{\rm CO} \sim 10^{-4}$, because the
OH abundance is $x_{\rm OH} \sim 10^{-7}$. As OH becomes less abundant
for higher FUV luminosities (when X-rays are fixed), the CO also
becomes less abundant and slightly more confined to the midplane. In
addition, the the average abundance of CO drops slowly with increasing
X-ray luminosity at the inner rim by destruction through ion-molecule
reactions, such as $\rm CO + He^+ \rightarrow C^+ + O + He$. This
could have an effect on the ro-vibrational lines that are
predominantly produced in these regions.

\begin{figure*}
  \centering
  \includegraphics[width=4cm,clip]{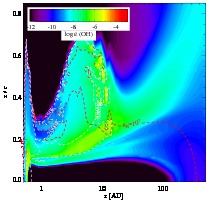}
  \includegraphics[width=4cm,clip]{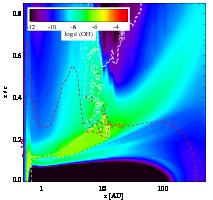}
  \includegraphics[width=4cm,clip]{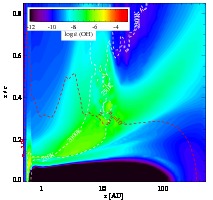}
  \includegraphics[width=4cm,clip]{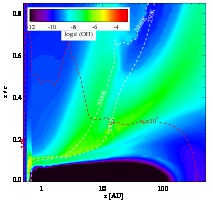}
  \includegraphics[width=4cm,clip]{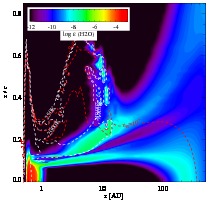}
  \includegraphics[width=4cm,clip]{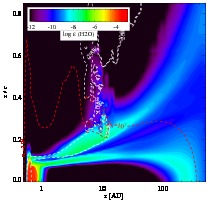}
  \includegraphics[width=4cm,clip]{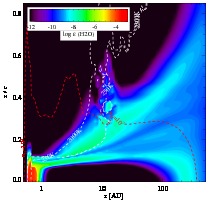}
  \includegraphics[width=4cm,clip]{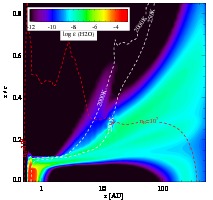}
  \includegraphics[width=4cm,clip]{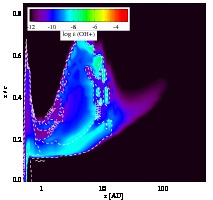}
  \includegraphics[width=4cm,clip]{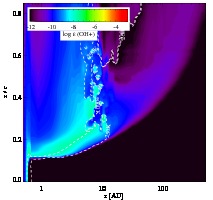}
  \includegraphics[width=4cm,clip]{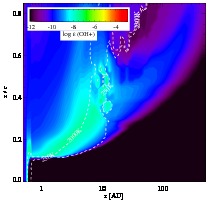}
  \includegraphics[width=4cm,clip]{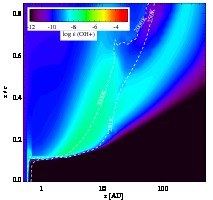}
  \includegraphics[width=4cm,clip]{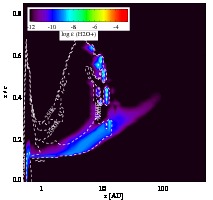}
  \includegraphics[width=4cm,clip]{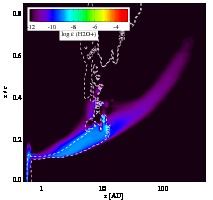}
  \includegraphics[width=4cm,clip]{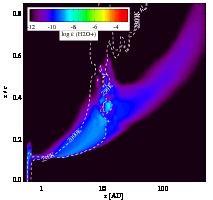}
  \includegraphics[width=4cm,clip]{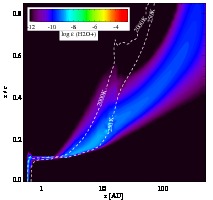}
  \includegraphics[width=4cm,clip]{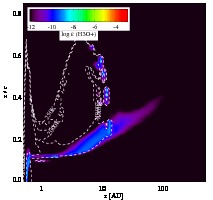}
  \includegraphics[width=4cm,clip]{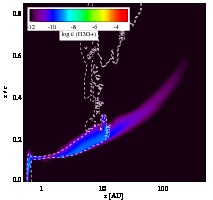}
  \includegraphics[width=4cm,clip]{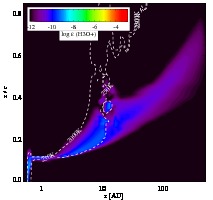}
  \includegraphics[width=4cm,clip]{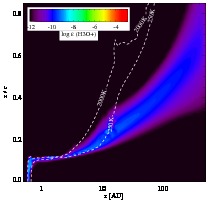}
  \caption{OH, H$_2$O, OH$^+$, H$_2$O$^+$, and H$_3$O$^+$
    abundances. Fluxes are the same as in
    Fig. \ref{model_electron_abundance}. The white contours indicate
    gas temperatures of 250 and 2000~K. The red contour indicates the
    number density $n_{\rm H}=10^6$~cm$^{-3}$.}
  \label{water_related_abundance_struct}
\end{figure*}

\begin{figure*}
  \centering
  \includegraphics[width=4cm,clip]{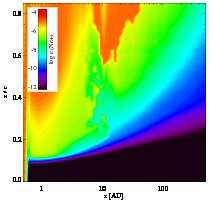}
  \includegraphics[width=4cm,clip]{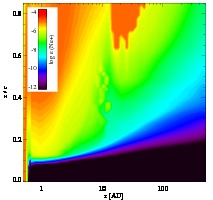}
  \includegraphics[width=4cm,clip]{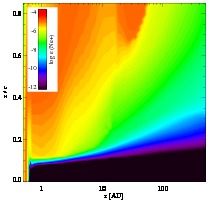}
  \includegraphics[width=4cm,clip]{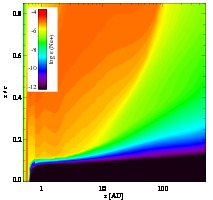}
  \includegraphics[width=4cm,clip]{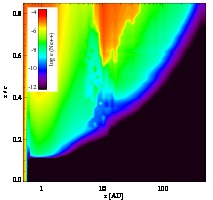}
  \includegraphics[width=4cm,clip]{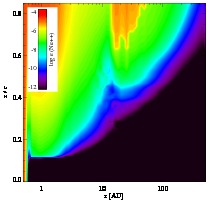}
  \includegraphics[width=4cm,clip]{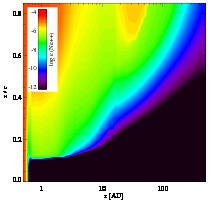}
  \includegraphics[width=4cm,clip]{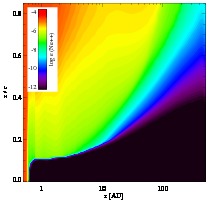}
  \includegraphics[width=4cm,clip]{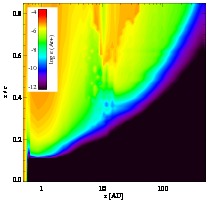}
  \includegraphics[width=4cm,clip]{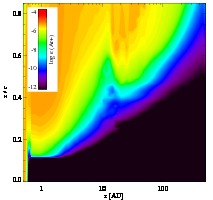}
  \includegraphics[width=4cm,clip]{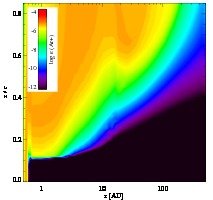}
  \includegraphics[width=4cm,clip]{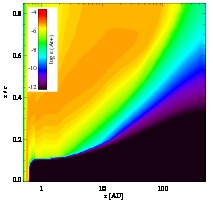}
  \includegraphics[width=4cm,clip]{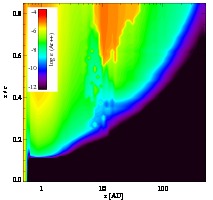}
  \includegraphics[width=4cm,clip]{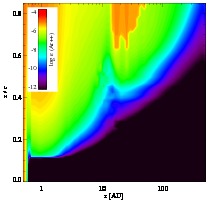}
  \includegraphics[width=4cm,clip]{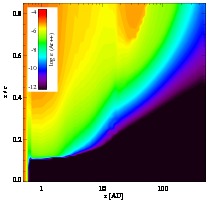}
  \includegraphics[width=4cm,clip]{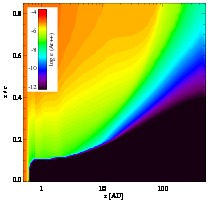}
  \caption{Abundance structure of selected heavy elements: Ne$^{+}$
    (top), Ne$^{2+}$, Ar$^+$, and Ar$^{2+}$ (bottom). The FUV
    luminosity is fixed at $L_{\rm FUV}=10^{31}$~erg\,s$^{-1}$. X-ray
    luminosities vary between $L_{\rm X}=10^{29}$ and
    $10^{32}$~erg\,s$^{-1}$.}
  \label{Heavy_element_abundance_struct}
\end{figure*}

\subsection{O$^{2+}$, O$^+$, and O abundances}

The ionization potentials of O and O$^+$ are $IP=13.68$ and
35.12~eV. Even the $IP$ of neutral oxygen is above the threshold of
$IP=13.6$~eV, where neutral hydrogen blocks the radiation
efficiently. The implications of this are seen in
Fig. \ref{oxygen_abundance_struct}, where the abundances of neutral,
singly and doubly ionized oxygen are shown for a FUV luminosity
$L_{\rm FUV} = 10^{31}$~erg\,s$^{-1}$ and all considered X-ray
luminosities except $L_{\rm X}=10^{32}$~erg\,s$^{-1}$. The second row
of Fig. \ref{oxygen_abundance_struct} (and also
Fig. \ref{O+abundance_struct_appendix} for variations with FUV
luminosity) shows the abundance structure of singly ionized
oxygen. Unlike the C$^+$, the abundances of O$^+$ do not exceed
$x_{\rm O^+}\sim 10^{-8} - 10^{-7}$ in the FUV-only case. The
abundance of ionized oxygen is mostly set by the charge exchange
balance with atomic hydrogen, while the ionization of both hydrogen
and oxygen is entirely due to cosmic ray ionization. When X-rays are
added, the abundances increase three to four orders of magnitude
compared to the FUV-only case, even at low X-ray luminosities ($L_{\rm
  X}=10^{29}$~erg\,s$^{-1}$). Important to note is that the O$^{+}$ is
confined to smaller radii for larger FUV luminosities, which is also
the case for C$^{2+}$. This was discussed earlier in
Sect. \ref{carbon_sub} and explained by the larger recombination rates
due to higher densities in the upper atmosphere of the disk and the
lower ionization rates due to shielding. The confinement to smaller
radii is even stronger for the case of O$^{2+}$, because this species
is not affected by cosmic rays.

The bottom panel of Fig. \ref{oxygen_abundance_struct} shows the
neutral oxygen abundance structure.  The critical density is $n_{\rm
  crit} = 5\times 10^5$~cm$^{-3}$ for the [OI] 63~$\mu$m line, which
is indicated by a black dashed contour. This species does not change
with X-ray luminosity. Neutral oxygen is the dominant oxygen carrier
throughout the disk, except for those regions, where large fractions
of the gas are frozen onto dust grains (i.e., the midplane). This
species is very insensitive to changes in FUV and the ambient
chemistry (see Fig. \ref{Oabundance_struct_appendix}). The bulk of
atomic oxygen is in LTE for the commonly observed [OI] 63 and
145~$\mu$m fine-structure lines and thus sensitive to the average
temperature of the region that it is probing. As a result, the [OI]
fine-structure lines have the potential to probe the total energy
budget of the gas (the combined FUV and X-ray luminosity irradiating
the disk). There will be an elaborate analysis of the oxygen
fine-structure line emission in paper II.

\subsection{OH, OH$^+$, H$_2$O, H$_2$O$^+$, and H$_3$O$^+$ abundances}

\noindent The water chemistry and its related species are
significantly affected by X-ray irradiation. The physical
circumstances in the disk change with radius $r$ and height $z$ due to
radiation shielding and large differences in chemistry, which means
that the X-ray and FUV energy deposition per particle, $H_{\rm X}/n$,
and $\chi/n$ \citep[with $\chi=1$ the interstellar radiation field as
defined by][and integrated between 91.2 and 205~nm]{Draine1978,
  Draine1996}, and their relative ratio change over orders of
magnitude in the disk. As a result, the main chemical pathways forming
the molecules OH and H$_2$O change throughout the disk and the grid.

X-rays heat the gas and therefore the reaction rates with activation
barriers increase. They also drive the ionization, which enhances
formation routes through ion-molecule reactions. The formation of the
neutral species OH and H$_2$O is either by the neutral-neutral
reactions, $\rm H_2 + O \rightarrow OH + H$, and $\rm H_2 + OH
\rightarrow H_2O + H$, or by recombination of ionized species, such as
$\rm H_3O^+ + e^- \rightarrow H_2O + H$ and $\rm H_2O^+ + e^-
\rightarrow OH + H$. Which of these pathways are dominating depends
thus strongly on the temperature and ionization fraction. The routes
through neutral-neutral reactions have activation barriers, making
them efficient only when gas temperatures are sufficiently high, i.e.,
$T > 200 - 300$~K. One way to increase the neutral-neutral route
without the need of higher temperatures is by lowering the activation
barrier, which can be done by exciting H$_2$ to a higher vibrational
state after absorption of a FUV photon. At high relative heights ($z/r
> 0.5$), this reaction will dominate when only FUV photons are
present. The ion-molecule chain is started with cosmic ray or X-ray
ionization of H$_2$ and thus requires significant ionization rates.

It turns out that the neutral-neutral reaction pathways dominate in
the inner part of the disk ($r < 5$~AU). The X-rays heat regions
deeper in the disk, resulting in the production of a thick warm water
layer. Although there is a larger fraction of X-rays ionizing the gas
than when only FUV photons are present, the X-ray heating causes the
neutral-neutral reactions to dominate. It is remarkable to note that
in the outer region ($r \gtrsim 20$~AU) of the disk the ion-neutral
reaction network dominates. This is because temperatures are too low
for the neutral-neutral reaction pathways to occur. These combined
effects become apparent in the distribution of the water throughout
the disk, as shown in the second panel from the top in
Fig. \ref{water_related_abundance_struct} (see also
Fig. \ref{H2Oabundance_struct_appendix} for a more elaborate view on
the effects of both FUV and X-rays). In the models without any X-rays
(left-hand side), the water shows a strong abundance peak at the inner
rim (all the way down to the midplane of the disk) and a warm water
layer at a relative height $z/r \approx 0.1$, with an abundance of
$x_{\rm H_2O}\sim 10^{-7}$, and a second water reservoir higher up in
the disk. This reservoir is located at a relative height $z/r \sim
0.015$ at small radii ($r \sim 1$~AU), and at a relative
height of $z/r\sim 0.5$ at a radius of $r \sim 100$~AU, thus
following the flaring of the disk. These different water reservoirs
were already noted by \citet{Woitke2009} for Herbig AeBe stars. An
elaborate discussion on the formation of warm water reservoirs through
(the dominant) neutral-neutral reactions and formation on dust grains
in the inner regions of an X-ray irradiated disk is also outlined by
\citet{Glassgold2009}.

In the models where only FUV is included, the second layer is confined
to the regions where temperatures are between $T\sim 300 -
1000$~K. The regions at higher temperatures are exposed to FUV fluxes
that essentially destroy the water faster than it is formed. When
X-rays are included, the second water reservoir extends to
increasingly larger radii. Although there is water present in the
FUV-only models out to radii $r\sim 200$~AU, the abundances are
ultimately one to two orders of magnitude higher when X-rays are
added. The additional water is increased since the ion-molecule
chemistry is very effective in forming the water. This is illustrated
in the third to fifth panel of
Fig. \ref{water_related_abundance_struct} (and also
Figs. \ref{OH+abundance_struct_appendix},
\ref{H2O+abundance_struct_appendix}, and
\ref{H3O+abundance_struct_appendix}), where the changing abundance
structures of the ionic species OH$^+$, H$_2$O$^+$, and H$_3$O$^+$ are
shown for various combinations of X-ray and FUV luminosities (appendix
A only). The ionic species extend to larger radii for larger X-ray
fluxes, but larger FUV fluxes decrease the extent. The fact that FUV
confines the ionic species to smaller radii is an indirect effect,
because FUV tends to puff up the upper layers of the disk and the
density becomes higher at larger relative heights (see
Fig. \ref{model_dens_struct}). These larger densities increase the
recombination rates and reduce the abundances of these ionic species.

The first layer described above is located in those regions of the
disk, where the conventional routes do not work anymore. Temperatures
are too low and the ionization fraction is small. This is the region
where more exotic reactions take over, such as $\rm NH_2 + NO
\rightarrow N_2 + H_2O$. This layer has a smaller vertical extent in
relative height for larger FUV fluxes. On the other hand, the second
separate layer becomes thicker when X-rays are added. When the
$L_X / L_{FUV} \gtrsim 1$ ratio, these two layers merge. The reason for
the merging is the enhancement of the transient species OH$^+$,
H$_2$O$^+$, and H$_3$O$^+$ (see also
Fig. \ref{OH+abundance_struct_appendix},
\ref{H2O+abundance_struct_appendix} and
\ref{H3O+abundance_struct_appendix}). As mentioned earlier, these
species react sensitively to the presence of an ionization source and
become increasingly abundant with higher X-ray fluxes.

The situation with OH is very similar. OH becomes more abundant in the
outer disk, when X-rays are added. OH is located at higher altitudes
in the disk (although there is also a large overlap with the regions
where the water is located). For this reason, the effect of the X-rays
on the abundance in the outer disk is not as pronounced as for
water. The H$_2$ used in the neutral-neutral formation route of OH is
on average at higher temperatures, so the relative contribution of the
ion-molecule route is smaller.

\subsection{\it Ar$^+$, Ar$^{2+}$, Ne$^{+}$, and Ne$^{2+}$ abundances} 

\noindent These species have ionization potentials that are larger
than that for atomic hydrogen, with $IP=21.56$, 40.96, 15.76, and
27.63~eV for Ne, Ne$^+$, Ar, and Ar$^+$, respectively. The only way to
produce these species is through direct (photon absorption and Auger
effect) or indirect (fast electron collisions) ionization by
X-rays. The presence of these ionized species is thus a direct result
of the X-ray irradiation of the disk.

The models include a thermal source of temperature, $T_{\rm X} =
1$~keV. As a result, the cross sections for neon are more favorable
for direct ionizations than for argon. The cross section for the
absorption is located at approximately $\sim 1$~keV for neon, while
for argon it is $\sim 4 $~keV. On the other hand, the rates for
secondary ionizations are a little higher for argon,
$\sigma(A)/\sigma({\rm H})=1.1$, 0.48, 3.7, and 1.8, with $A$ equal to
Ne, Ne$^+$, Ar, Ar$^+$, respectively. Consequently, the total
ionization rates are very comparable. However, there is one big
difference in the chemical network of these species. The charge
transfer rate of Ne$^+$ with H$_2$ is very low
($k<10^{-14}$~cm$^{3}$~s$^{-1}$), while the other ions have
significant rates for charge exchange with molecular hydrogen. The
Ne$^+$ abundance structure thus extends to much smaller relative
heights, down to $z/r \sim 0.01$, than Ar$^+$, which is abundant above
$z/r \sim 0.01 - 0.06$ at radii $r \gtrsim 10 - 200$~AU. The abundance
structure of Ar$^+$ is very similar to those for Ne$^{2+}$ and
Ar$^{2+}$.

\begin{figure*}
  \centering
  \includegraphics[width=16cm,clip]{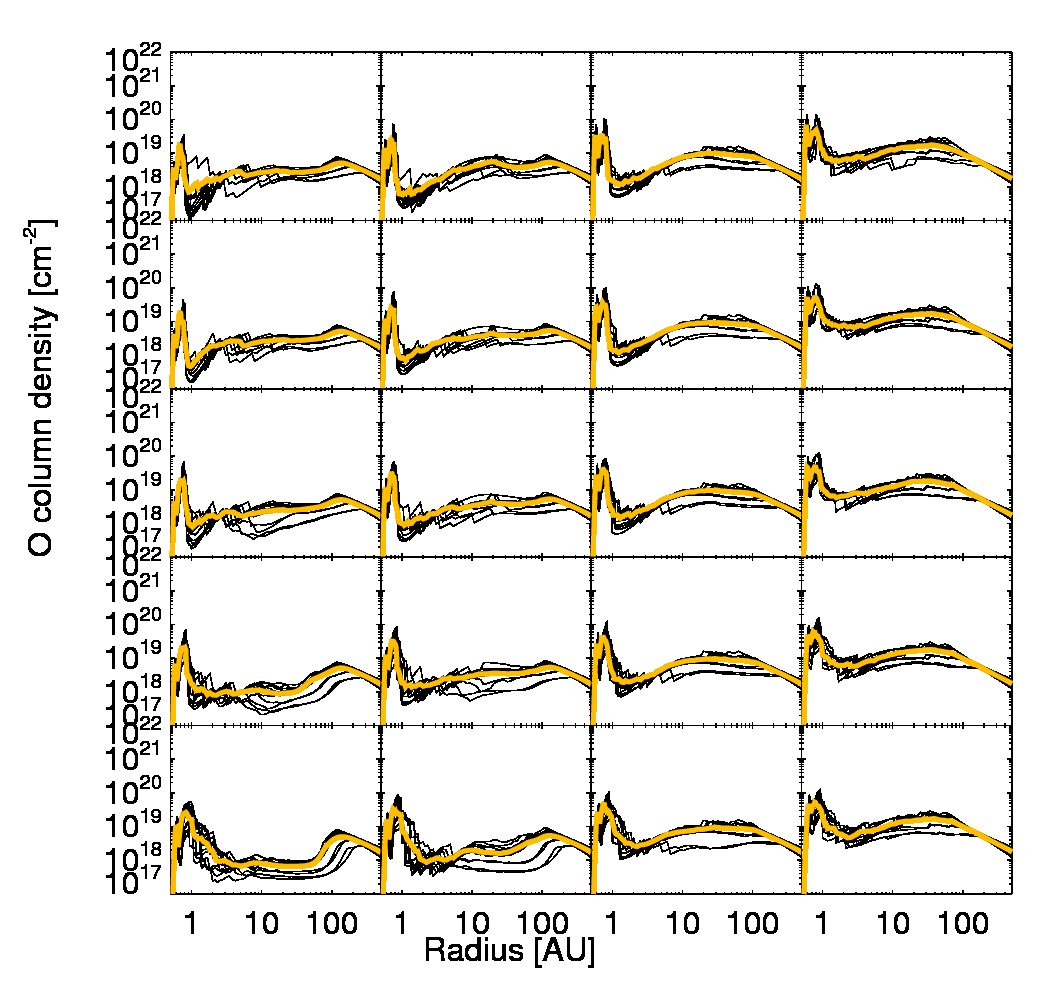}
  \caption{Radial column density distribution of O. FUV and X-ray fluxes are the same as Fig. \ref{model_dens_struct}.}
  \label{Oradial}
\end{figure*}

\begin{figure*}
  \centering
  \includegraphics[width=16cm,clip]{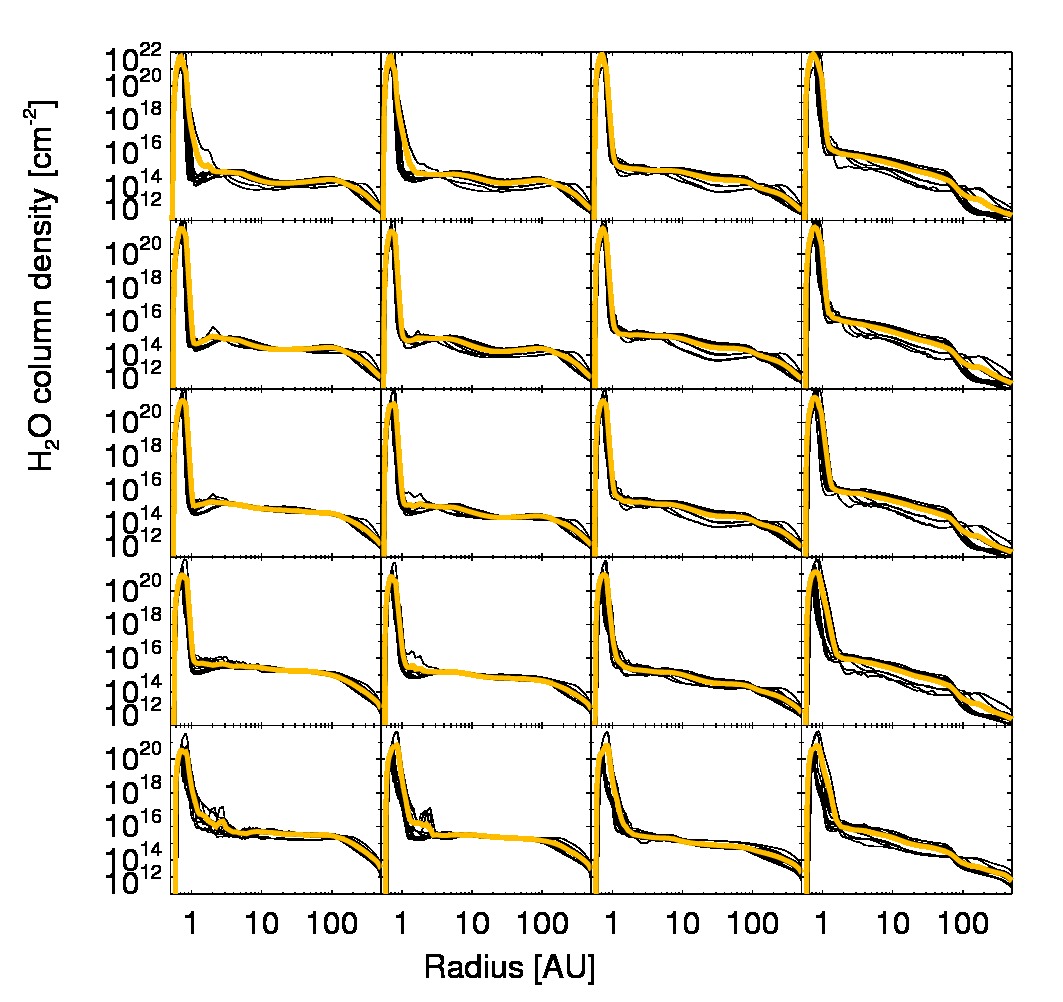}
  \caption{Radial column density distribution of H$_2$O. FUV and X-ray fluxes are the same as Fig. \ref{model_dens_struct}.}
  \label{H2Oradial}
\end{figure*}

\section{Radial column density profiles}\label{rad_col}

Although the abundance distributions of species well illustrate the
change in the chemistry due to different combinations of X-ray and FUV
luminosities, it is necessary to look at the integrated properties of
the disk that are observed by our telescopes. The focus in this is on
the species C$^+$, O, H$_2$O, and Ne$^+$, as their line intensities
and line profiles are extensively discussed the paper II. The figures
show the average column density in yellow, while the black lines
represent the 12 models at fixed FUV and X-ray luminosities. Changing
parameters other than those for FUV and X-rays does not affect the
results significantly.

{\it C$^+$ column density profile (Fig. \ref{C+radial}):} The column
density of C$^+$ is strongly related to the inicident FUV flux at the
inner rim, where it increases from $N_{\rm C^+} \sim 10^{17}$ to
$10^{19}$~cm$^{-2}$, when the FUV luminosity increases from $L_{\rm FUV} =
10^{29}$ to $10^{32}$~erg\,s$^{-1}$. Right after the peak at the inner
rim, there is a sharp drop in the column density, which is caused by
the inner rim casting a shadow and therefore reducing ionizing
radiation. A second peak in the column density profile is seen between
$r \sim 5 - 10$~AU. The peaks moves to larger radii for higher FUV
luminosities. The X-rays affect the column density profile in a
different way. They reduce the minimum column density right after the
inner rim and smooth the radial distribution. The combination of the
highest FUV and X-ray fluxes gives the flattest distribution.

{\it O column density profile (Fig. \ref{Oradial}):} The column
density profiles of neutral oxygen do not change as much as those for
C$^+$. The maximum column density, $N_{\rm O}\sim 3 - 10\times
10^{19}$~cm$^{-2}$, is located right behind the inner rim and only
varies with a factor of at most ten, which is again caused by the FUV
irradiation. The width of the oxygen column density peak broadens a
bit for higher FUV and X-ray fluxes. This makes the mininum in the
column density less prominent in the radial column density
profile. Because the column density is not very affected by
radiation, it is a clean probe of the properties of a disk
temperature ([OI] fine-structure lines), since there are no strong
dependencies on uncertainties in the chemical network.

{\it H$_2$O column density profile (Fig. \ref{H2Oradial}):} The
highest water column densities, $N_{\rm H_2O}\sim 10^{20} -
10^{22}$~cm$^{-2}$, are found at the inner rim, $r\sim 0.5$~AU. As
discussed, the water abundances are enhanced by higher temperatures
and higher ionization fractions throughout the disk. The FUV
counteracts this to some extent, and the most favorable situation is a
high X-ray to FUV luminosity ratio. FUV irradiation puffs up the inner
rim, and shields the outer disk, thus allowing less ionizing radiation
to penetrate into the outer regions of the disk. As a result, it
confines the water to smaller regions of the disk. The bottom panel
($L_{\rm X}=10^{32}$~erg\,s$^{-1}$) of Fig. \ref{H2Oradial} shows that
the column density distribution is flat for the lowest FUV flux,
dropping steadily as a function of radius for the highest FUV
flux. The only region in the disk where the column density becomes
smaller is at the inner rim due to reactions with ions such as $\rm
H_2O + He^+ \rightarrow H^+ + He + OH$. The water column density is
almost two orders of magnitude smaller for the models with the highest
X-ray luminosities compared to those with only FUV.

{\it Ne$^+$ column density profile (Fig. \ref{Ne+radial}):} Ne$^+$ is
only produced by X-rays and therefore only the models that include
X-rays show significant column densities. The profiles show a peak in
the column density at the inner rim and a second bump at a few
AU. The second bump smoothes out for larger X-ray fluxes, while the
maximum column density increases from $N_{\rm Ne^+}\sim 10^{15}$ to
$10^{17}$~cm$^{-2}$. The FUV tends to confine the Ne$^+$ to smaller
radii for the lower X-ray luminosities, which becomes apparent in
the line profile (see paper II).

\section{Conclusions}\label{conclusions}

In this paper, we discussed the combined effects of FUV and X-rays on
both the density structure and the thermal and chemical balance of
disks around T Tauri stars for an expected range of parameters (dust
size distribution, density profile, etc.), yielding a total of 240
models. Here we highlight the main results and a few implications:

\subsection{The disk thermal and chemical structure}

{\it Temperature structure:} The extent of the disk where the
temperature is higher than $T > 1000$~K is much larger when X-rays
are included. X-rays have a much higher heating efficiency than FUV,
$30 - 50$\% compared to $< 3$\%, respectively.

{\it Density structure:} Increasing FUV luminosities does not change
the scale height of the inner rim; it only alters the width and height
of the second bump in the disk that is created at intermediate radii
($r \sim 3 - 10$~AU), behind the region shielded by the puffed-up
inner rim. Gas temperatures at the inner rim are much higher when
X-rays are included and, as a result, the inner rim is puffed up to
higher and higher altitudes for increasing X-ray luminosities. As a
direct extension of this theoretical work, the existence of the second
bump could potentially be tested by continuum imaging face-on
protoplanetary disks in the near-infrared with, e.g., VLT or Keck.

{\it Scale height:} Considering only FUV, we see that the scale height
shows a maximum in the unattenuated parts ($z/r > 0.5$) of the
disk. When X-rays are added we find that this maximum is smoothed over
a larger region (out to $r \sim 10$~AU). The scale height in these
regions is larger than one would expect from the flaring index in the
outer regions of the disk. When moving to smaller relative height,
e.g, $z/r = 0.1$, the break in the flaring index disappears. Another
observational possibility would be to do continuum interferometry in
the near-IR (VLTI) to directly measure the physical height of the
inner rim.

\subsection{Chemical balance}

{\it Ionization fraction:} The ionization fraction reaches values as
high as $x_{\rm e^-} \sim 10^{-2}$ in exceptional cases in our
FUV-only models, whereas X-rays easily maintain these ionization
fractions throughout large portions of the disk. Even when the gas
becomes partially shielded, it can still maintain a significant
ionization fraction, leading to an ion-molecule chemistry that can
form molecules at low temperatures, which is not possible with
neutral-neutral reactions as they usually have temperature barriers.

{\it Formation of H$_2$ through the H$^-$ route:} Formation of H$_2$
dust is usually much more efficient in environments that have solar
metallicities. Because of the high ionization fraction due to X-ray
irradiation, the formation route $\rm H^- + H \rightarrow H_2 + e^-$
is able to provide a significant addition of order 50\% percent to
the H$_2$ on dust formation route. Overall, the H$_2$ to H abundance
ratio is increased by at least two orders of magnitude when X-rays are
present.

{\it Formation of water and OH:} The OH and H$_2$O abundances are more
concentrated toward the inner regions of the disk, when only FUV is
irradiating the disk. This is because the temperature is only there
sufficient to drive the neutral-neutral formation route. The outer
disk shows significant enhancements (up to two orders of magnitude)
when X-rays are added. The higher ionization fractions make it
possible to form the species through ion-molecule reactions and
sustain abundance levels of $x_{\rm H_2O}\sim 10^{-6} - 10^{-7}$. Such
abundance levels cannot be reached in outer disk models without
X-rays. Only the warm inner disks allow even higher levels of water
abundance through warm neutral-neutral chemistry.

{\it Resulting abundance structures and radial column density
  profiles:} Whereas neutral oxygen and CO are very stable to both FUV
and X-rays, this is not the case for, e.g., C$^+$, Ne$^+$, and
H$_2$O. Ne$^+$ is strongly enhanced by X-rays and confined to the
inner regions by larger FUV luminosities. This latter aspect will
certainly affect the line widths. A high $L_{\rm X}/L_{\rm FUV}$ ratio
is favorable for water formation, especially in the outer disk.

\subsection{Outlook} 

In paper II, we will perform a radiation transfer analysis of the
aforementioned species and correlate line fluxes and line widths
directly to the FUV and X-ray luminosities. This will allow a
discussion of the diagnostic value of these species and provide a
theoretical framework for the interpretation of observational data. We
will also discuss our results in the context of data obtained within
several observational efforts (Spitzer, Herschel, and ground-based
observing programs).

\acknowledgements{The research leading to these results received
  funding from the EU Seventh Framework Programme (FP7) in 2011 under
  grant agreement no 284405.}

\bibliographystyle{aa}
\bibliography{ms.bib}

\clearpage

\begin{appendix}

\section{X-ray chemistry}\label{xray_chem}

Our chemical network is composed of $N_{\rm sp}$=110 species (see Table \ref{species}), linked
by $\sim$1500 reactions. 

\begin{table}[h]
\centering
\caption{Species included in the chemical network.}
 \begin{tabular}{l}
 \hline
 110 species \\
 \hline
 \hline
  H, H$^+$, H$^-$, H$_2$, H$_2^+$, H$_2^*$, H$_3^+$, He, He$^+$, O, O$^+$, O$^{2+}$, O$_2$, O$_{2}^+$,\\
  OH, OH$^+$, H$_2$O, H$_2$O$^+$, H$_3$O$^+$, CO, CO$^+$,CO$_2$, CO$_2^+$, HCO, HCO$^+$, \\
  H$_2$CO,\\
  N, N$^+$, N$^{2+}$, NO, NO$^+$, C, C$^+$, C$^{2+}$, CH, CH$^+$, CH$_2$, CH$_{2}^+$, CH$_3$, CH$_3^+$,\\
  CH$_4$, CH$^+_4$, CH$^+_5$, Si, Si$^+$, Si$^{2+}$, SiO, SiO$^+$, SiH, SiH$^+$, SiH$_2^+$, SiOH$^+$,  \\
  S, S$^+$, S$^{2+}$, SO, SO$^+$, SO$_2$, SO$_2^+$, OCS, CS, CS$^+$, HS, HS$^+$, HCS$^+$, \\
  H$_2$S$^+$, H$_3$S$^+$, \\
  Mg, Mg$^+$, Mg$^{2+}$, Fe, Fe$^+$, Fe$^{2+}$, Ne, Ne$^+$, Ne$^{2+}$, Ar, Ar$^+$, Ar$^{2+}$, Na, \\
  Na$^+$, Na$^{2+}$,  \\
  NH, NH$^+$, NH$_2$, NH$_2^+$, N$_2$H$^+$,  NH$_3$,  NH$_3^+$, NH$_4^+$, N$_2$, HN$_2^+$, CN, CN$^+$, \\
  HCN, HCN$^+$, HNC, C$_2$H$_2$, C$_2$H$_3^+$, HCNH$^+$, CO\#, H$_2$O\#, CO$_2$\#, \\ 
  CH$_4$\#, NH$_3$\#, \\ 
  PAH, PAH$^-$, PAH$^{+}$, PAH$^{2+}$, PAH$^{3+}$\\
  \hline
 \end{tabular}
 References for the rate coefficients:
 \citet{Mil86,Len88,Lan91,Ani93,Bad06}
 \label{species}
\end{table}

\noindent For a given species \emph{i}, the net
formation rate reads (following Woitke et al. 2009):

\begin{eqnarray}
\frac{dn_i}{dt} &=& \sum_{jkl}R_{jk\rightarrow il} (T_{\rm g})n_j n_k + \sum_{jl} \left( R_{j\rightarrow il}^{\rm ph} + R_{j\rightarrow il}^{\rm cr} + R_{j\rightarrow il}^{\rm Xpr}+ R_{j\rightarrow il}^{\rm Xsec}\right)n_j \nonumber \\
&-&  n_i \left(\sum_{jkl}R_{il\rightarrow jk}(T_g) + \sum_{jk} \left( R_{i\rightarrow jk}^{\rm ph} + R_{i\rightarrow jk}^{\rm cr} + R_{i\rightarrow jk}^{\rm Xpr}+ R_{i\rightarrow jk}^{\rm Xsec}\right)\right).
\end{eqnarray}

\noindent The terms involved are

\begin{itemize}
\item $R_{jk\rightarrow il}$, the temperature-dependent rate for a two-body reaction where species \emph{i} and \emph{l} are formed out of species \emph{j} and \emph{k}
\item $R_{i\rightarrow jk}^{\rm ph}$, a photo-reaction rate that depends on the local strength of the FUV radiation field
\item $R_{i\rightarrow jk}^{\rm cr}$, a reaction that depends on the cosmic ray ionization rate
\item $R_{j\rightarrow il}^{\rm Xpr}$ and $R_{j\rightarrow i}^{\rm Xsec}$, the X-ray primary and secondary ionization reaction rates. 
\end{itemize}

\noindent Assuming statistical equilibrium
$\left(\frac{dn}{dt}=0\right)$, we obtain $N_{\rm sp}$ non-linear
equations for the $N_{\rm sp}$ unknown particle densities $n_k$,

\begin{equation}
F_i(n_k)=0.
\end{equation}

\noindent For the
species densities $n_k$, this system of nonlinear equations is solved through a Newton-Raphson iterative method,
which expands $F_i$ into a Taylor series in the neighborhood of $n_k$,

\begin{equation}
F_i(n_k+\delta n_k) = F_i(n_k) + \sum_j \frac{\partial F_i}{\partial n_k}\delta n_k + O(\delta n^2_k). \label{taylor}
\end{equation}

\noindent The code then needs to calculate the Jacobi matrix:

\begin{equation}
J_{ik} = \frac{\partial F_i}{\partial n_k}.
\end{equation}

\noindent Neglecting terms of the order of $\delta n^2_k$, the math
SLATEC routines are used to find a set of $n_k$ that satisfies
$F_i(n_k+\delta n_k)=0$:

\begin{equation}
J_{ik}\cdot \delta n_k = -F_i.
\end{equation}

\noindent When the rate of a given reaction does not depend on the
particle density $n_k$, the derivative of $F_i$ is straightforward,
e.g, a photo-reaction that destroys species \emph{i}:

\begin{equation}\label{c0}
J_{ik} = \frac{\partial F_i}{\partial n_k} = -R_{i\rightarrow j}^{\rm ph}\cdot \delta_{ik}.  
\end{equation}

\noindent X-ray reaction rates, however, depend on the electron
density $n_{\rm e}$, atomic hydrogen density $n_{\rm H}$, and/or
molecular hydrogen density $n_{\rm H_2}$. These densities determine
how much of the absorbed X-ray photons will go into the different
channels for heating, ionization, and excitation. If the rate $R$
shows a dependency on the particle densities, for example, the electron
density ($n_{\rm el}$), atomic hydrogen density ($n_{\rm H}$), and/or
molecular hydrogen density ($n_{\rm H_2}$), the Jacobian can be
expressed as:

\begin{equation}\label{tutte}
 J_{ik} = \frac{dF_i}{dn_k} = \frac{\partial F_i}{\partial n_k} + \frac{\partial F_i}{\partial n_{\rm el}}\frac{\partial n_{\rm el}}{\partial n_k} + \frac{\partial F_i}{\partial n_{\rm H}}\frac{\partial n_{\rm H}}{\partial n_k} + \frac{\partial F_i}{\partial n_{\rm H_2}}\frac{\partial n_{\rm H_2}}{\partial n_k}.
\end{equation}

\noindent ProDiMo calculates this term analytically to ensure an
accurate chemical solution. Here we describe how the Jacobian terms
for the X-ray reactions rates are calculated. We define the following
quantities, which will be used below:

\begin{eqnarray}
\ntot &=& \sum_j n_j Q(j,{\rm H}) \\
n_{\rm el} &=& \sum_j n_j q_j,
\end{eqnarray}

\noindent where $\ntot$ is the total hydrogen nuclei density, $Q$(j,H)
is the stoichiometric coefficient for the hydrogen nuclei (which is 0 for
all those species that do not contain hydrogen, while for species
containing hydrogen we have: $Q$(H,H)=1, $Q$(H$_2$,H)=2,
$Q$(H$^+_3$,H)=3, etc.). The electron density is denoted as $n_{\rm
  el}$ and $q_j$ is the charge of the particle \emph{j}.

\subsection{Primary ionization}

\noindent For the generic atomic species A, A$^+$ and molecular species AB, these are reactions of the kind:

\begin{eqnarray}
{\rm A} + X_{\rm ph} &\rightarrow& {\rm A}^{2+} + 2{\rm e}^- \label{ato1}\\
{\rm A}^+ + X_{\rm ph} &\rightarrow& {\rm A}^{2+} + {\rm e}^- \label{ato2}\\
{\rm AB} + X_{\rm ph} &\rightarrow& {\rm A}^{2+} + {\rm B} + 2{\rm e}^- \label{mol}\\
             &\rightarrow& {\rm A}^+ + {\rm B}^+ + 2{\rm e}^- \nonumber\\
             &\rightarrow& {\rm A} + {\rm B}^{2+} + 2{\rm e}^-\nonumber,
\end{eqnarray}

\noindent where $X_{\rm ph}$ is an X-ray photon. When the $i$-th
species is molecular (\ref{mol}), the reaction rate $R_{i\rightarrow
  jl}^{\rm Xpr}$ for the reaction that destroys the species \emph{i}
has two indices \emph{j}, and \emph{l}, for the resulting species.
Otherwise the reaction rate should simply read $R_{i\rightarrow
  j}^{\rm Xpr}$ as only one species is produced out of reactions
\ref{ato1} and \ref{ato2}. For simplicity, from now on, we use
the molecular reaction rate as an example for X-ray primary
ionization. The rate is calculated as

\begin{equation}
\rpr = \int_{E_i}^{\infty} \sigma_i(E)F(E,r)dE \qquad[\rm s^{-1}].
\end{equation}

\noindent It depends on the X-ray radiation field $F(E,r)$ at the
point $r$ where it is computed and on the cross section $\sigma_i$ of
the species \emph{i}. It does not depend on the local particle
density. From equations \ref{taylor} and \ref{c0}, we see that the
contribution to the equilibrium equation for the species \emph{i} can
be written as

\begin{equation}
F_i = F_{\rm oth} - \sum_{jl}\rpr\cdot n_i + \sum_{jl}R_{j\rightarrow il}^{\rm Xpr}\cdot n_j.
\end{equation}
The contribution to the Jacobi element of this reaction is then
\begin{equation}
J_{ik} = \frac{\partial F_{\rm oth}}{\partial n_k} - \sum_{jl}\rpr\cdot \delta_{ik} 
+ \sum_{jl}R_{j\rightarrow il}^{\rm Xpr}\cdot \delta_{jk}.
\end{equation}

\subsection{Secondary ionization}

\noindent We consider X-ray secondary ionization for all the atomic
species A and only for a single molecule H$_2$:

\begin{eqnarray}
{\rm A} + {\rm e}^{-}&\rightarrow& {\rm A}^{+} + 2{\rm e}^- \\ 
{\rm H_2} + {\rm e}^{-}&\rightarrow& {\rm H_2}^{+} + 2{\rm e}^- \\
              &\rightarrow& {\rm H} + {\rm H}^+ + 2{\rm e}^-.
\end{eqnarray}

\noindent The reaction rate for the X-ray secondary ionization of species
\emph{i} is

\begin{equation}\label{second}
 R_{i\rightarrow j}^{\rm Xsec} = r_i \frac{H_X\ntot}{Wn_H} \qquad[\rm s^{-1}],
\end{equation}

\noindent where $r_i$ is a parameter that takes into account the
geometrical cross section of species $i$ compared to the hydrogen
cross section, $\ntot$\,is the total hydrogen nuclei density, $H_{\rm
  X}$ is the X-ray energy deposition \citep{Maloney1996}, and $W$ [eV]
is the mean energy consumed per ion pair \citep{Dalgarno1999}.

$W$ for hydrogen and all the other atomic elements except He is

\begin{equation}\label{W}
W_{\rm H} = W_{0,{\rm H}} \left(1+c_1\left(\frac{n_{\rm el}}{n_{\rm \left<\rm  H\right>}}\right)^{\alpha}\right)\left(1+c_2\frac{n_{\rm H_2}}{n_H}\right) \qquad[{\rm eV}],
\end{equation}

\noindent where $W_{0,{\rm H}}$ is the collisional ionization rate for
H in a pure neutral atomic gas (13.6 eV), $n_{\rm el}$ is the electron
density, $n_{\rm H_2}$ is the molecular hydrogen density, and $\alpha$,
$c_1$ and $c_2$ are fitting parameters.

$W_{\rm H}$ is then the energy needed to collisionally ionize hydrogen
in a gas mixture with $n_{\rm el}$, $n_{\rm H}$ and $n_{\rm H_2}$. The
values for other atomic elements are scaled up, considering the
geometrical factor $r_i = \sigma_i ^{coll}/\sigma_H^{coll}$
\citep{Adamkovics2011}. $W$ has been calculated for He and H$_2$ as
well; these cases will be treated separately further on.

\subsection{Hydrogen} 

\noindent The X-ray energy deposition is a quantity defined per unit
of hydrogen nuclei as follows:

\begin{equation}\label{hx}
 H_X = \int \sigma_{\rm tot}(E)F(E,r) dE \qquad\rm{[erg\,\left<H\right>^{-1}]},
\end{equation}

\noindent where $\sigma_{\rm tot}$ is given by

\begin{equation}
\sigma_{\rm tot}(E) = \sum_{i=1}^{N_{\rm sp}} \sigma_{i}(E) n_i/\ntot \qquad\rm{[cm^{2}]}, 
\end{equation}

\noindent and $F(E,r)$ is the radiation field in unit of erg s$^{-1}$
cm$^{-2}$ eV$^{-1}$. For a given atomic species \emph{i}, at the
equilibrium, the contribution of the secondary ionization to the
volumetric rate is

\begin{eqnarray}\label{Fhy}
F_i &=& F_{\rm oth} + F_{i}^{\rm Xsec} \nonumber\\
    &=& F_{\rm oth} - \rse\cdot n_i + R_{j\rightarrow i}^{\rm Xsec}\cdot n_j\nonumber\\
    &=& F_{\rm oth} - r_i \frac{H_X\ntot}{W_Hn_H} n_i + r_j \frac{H_X\ntot}{W_Hn_H} n_j.
\end{eqnarray}

\noindent Following Eq. \ref{tutte}, the derivative of $F_i^{\rm Xsec}$
with respect to $n_k$ is

\begin{eqnarray}\label{big}
\frac{dF_i^{\rm Xsec}}{dn_k} &=& -r_i \times \\ & & \left[ \frac{n_i}{W_{\rm H}n_{\rm H}}\frac{\partial \left( H_X\ntot\right)}{\partial n_k} - \frac{H_x\ntot n_i}{W_{\rm H}^2n_{\rm H}}\frac{\partial W_{\rm H}}{\partial n_k}- \frac{H_x\ntot n_i}{W_{\rm H}n_{\rm H}^2}\delta_{{\rm H},k}+ \frac{H_x\ntot}{W_{\rm H}n_{\rm H}}\delta_{ik} \right]. \nonumber
\end{eqnarray}

\noindent Here we develop separately the partial derivative for all involved terms:
\begin{eqnarray}\label{hxder}
 \frac{\partial \left(H_X\ntot \right)}{\partial n_k} &=& \frac{\partial}{\partial n_k} \int \Sigma_j n_j \sigma_j(E) F(E,r)dE \nonumber \\
                                   &=& \int \sigma_k(E) F(E,r)dE \nonumber\\
                                   &=& D_K
\end{eqnarray}
\begin{eqnarray}
 \frac{\partial W_{\rm H}}{\partial n_k} &=& \frac{\partial}{\partial n_k} \left[W_{0,{\rm H}}\left(1+c_1\left(\frac{n_{\rm el}}{\ntot} \right) ^{\alpha}\right)\left(1+c_2\frac{n_{\rm H_2}}{n_{\rm H}}\right) \right]  \nonumber\\
                                   &=&  W_{0,{\rm H}}\,c_1 \frac{\alpha q_k n_{\rm el}^{\alpha-1}}{\ntot^\alpha} \left(1+c_2\frac{n_{\rm H_2}}{n_{\rm H}}\right)\nonumber\\
                                   &-& W_{0,{\rm H}}\,c_1 \frac{\alpha n_{\rm el}^{\alpha} Q_{k,{\rm H}}}{\ntot^{\alpha+1}} \left(1+c_2\frac{n_{\rm H_2}}{n_{\rm H}}\right)\nonumber\\
                                   &+& W_{0,{\rm H}}\left(1+c_1\frac{n_{\rm el}^{\alpha}}{\ntot^{\alpha}}\right)\frac{c_2}{n_{\rm H}} \delta_{{\rm H_2},k}\nonumber\\
                                   &-& W_{0,{\rm H}}\left(1+c_1\frac{n_{\rm el}^{\alpha}}{\ntot^{\alpha}}\right)c_2\frac{n_{\rm H_2}}{n_{\rm H}^2} \delta_{{\rm H},k}.\nonumber
\end{eqnarray}

\noindent Substituting these results in Eq. \ref{big} we obtain seven terms:
\begin{eqnarray}\label{WfullH} 
\frac{dF_i^{\rm Xsec}}{dn_k} &=& -r_i \frac{n_i}{W_Hn_H}D_K\\ \nonumber
&-& r_i \frac{H_X\ntot}{W_{\rm H}n_{\rm H}}\delta_{ik}\\ \nonumber
&-& r_i \frac{H_X}{W^2_{\rm H}n_{\rm H}}n_iW_{0,{\rm H}}c_1\alpha \left(\frac{n_{\rm el}}{\ntot}\right)^{\alpha} \left( 1+c_2\frac{n_{\rm H_2}}{n_{\rm H}}\right) Q(k,{\rm H}) \\  \nonumber
&-& r_i \frac{H_X\ntot n_i}{W^2_{\rm H}n_{\rm H}^2} W_{0,{\rm H}}\left(1+c_1\left(\frac{n_{\rm el}}{\ntot}\right)^{\alpha}\right)c_2\frac{n_{\rm H_2}}{n_{\rm H}}\delta_{{\rm H},k} \\ \nonumber
&+& r_i \frac{H_X}{W^2_{\rm H}n_{\rm H}}n_iW_{0,{\rm H}}c_1\alpha \left(\frac{n_{\rm el}}{\ntot}\right)^{\alpha-1} \left( 1+c_2\frac{n_{\rm H_2}}{n_{\rm H}}\right) q_k \\ \nonumber
&+& r_i \frac{H_X\ntot n_i}{W^2_{\rm H}n_{\rm H}^2} W_{0,{\rm H}}\left(1+c_1\left(\frac{n_{\rm el}}{\ntot}\right)^{\alpha}\right)c_2\delta_{{\rm H}_{2,k}} \\ \nonumber
&+& r_i \frac{H_X\ntot n_i}{W_{\rm H}n_{\rm H^2}}\delta_{{\rm H},k}. 
\end{eqnarray}

\noindent The seven terms can be interpreted as follows:
\begin{enumerate}
\item The first term represents the variation of the energy deposition $H_{\rm X}$
  when $n_k$ is increased.  According to this term only, increasing
  $n_k$ brings more fast electrons in the gas phase via primary
  ionization of the species \emph{k}. Hence more electrons are
  available for the secondary ionization of the species \emph{i}.
\item The second term directly impacts $\rse$ if \emph{i=k} more particles
  are available for secondary ionization.
\item The third calculates the variations in $W$ if the species
  considered contains hydrogen.
\item The fourth term comes from the dependency of $W_{\rm H}$ on
  $n_{\rm H}$ (the higher $n_{\rm H}$ the closer is $W_{\rm H}$ to
  $W_{\rm 0,H}$).
\item Analogous to the previous one, the fifth term shows that if the
  molecular hydrogen density increases, atoms are less likely to be
  ionized by fast electrons because $W_{\rm H}$ will be higher.
\item If the density of a positively charged particle increases, the
  electron density must increase at the same time as well. This
  favors Coulomb losses in energy of the incoming fast electrons over
  secondary ionization of species \emph{i}. If $k=$H$^-$, this sixth
  term becomes negative because fewer electrons will be in the gas
  phase, favouring secondary ionization over Coulomb heating.
\item The seventh term comes directly from the secondary ionization
  rate. The higher the atomic hydrogen density, the lower is $\rse$.
\end{enumerate}

\subsection{Molecular Hydrogen and Helium} 

\noindent The mean energy per ion pair for H$_2$ is

\begin{equation}
W_{\rm H_2} = W_{0,{\rm H_2}}\left(1 + c_1 \left(\epsilon_{\rm el}^{*}\right)^{\alpha}\right)\left(1 + c_2\frac{n_{\rm H}}{n_{\rm H_2}}\right),
\end{equation}

\noindent while for helium, it is

\begin{equation}
W_{\rm He} = W_{0,{\rm He}}\left(1 + c_1 \left(\epsilon_{\rm el}\right)^{\alpha}\right)
\end{equation}

\noindent where $\epsilon^*_{\rm el}$ is

\begin{equation}
\epsilon^*_{\rm el} = \frac{1.83\epsilon_{\rm el}}{1+0.83\epsilon_{\rm el}},
\end{equation}

\noindent with $\epsilon_{\rm el}$ is the electron fraction ($n_{\rm
  el}/\ntot$). Inserting these terms into \ref{Fhy} and generating the
derivative with respect to $n_k$ gives the Jacobi elements for
molecular hydrogen and helium respectively. Since they look very
similar to Eq. \ref{WfullH}, we will not write them out.

\onecolumn

\section{Additional figures}

\begin{figure*}
  \centering
  \includegraphics[width=4cm,clip]{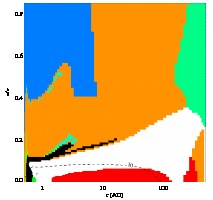}
  \includegraphics[width=4cm,clip]{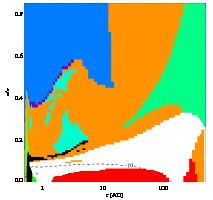}
  \includegraphics[width=4cm,clip]{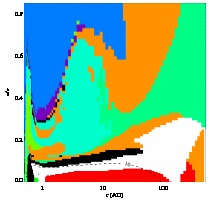}
  \includegraphics[width=4cm,clip]{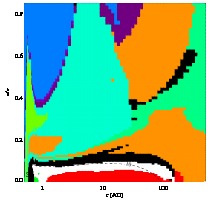}
  \includegraphics[width=4cm,clip]{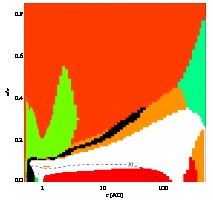}
  \includegraphics[width=4cm,clip]{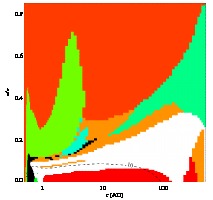}
  \includegraphics[width=4cm,clip]{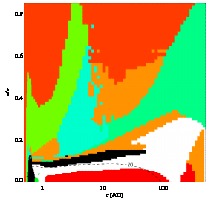}
  \includegraphics[width=4cm,clip]{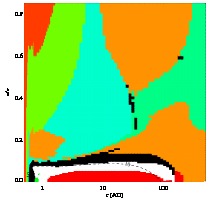}
  \includegraphics[width=4cm,clip]{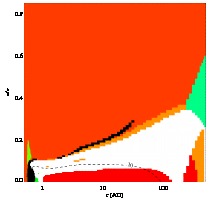}
  \includegraphics[width=4cm,clip]{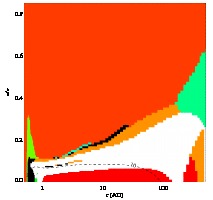}
  \includegraphics[width=4cm,clip]{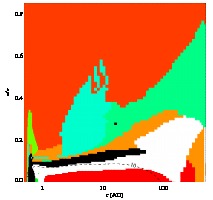}
  \includegraphics[width=4cm,clip]{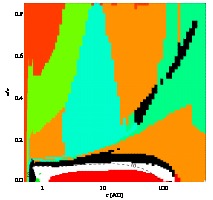}
  \includegraphics[width=4cm,clip]{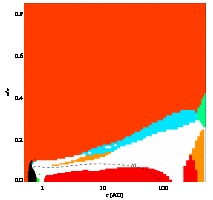}
  \includegraphics[width=4cm,clip]{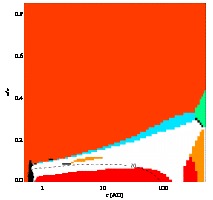}
  \includegraphics[width=4cm,clip]{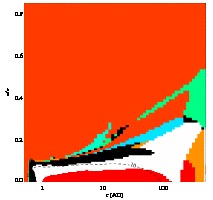}
  \includegraphics[width=4cm,clip]{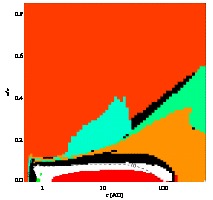}
  \includegraphics[width=4cm,clip]{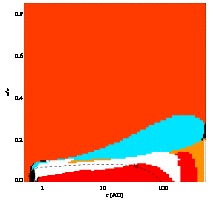}
  \includegraphics[width=4cm,clip]{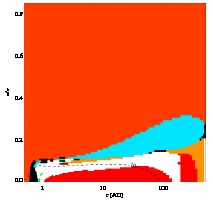}
  \includegraphics[width=4cm,clip]{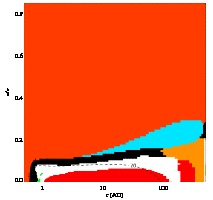}
  \includegraphics[width=4cm,clip]{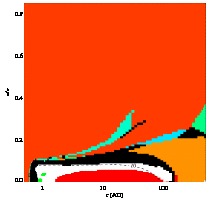}
  \caption{Main heating sources throughout the disk: FUV luminosity
    increasing from $L_{FUV}=10^{29}$ (left) to $10^{32}$~erg/s. X-ray
    luminosity increasing from $L_x=0$ to $10^{32}$~erg/s. background
    heating by [CII] (blue), PAH heating (orange), photo-electric
    heating (dark purple), X-ray Coulomb heating (light red), heating
    by collisional de-exciation of H$_2$ (blue-green), CI ionization
    heating (green-blue), infrared background by CO ro-vibrational
    lines (black), heating by thermal accomodation grains (white),
    cosmic ray heating (red), X-ray H$_2$ dissociation heating (light
    blue), free-free absorption (green), background heating by FeII
    (light green), background heating by SiII (green-yellow), infrared
    background heating by H$_2$O rotational transitions (yellow),
    heating by H$_2$ formation on dust (dark blue), and background
    heating by [OI] (purple). }
  \label{heating}
\end{figure*}

\begin{figure*}
  \centering
  \includegraphics[width=4cm,clip]{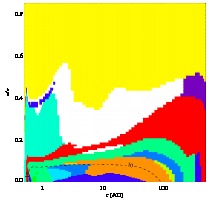}
  \includegraphics[width=4cm,clip]{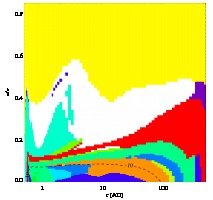}
  \includegraphics[width=4cm,clip]{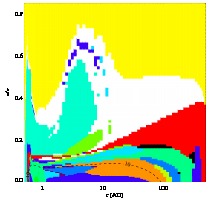}
  \includegraphics[width=4cm,clip]{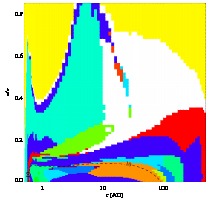}
  \includegraphics[width=4cm,clip]{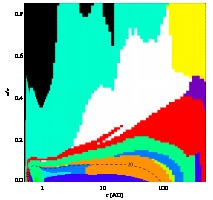}
  \includegraphics[width=4cm,clip]{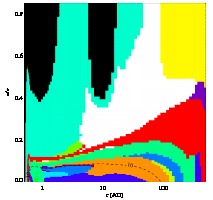}
  \includegraphics[width=4cm,clip]{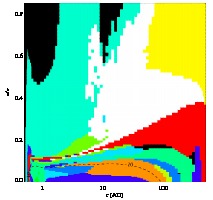}
  \includegraphics[width=4cm,clip]{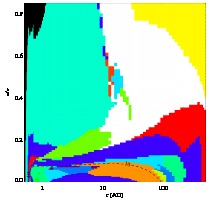}
  \includegraphics[width=4cm,clip]{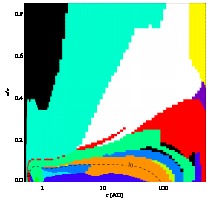}
  \includegraphics[width=4cm,clip]{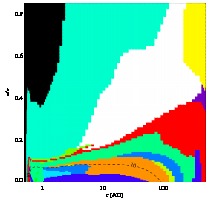}
  \includegraphics[width=4cm,clip]{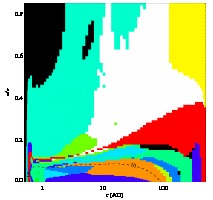}
  \includegraphics[width=4cm,clip]{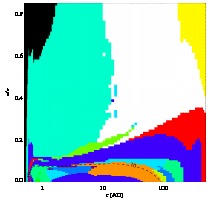}
  \includegraphics[width=4cm,clip]{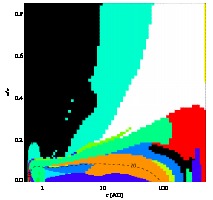}
  \includegraphics[width=4cm,clip]{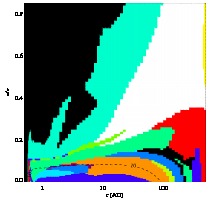}
  \includegraphics[width=4cm,clip]{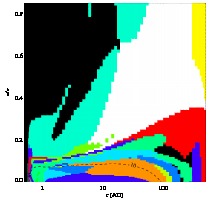}
  \includegraphics[width=4cm,clip]{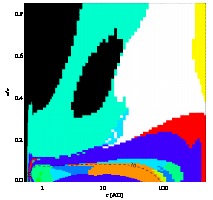}
  \includegraphics[width=4cm,clip]{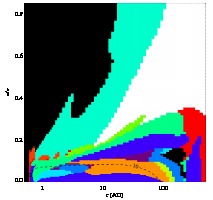}
  \includegraphics[width=4cm,clip]{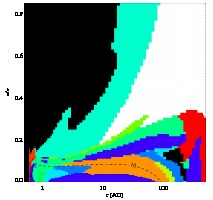}
  \includegraphics[width=4cm,clip]{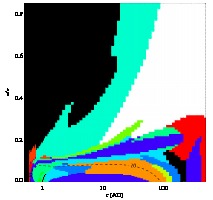}
  \includegraphics[width=4cm,clip]{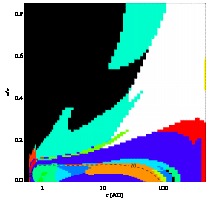}
  \caption{Main cooling sources throughout the disk: FUV luminosity
    increasing from $L_{FUV}=10^{29}$ (left) to $10^{32}$~erg/s. X-ray
    luminosity increasing from $L_x=0$ to $10^{32}$~erg/s. Lyman
    $\alpha$ cooling (black), [FeII] line cooling (blue-green), [OI]
    line cooling (white), [CII] line cooling (yellow), CO rotational and
    ro-vibrational cooling (red), H$_2$O rotational cooling
    (green-blue), OH rotational cooling (light green), HCN line
    cooling (dark purple), [CI] line cooling (light purple), HNC line
    cooling (blue), cooling by thermal accomodation on grains (dark
    blue), CS line cooling (yellow-green), H$_2$ line cooling (light
    blue), chemical cooling (red-orange), and free-free emission
    (green).}
  \label{cooling}
\end{figure*}

\begin{figure*}
  \centering
  \includegraphics[width=4cm,clip]{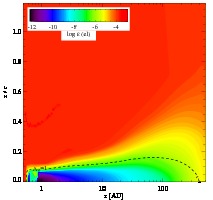}
  \includegraphics[width=4cm,clip]{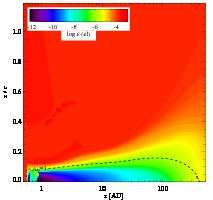}
  \includegraphics[width=4cm,clip]{electron_abundance_00034.jpg}
  \includegraphics[width=4cm,clip]{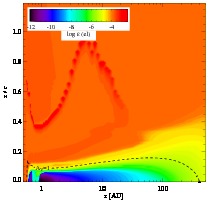}
  \includegraphics[width=4cm,clip]{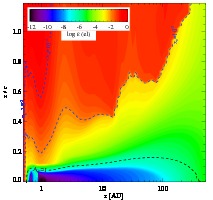}
  \includegraphics[width=4cm,clip]{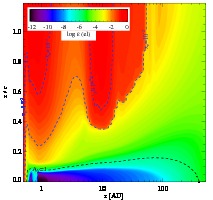}
  \includegraphics[width=4cm,clip]{electron_abundance_00082.jpg}
  \includegraphics[width=4cm,clip]{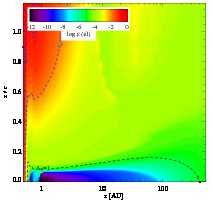}
  \includegraphics[width=4cm,clip]{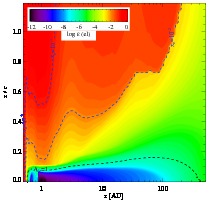}
  \includegraphics[width=4cm,clip]{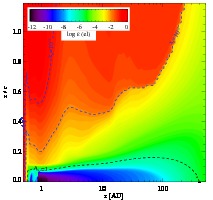}
  \includegraphics[width=4cm,clip]{electron_abundance_00130.jpg}
  \includegraphics[width=4cm,clip]{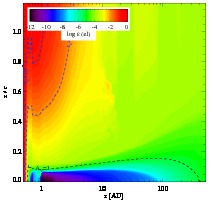}
  \includegraphics[width=4cm,clip]{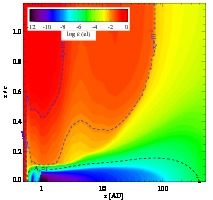}
  \includegraphics[width=4cm,clip]{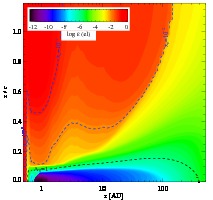}
  \includegraphics[width=4cm,clip]{electron_abundance_00178.jpg}
  \includegraphics[width=4cm,clip]{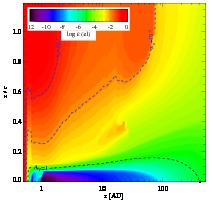}
  \includegraphics[width=4cm,clip]{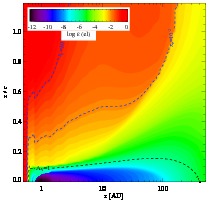}
  \includegraphics[width=4cm,clip]{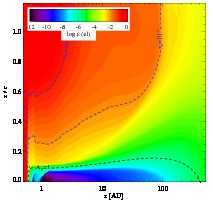}
  \includegraphics[width=4cm,clip]{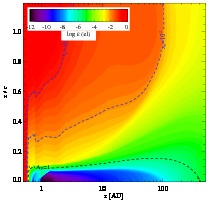}
  \includegraphics[width=4cm,clip]{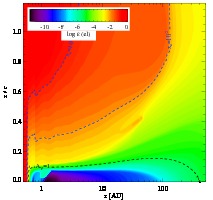}
  \caption{Electron abundance: FUV luminosity increasing from
    $L_{FUV}=10^{29}$ (left) to $10^{32}$~erg/s (right). X-ray luminosity
    increasing from $L_x=0$ (top) to $10^{32}$~erg/s (bottom).}
  \label{model_electron_abundance_appendix}
\end{figure*}

\begin{figure*}
  \centering
  \includegraphics[width=4cm,clip]{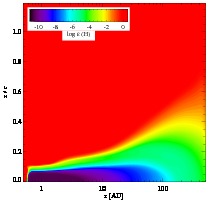}
  \includegraphics[width=4cm,clip]{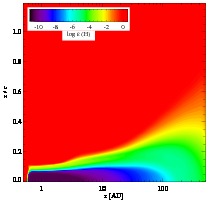}
  \includegraphics[width=4cm,clip]{Habundance_00034.jpg}
  \includegraphics[width=4cm,clip]{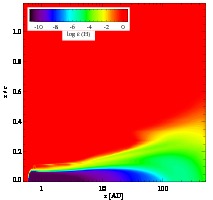}
  \includegraphics[width=4cm,clip]{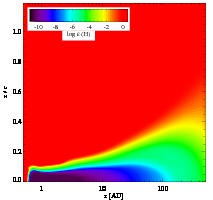}
  \includegraphics[width=4cm,clip]{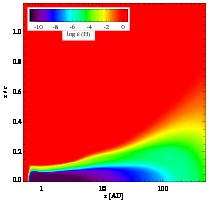}
  \includegraphics[width=4cm,clip]{Habundance_00082.jpg}
  \includegraphics[width=4cm,clip]{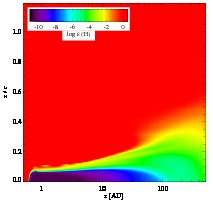}
  \includegraphics[width=4cm,clip]{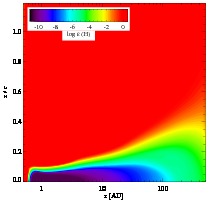}
  \includegraphics[width=4cm,clip]{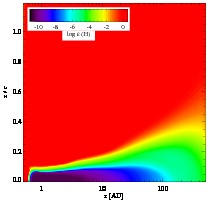}
  \includegraphics[width=4cm,clip]{Habundance_00130.jpg}
  \includegraphics[width=4cm,clip]{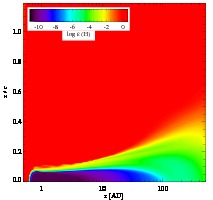}
  \includegraphics[width=4cm,clip]{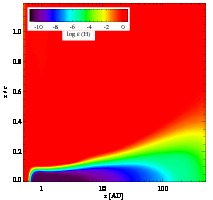}
  \includegraphics[width=4cm,clip]{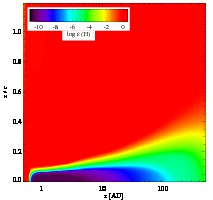}
  \includegraphics[width=4cm,clip]{Habundance_00178.jpg}
  \includegraphics[width=4cm,clip]{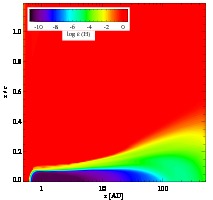}
  \includegraphics[width=4cm,clip]{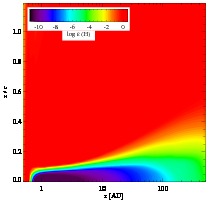}
  \includegraphics[width=4cm,clip]{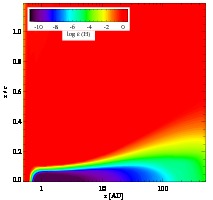}
  \includegraphics[width=4cm,clip]{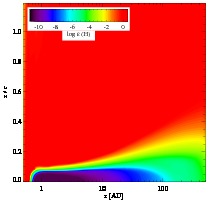}
  \includegraphics[width=4cm,clip]{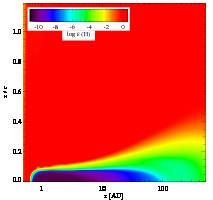}
  \caption{H abundances. FUV and X-ray fluxes are the same as Fig. \ref{model_electron_abundance_appendix}.}
  \label{Habundance_struct_appendix}
\end{figure*}

\begin{figure*}
  \centering
  \includegraphics[width=4cm,clip]{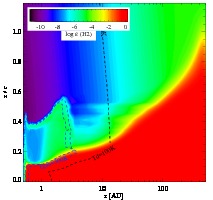}
  \includegraphics[width=4cm,clip]{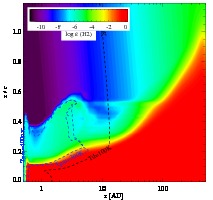}
  \includegraphics[width=4cm,clip]{H2abundance_00034.jpg}
  \includegraphics[width=4cm,clip]{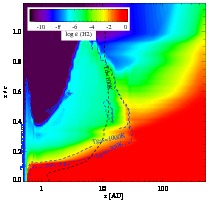}
  \includegraphics[width=4cm,clip]{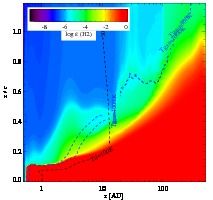}
  \includegraphics[width=4cm,clip]{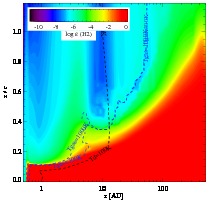}
  \includegraphics[width=4cm,clip]{H2abundance_00082.jpg}
  \includegraphics[width=4cm,clip]{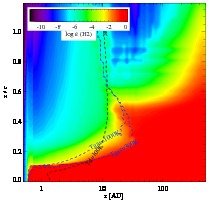}
  \includegraphics[width=4cm,clip]{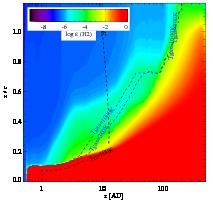}
  \includegraphics[width=4cm,clip]{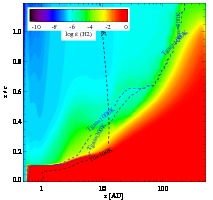}
  \includegraphics[width=4cm,clip]{H2abundance_00130.jpg}
  \includegraphics[width=4cm,clip]{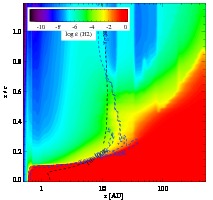}
  \includegraphics[width=4cm,clip]{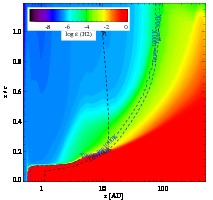}
  \includegraphics[width=4cm,clip]{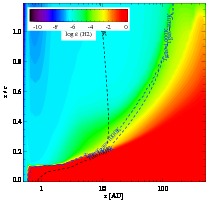}
  \includegraphics[width=4cm,clip]{H2abundance_00178.jpg}
  \includegraphics[width=4cm,clip]{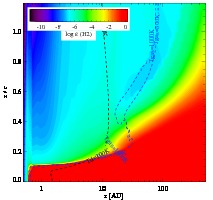}
  \includegraphics[width=4cm,clip]{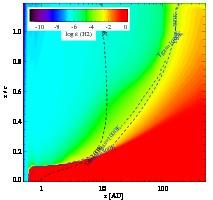}
  \includegraphics[width=4cm,clip]{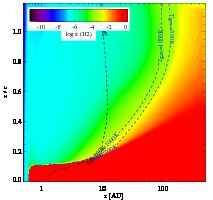}
  \includegraphics[width=4cm,clip]{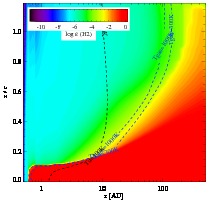}
  \includegraphics[width=4cm,clip]{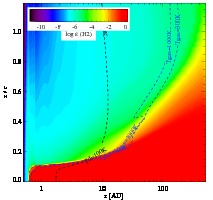}
  \caption{H$_2$ abundances. FUV and X-ray fluxes are the same as Fig. \ref{model_electron_abundance_appendix}.}
  \label{H2abundance_struct_appendix}
\end{figure*}

\begin{figure*}
  \centering
  \includegraphics[width=4cm,clip]{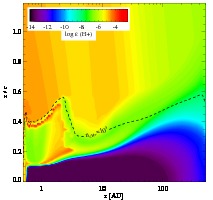}
  \includegraphics[width=4cm,clip]{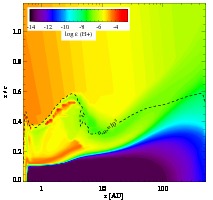}
  \includegraphics[width=4cm,clip]{H+abundance_00034.jpg}
  \includegraphics[width=4cm,clip]{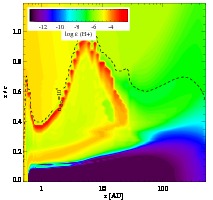}
  \includegraphics[width=4cm,clip]{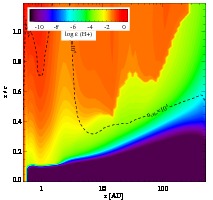}
  \includegraphics[width=4cm,clip]{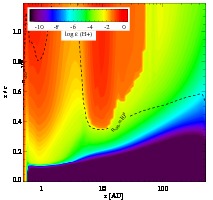}
  \includegraphics[width=4cm,clip]{H+abundance_00082.jpg}
  \includegraphics[width=4cm,clip]{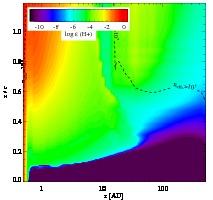}
  \includegraphics[width=4cm,clip]{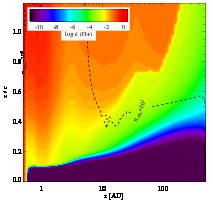}
  \includegraphics[width=4cm,clip]{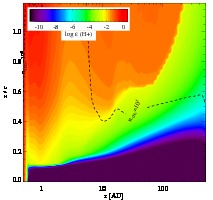}
  \includegraphics[width=4cm,clip]{H+abundance_00130.jpg}
  \includegraphics[width=4cm,clip]{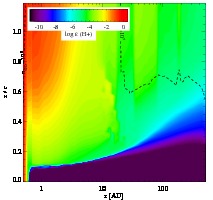}
  \includegraphics[width=4cm,clip]{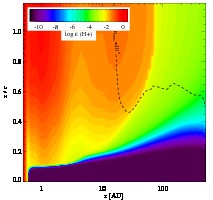}
  \includegraphics[width=4cm,clip]{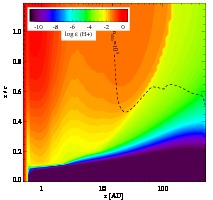}
  \includegraphics[width=4cm,clip]{H+abundance_00178.jpg}
  \includegraphics[width=4cm,clip]{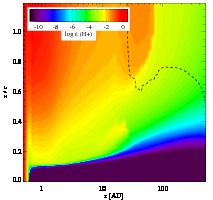}
  \includegraphics[width=4cm,clip]{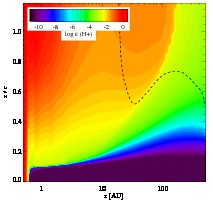}
  \includegraphics[width=4cm,clip]{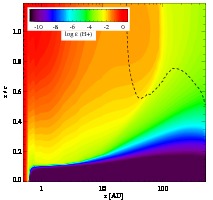}
  \includegraphics[width=4cm,clip]{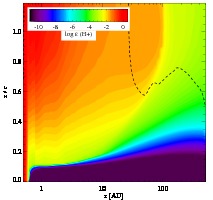}
  \includegraphics[width=4cm,clip]{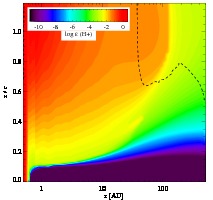}
  \caption{H$^+$ abundances. FUV and X-ray fluxes are the same as Fig. \ref{model_electron_abundance_appendix}.}
  \label{H+abundance_struct_appendix}
\end{figure*}

\begin{figure*}
  \centering
  \includegraphics[width=4cm,clip]{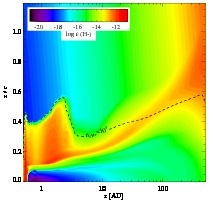}
  \includegraphics[width=4cm,clip]{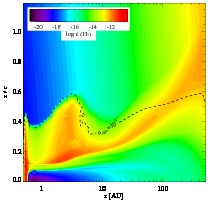}
  \includegraphics[width=4cm,clip]{H-abundance_00034.jpg}
  \includegraphics[width=4cm,clip]{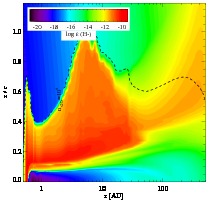}
  \includegraphics[width=4cm,clip]{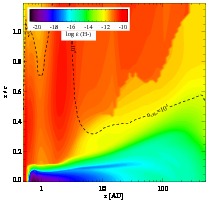}
  \includegraphics[width=4cm,clip]{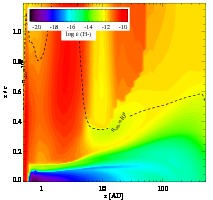}
  \includegraphics[width=4cm,clip]{H-abundance_00082.jpg}
  \includegraphics[width=4cm,clip]{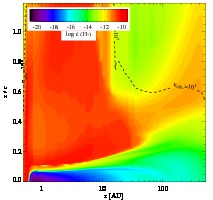}
  \includegraphics[width=4cm,clip]{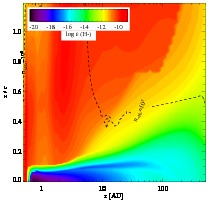}
  \includegraphics[width=4cm,clip]{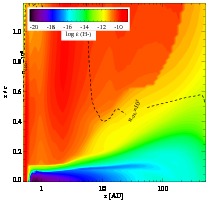}
  \includegraphics[width=4cm,clip]{H-abundance_00130.jpg}
  \includegraphics[width=4cm,clip]{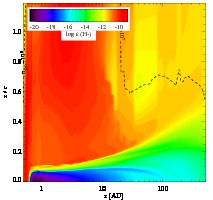}
  \includegraphics[width=4cm,clip]{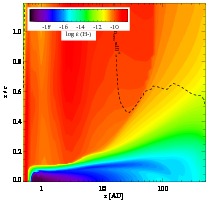}
  \includegraphics[width=4cm,clip]{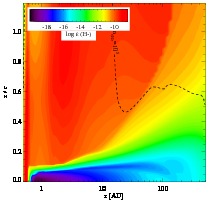}
  \includegraphics[width=4cm,clip]{H-abundance_00178.jpg}
  \includegraphics[width=4cm,clip]{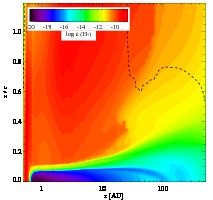}
  \includegraphics[width=4cm,clip]{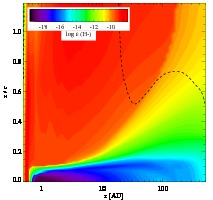}
  \includegraphics[width=4cm,clip]{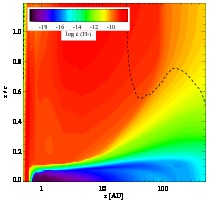}
  \includegraphics[width=4cm,clip]{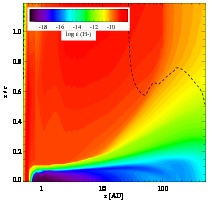}
  \includegraphics[width=4cm,clip]{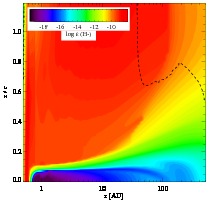}
  \caption{H$^-$ abundances. FUV and X-ray fluxes are the same as Fig. \ref{model_electron_abundance_appendix}.}
  \label{H-abundance_struct_appendix}
\end{figure*}

\begin{figure*}
  \centering
  \includegraphics[width=4cm,clip]{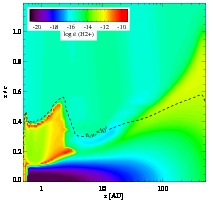}
  \includegraphics[width=4cm,clip]{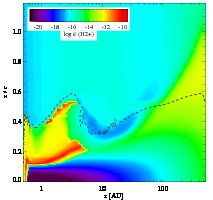}
  \includegraphics[width=4cm,clip]{H2+abundance_00034.jpg}
  \includegraphics[width=4cm,clip]{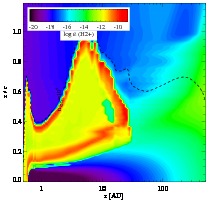}
  \includegraphics[width=4cm,clip]{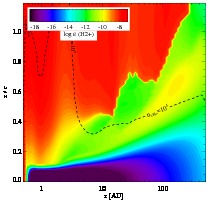}
  \includegraphics[width=4cm,clip]{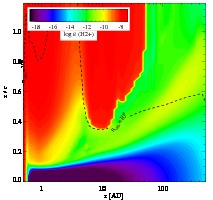}
  \includegraphics[width=4cm,clip]{H2+abundance_00082.jpg}
  \includegraphics[width=4cm,clip]{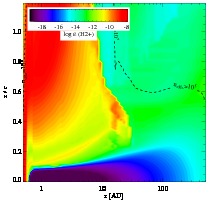}
  \includegraphics[width=4cm,clip]{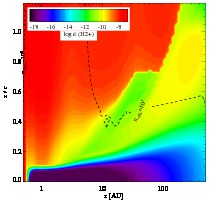}
  \includegraphics[width=4cm,clip]{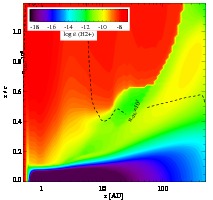}
  \includegraphics[width=4cm,clip]{H2+abundance_00130.jpg}
  \includegraphics[width=4cm,clip]{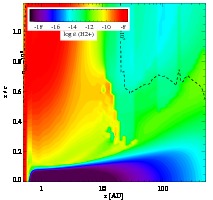}
  \includegraphics[width=4cm,clip]{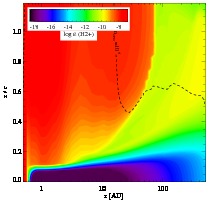}
  \includegraphics[width=4cm,clip]{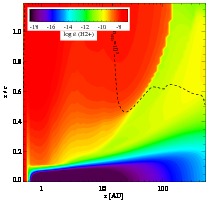}
  \includegraphics[width=4cm,clip]{H2+abundance_00178.jpg}
  \includegraphics[width=4cm,clip]{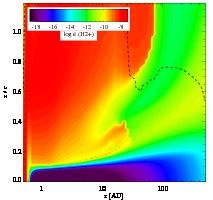}
  \includegraphics[width=4cm,clip]{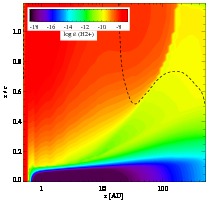}
  \includegraphics[width=4cm,clip]{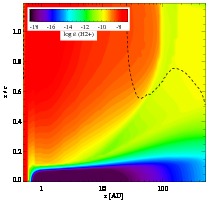}
  \includegraphics[width=4cm,clip]{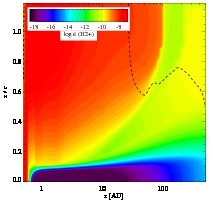}
  \includegraphics[width=4cm,clip]{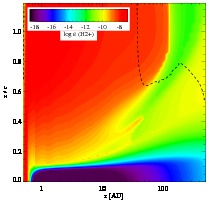}
  \caption{H$_2^+$ abundances. FUV and X-ray fluxes are the same as Fig. \ref{model_electron_abundance_appendix}.}
  \label{H2+abundance_struct_appendix}
\end{figure*}

\begin{figure*}
  \centering
  \includegraphics[width=4cm,clip]{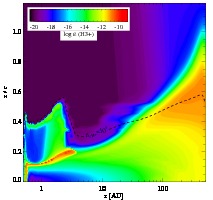}
  \includegraphics[width=4cm,clip]{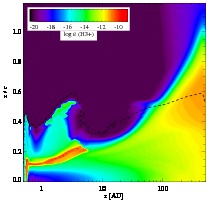}
  \includegraphics[width=4cm,clip]{H3+abundance_00034.jpg}
  \includegraphics[width=4cm,clip]{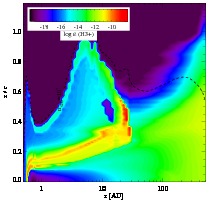}
  \includegraphics[width=4cm,clip]{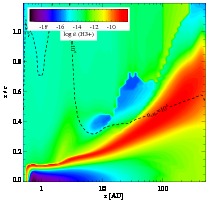}
  \includegraphics[width=4cm,clip]{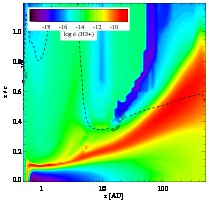}
  \includegraphics[width=4cm,clip]{H3+abundance_00082.jpg}
  \includegraphics[width=4cm,clip]{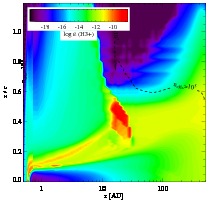}
  \includegraphics[width=4cm,clip]{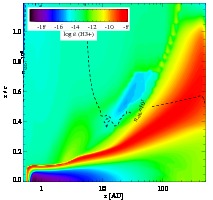}
  \includegraphics[width=4cm,clip]{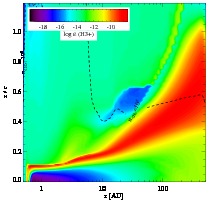}
  \includegraphics[width=4cm,clip]{H3+abundance_00130.jpg}
  \includegraphics[width=4cm,clip]{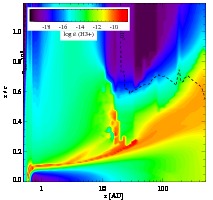}
  \includegraphics[width=4cm,clip]{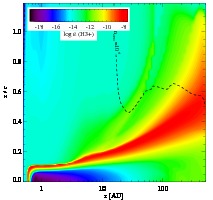}
  \includegraphics[width=4cm,clip]{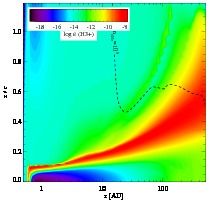}
  \includegraphics[width=4cm,clip]{H3+abundance_00178.jpg}
  \includegraphics[width=4cm,clip]{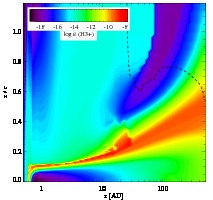}
  \includegraphics[width=4cm,clip]{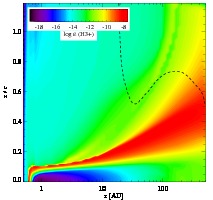}
  \includegraphics[width=4cm,clip]{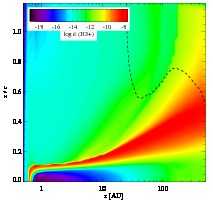}
  \includegraphics[width=4cm,clip]{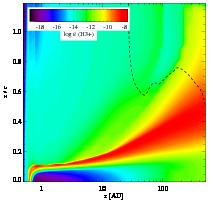}
  \includegraphics[width=4cm,clip]{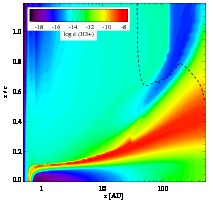}
  \caption{H$_3^+$ abundances. FUV and X-ray fluxes are the same as Fig. \ref{model_electron_abundance_appendix}.}
  \label{H3+abundance_struct_appendix}
\end{figure*}

\begin{figure*}
  \centering
  \includegraphics[width=4cm,clip]{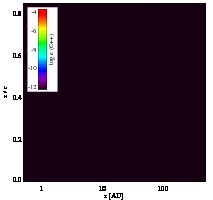}
  \includegraphics[width=4cm,clip]{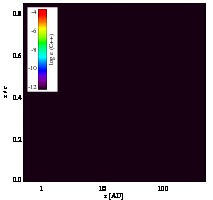}
  \includegraphics[width=4cm,clip]{C2+abundance_00034.jpg}
  \includegraphics[width=4cm,clip]{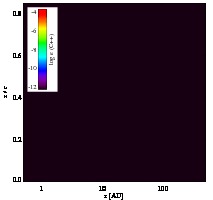}
  \includegraphics[width=4cm,clip]{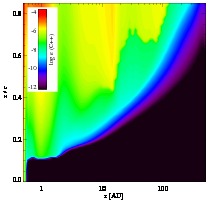}
  \includegraphics[width=4cm,clip]{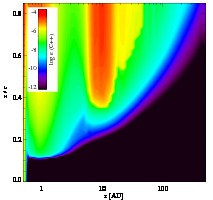}
  \includegraphics[width=4cm,clip]{C2+abundance_00082.jpg}
  \includegraphics[width=4cm,clip]{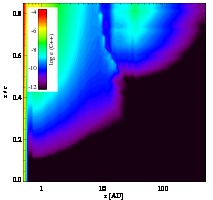}
  \includegraphics[width=4cm,clip]{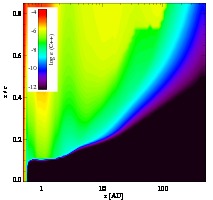}
  \includegraphics[width=4cm,clip]{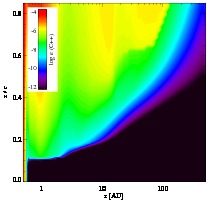}
  \includegraphics[width=4cm,clip]{C2+abundance_00130.jpg}
  \includegraphics[width=4cm,clip]{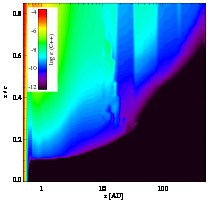}
  \includegraphics[width=4cm,clip]{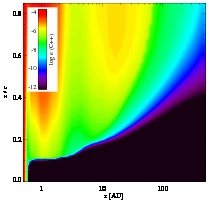}
  \includegraphics[width=4cm,clip]{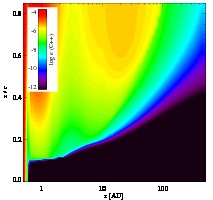}
  \includegraphics[width=4cm,clip]{C2+abundance_00178.jpg}
  \includegraphics[width=4cm,clip]{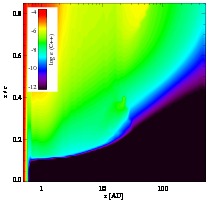}
  \includegraphics[width=4cm,clip]{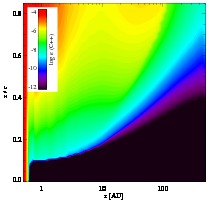}
  \includegraphics[width=4cm,clip]{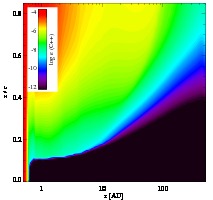}
  \includegraphics[width=4cm,clip]{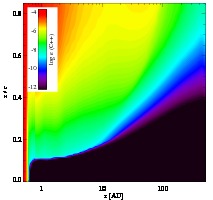}
  \includegraphics[width=4cm,clip]{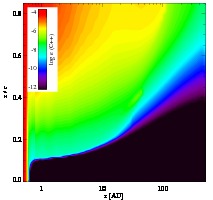}
  \caption{C$^{2+}$ abundances. FUV and X-ray fluxes are the same as Fig. \ref{model_electron_abundance_appendix}.}
  \label{C2+abundance_struct_appendix}
\end{figure*}

\begin{figure*}
  \centering
  \includegraphics[width=4cm,clip]{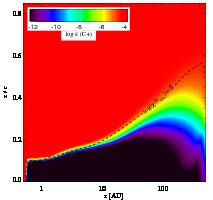}
  \includegraphics[width=4cm,clip]{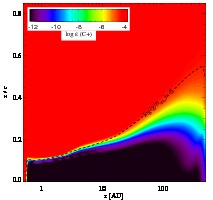}
  \includegraphics[width=4cm,clip]{C+abundance_00034.jpg}
  \includegraphics[width=4cm,clip]{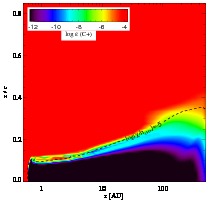}
  \includegraphics[width=4cm,clip]{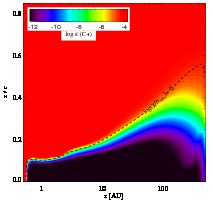}
  \includegraphics[width=4cm,clip]{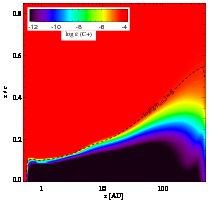}
  \includegraphics[width=4cm,clip]{C+abundance_00082.jpg}
  \includegraphics[width=4cm,clip]{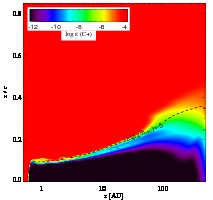}
  \includegraphics[width=4cm,clip]{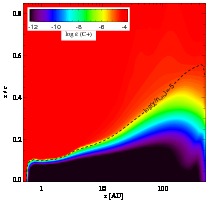}
  \includegraphics[width=4cm,clip]{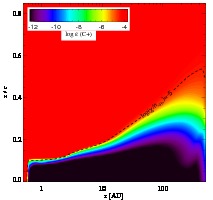}
  \includegraphics[width=4cm,clip]{C+abundance_00130.jpg}
  \includegraphics[width=4cm,clip]{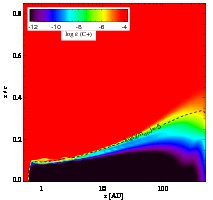}
  \includegraphics[width=4cm,clip]{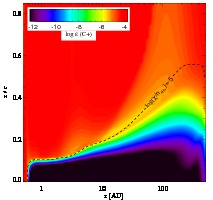}
  \includegraphics[width=4cm,clip]{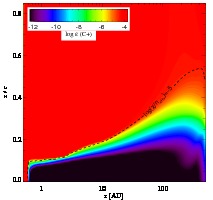}
  \includegraphics[width=4cm,clip]{C+abundance_00178.jpg}
  \includegraphics[width=4cm,clip]{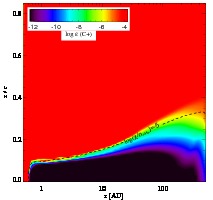}
  \includegraphics[width=4cm,clip]{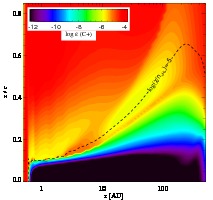}
  \includegraphics[width=4cm,clip]{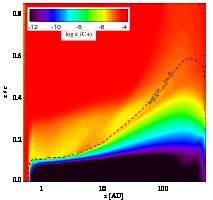}
  \includegraphics[width=4cm,clip]{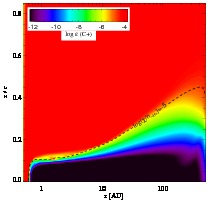}
  \includegraphics[width=4cm,clip]{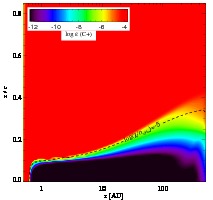}
  \caption{C$^+$ abundances. FUV and X-ray fluxes are the same as Fig. \ref{model_electron_abundance_appendix}.}
  \label{C+abundance_struct_appendix}
\end{figure*}

\begin{figure*}
  \centering
  \includegraphics[width=4cm,clip]{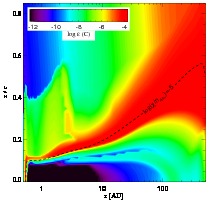}
  \includegraphics[width=4cm,clip]{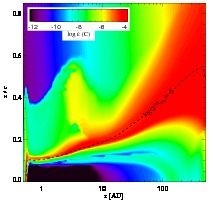}
  \includegraphics[width=4cm,clip]{Cabundance_00034.jpg}
  \includegraphics[width=4cm,clip]{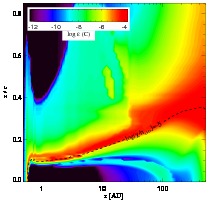}
  \includegraphics[width=4cm,clip]{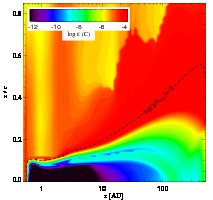}
  \includegraphics[width=4cm,clip]{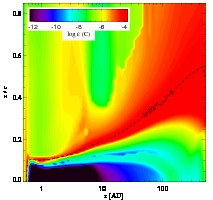}
  \includegraphics[width=4cm,clip]{Cabundance_00082.jpg}
  \includegraphics[width=4cm,clip]{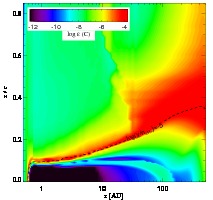}
  \includegraphics[width=4cm,clip]{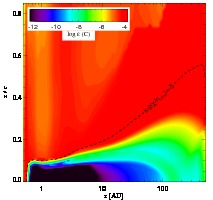}
  \includegraphics[width=4cm,clip]{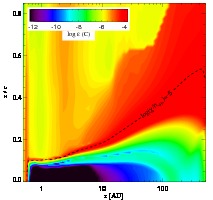}
  \includegraphics[width=4cm,clip]{Cabundance_00130.jpg}
  \includegraphics[width=4cm,clip]{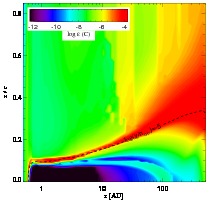}
  \includegraphics[width=4cm,clip]{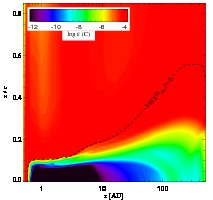}
  \includegraphics[width=4cm,clip]{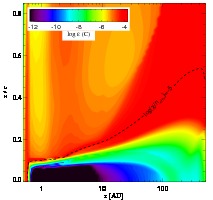}
  \includegraphics[width=4cm,clip]{Cabundance_00178.jpg}
  \includegraphics[width=4cm,clip]{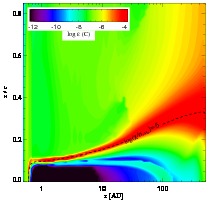}
  \includegraphics[width=4cm,clip]{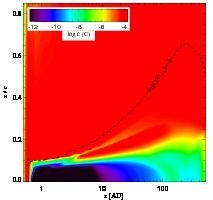}
  \includegraphics[width=4cm,clip]{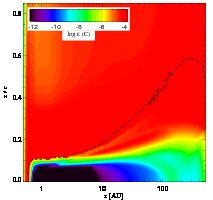}
  \includegraphics[width=4cm,clip]{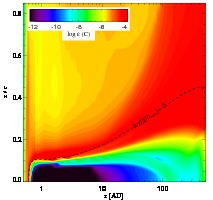}
  \includegraphics[width=4cm,clip]{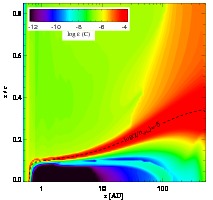}
  \caption{C abundances. FUV and X-ray fluxes are the same as Fig. \ref{model_electron_abundance_appendix}.}
  \label{Cabundance_struct_appendix}
\end{figure*}

\clearpage

\begin{figure*}
  \centering
  \includegraphics[width=4cm,clip]{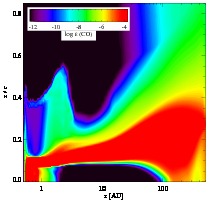}
  \includegraphics[width=4cm,clip]{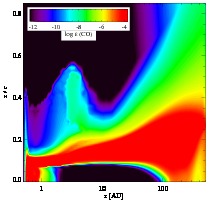}
  \includegraphics[width=4cm,clip]{COabundance_00034.jpg}
  \includegraphics[width=4cm,clip]{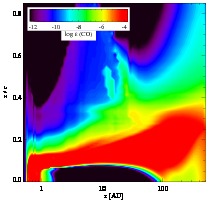}
  \includegraphics[width=4cm,clip]{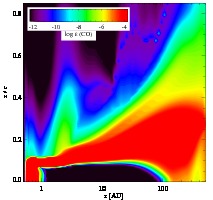}
  \includegraphics[width=4cm,clip]{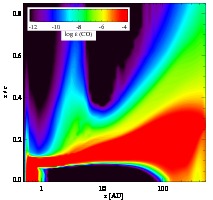}
  \includegraphics[width=4cm,clip]{COabundance_00082.jpg}
  \includegraphics[width=4cm,clip]{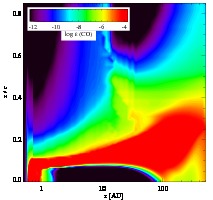}
  \includegraphics[width=4cm,clip]{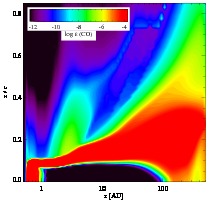}
  \includegraphics[width=4cm,clip]{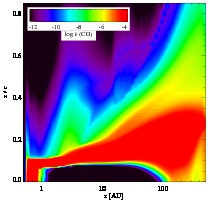}
  \includegraphics[width=4cm,clip]{COabundance_00130.jpg}
  \includegraphics[width=4cm,clip]{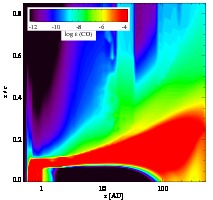}
  \includegraphics[width=4cm,clip]{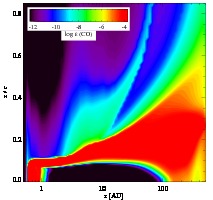}
  \includegraphics[width=4cm,clip]{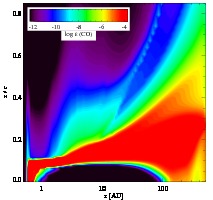}
  \includegraphics[width=4cm,clip]{COabundance_00178.jpg}
  \includegraphics[width=4cm,clip]{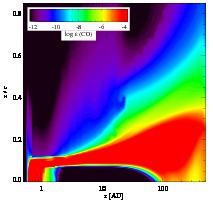}
  \includegraphics[width=4cm,clip]{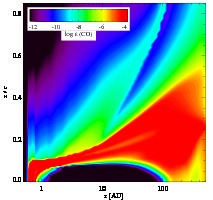}
  \includegraphics[width=4cm,clip]{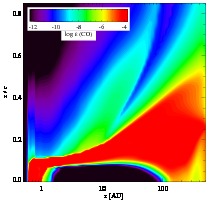}
  \includegraphics[width=4cm,clip]{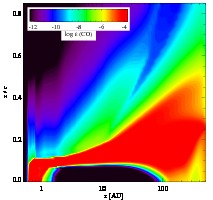}
  \includegraphics[width=4cm,clip]{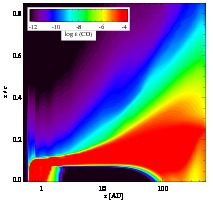}
  \caption{CO abundances. FUV and X-ray fluxes are the same as Fig. \ref{model_electron_abundance_appendix}.}
  \label{COabundance_struct_appendix}
\end{figure*}

\begin{figure*}
  \centering
  \includegraphics[width=4cm,clip]{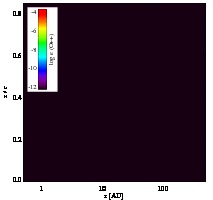}
  \includegraphics[width=4cm,clip]{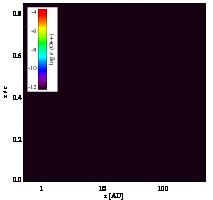}
  \includegraphics[width=4cm,clip]{O2+abundance_00034.jpg}
  \includegraphics[width=4cm,clip]{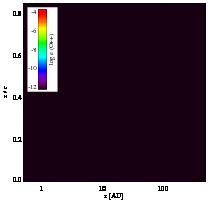}
  \includegraphics[width=4cm,clip]{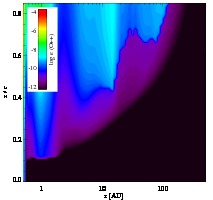}
  \includegraphics[width=4cm,clip]{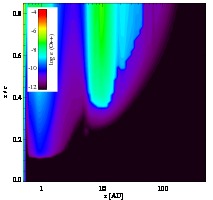}
  \includegraphics[width=4cm,clip]{O2+abundance_00082.jpg}
  \includegraphics[width=4cm,clip]{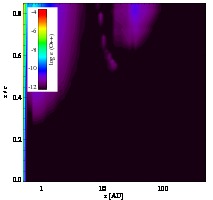}
  \includegraphics[width=4cm,clip]{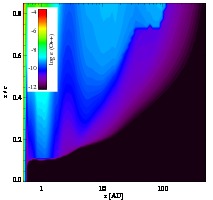}
  \includegraphics[width=4cm,clip]{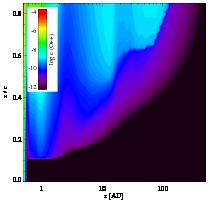}
  \includegraphics[width=4cm,clip]{O2+abundance_00130.jpg}
  \includegraphics[width=4cm,clip]{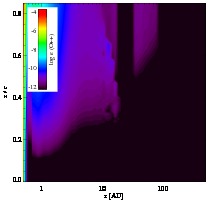}
  \includegraphics[width=4cm,clip]{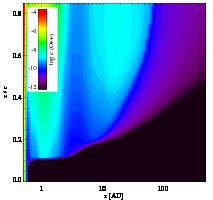}
  \includegraphics[width=4cm,clip]{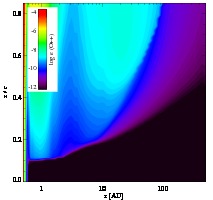}
  \includegraphics[width=4cm,clip]{O2+abundance_00178.jpg}
  \includegraphics[width=4cm,clip]{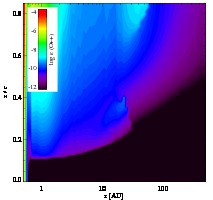}
  \includegraphics[width=4cm,clip]{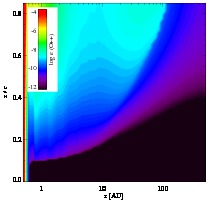}
  \includegraphics[width=4cm,clip]{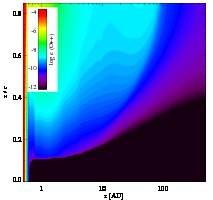}
  \includegraphics[width=4cm,clip]{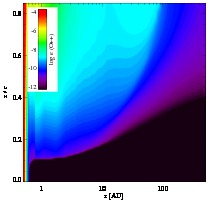}
  \includegraphics[width=4cm,clip]{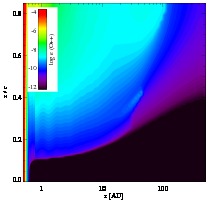}
  \caption{O$^{2+}$ abundances. FUV and X-ray fluxes are the same as Fig. \ref{model_electron_abundance_appendix}.}
  \label{O2+abundance_struct_appendix}
\end{figure*}

\begin{figure*}
  \centering
  \includegraphics[width=4cm,clip]{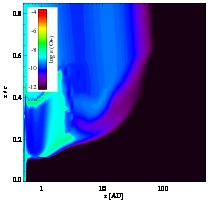}
  \includegraphics[width=4cm,clip]{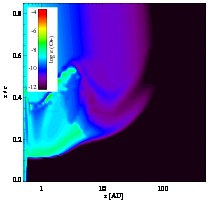}
  \includegraphics[width=4cm,clip]{O+abundance_00034.jpg}
  \includegraphics[width=4cm,clip]{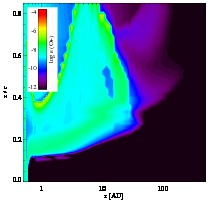}
  \includegraphics[width=4cm,clip]{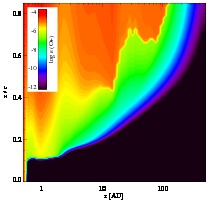}
  \includegraphics[width=4cm,clip]{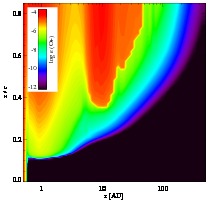}
  \includegraphics[width=4cm,clip]{O+abundance_00082.jpg}
  \includegraphics[width=4cm,clip]{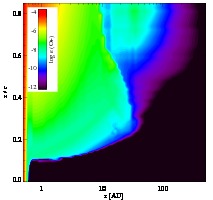}
  \includegraphics[width=4cm,clip]{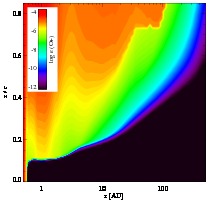}
  \includegraphics[width=4cm,clip]{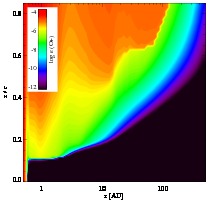}
  \includegraphics[width=4cm,clip]{O+abundance_00130.jpg}
  \includegraphics[width=4cm,clip]{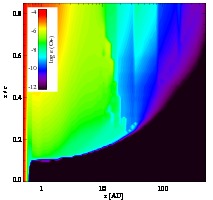}
  \includegraphics[width=4cm,clip]{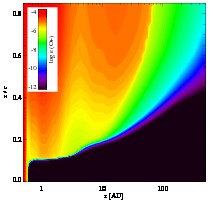}
  \includegraphics[width=4cm,clip]{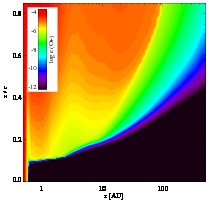}
  \includegraphics[width=4cm,clip]{O+abundance_00178.jpg}
  \includegraphics[width=4cm,clip]{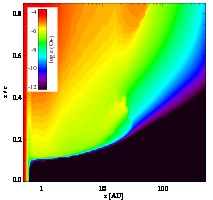}
  \includegraphics[width=4cm,clip]{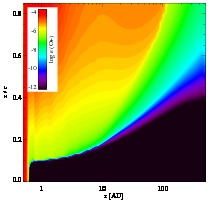}
  \includegraphics[width=4cm,clip]{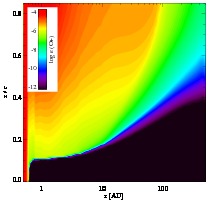}
  \includegraphics[width=4cm,clip]{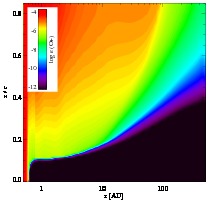}
  \includegraphics[width=4cm,clip]{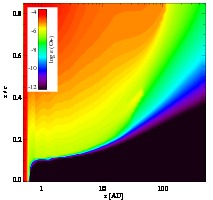}
  \caption{O$^+$ abundances. FUV and X-ray fluxes are the same as Fig. \ref{model_electron_abundance_appendix}.}
  \label{O+abundance_struct_appendix}
\end{figure*}

\begin{figure*}
  \centering
  \includegraphics[width=4cm,clip]{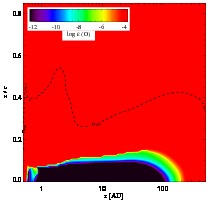}
  \includegraphics[width=4cm,clip]{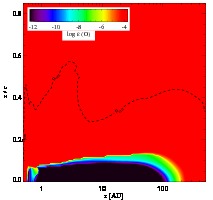}
  \includegraphics[width=4cm,clip]{Oabundance_00034.jpg}
  \includegraphics[width=4cm,clip]{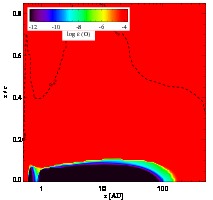}
  \includegraphics[width=4cm,clip]{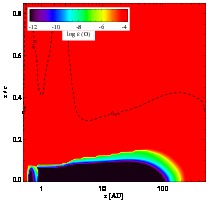}
  \includegraphics[width=4cm,clip]{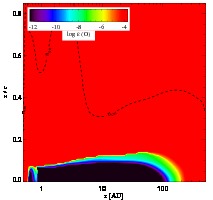}
  \includegraphics[width=4cm,clip]{Oabundance_00082.jpg}
  \includegraphics[width=4cm,clip]{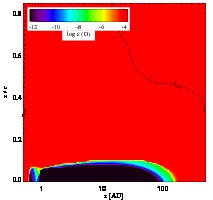}
  \includegraphics[width=4cm,clip]{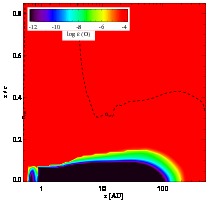}
  \includegraphics[width=4cm,clip]{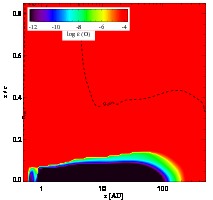}
  \includegraphics[width=4cm,clip]{Oabundance_00130.jpg}
  \includegraphics[width=4cm,clip]{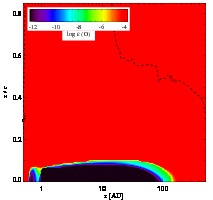}
  \includegraphics[width=4cm,clip]{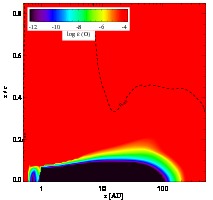}
  \includegraphics[width=4cm,clip]{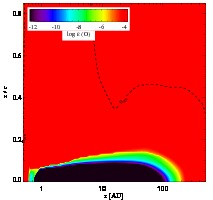}
  \includegraphics[width=4cm,clip]{Oabundance_00178.jpg}
  \includegraphics[width=4cm,clip]{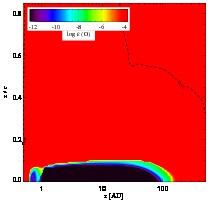}
  \includegraphics[width=4cm,clip]{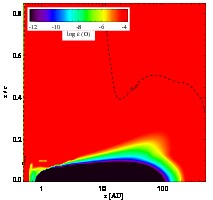}
  \includegraphics[width=4cm,clip]{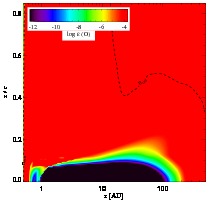}
  \includegraphics[width=4cm,clip]{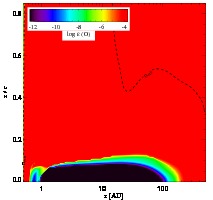}
  \includegraphics[width=4cm,clip]{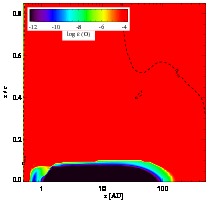}
  \caption{O abundances. FUV and X-ray fluxes are the same as Fig. \ref{model_electron_abundance_appendix}.}
  \label{Oabundance_struct_appendix}
\end{figure*}

\begin{figure*}
  \centering
  \includegraphics[width=4cm,clip]{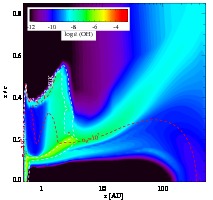}
  \includegraphics[width=4cm,clip]{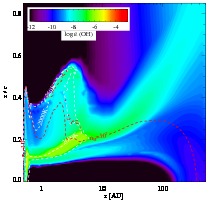}
  \includegraphics[width=4cm,clip]{OHabundance_00034.jpg}
  \includegraphics[width=4cm,clip]{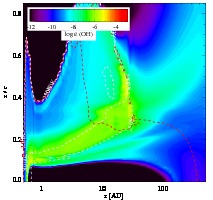}
  \includegraphics[width=4cm,clip]{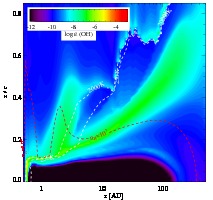}
  \includegraphics[width=4cm,clip]{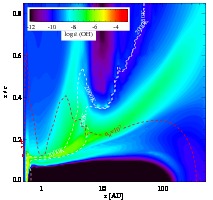}
  \includegraphics[width=4cm,clip]{OHabundance_00082.jpg}
  \includegraphics[width=4cm,clip]{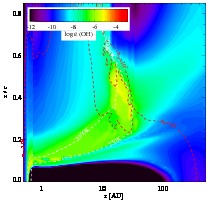}
  \includegraphics[width=4cm,clip]{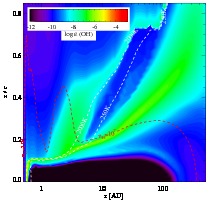}
  \includegraphics[width=4cm,clip]{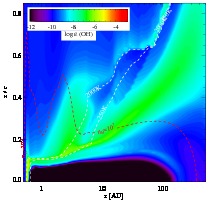}
  \includegraphics[width=4cm,clip]{OHabundance_00130.jpg}
  \includegraphics[width=4cm,clip]{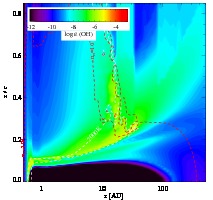}
  \includegraphics[width=4cm,clip]{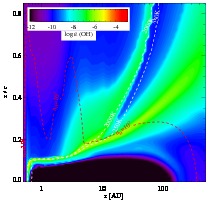}
  \includegraphics[width=4cm,clip]{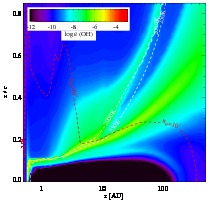}
  \includegraphics[width=4cm,clip]{OHabundance_00178.jpg}
  \includegraphics[width=4cm,clip]{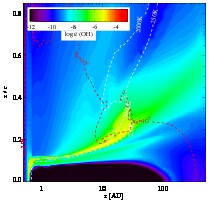}
  \includegraphics[width=4cm,clip]{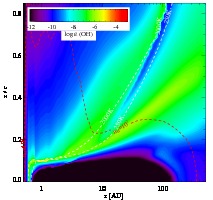}
  \includegraphics[width=4cm,clip]{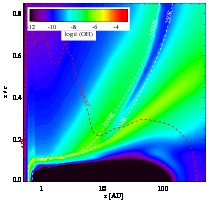}
  \includegraphics[width=4cm,clip]{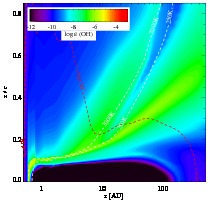}
  \includegraphics[width=4cm,clip]{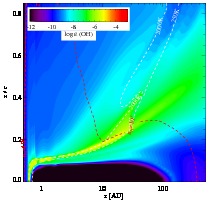}
  \caption{OH abundances. FUV and X-ray fluxes are the same as Fig. \ref{model_electron_abundance_appendix}.}
  \label{OHabundance_struct_appendix}
\end{figure*}

\clearpage

\begin{figure*}
  \centering
  \includegraphics[width=4cm,clip]{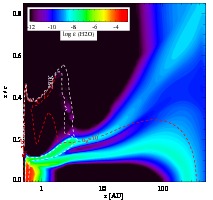}
  \includegraphics[width=4cm,clip]{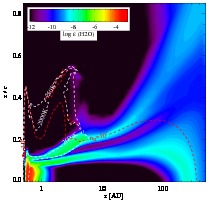}
  \includegraphics[width=4cm,clip]{H2Oabundance_00034.jpg}
  \includegraphics[width=4cm,clip]{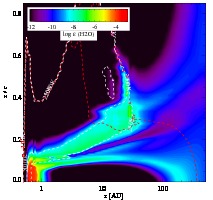}
  \includegraphics[width=4cm,clip]{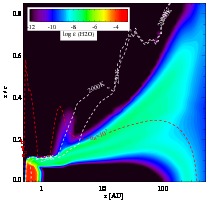}
  \includegraphics[width=4cm,clip]{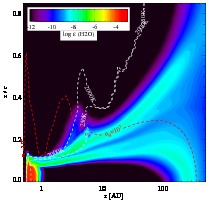}
  \includegraphics[width=4cm,clip]{H2Oabundance_00082.jpg}
  \includegraphics[width=4cm,clip]{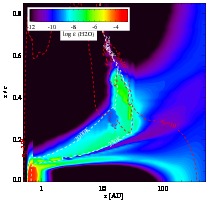}
  \includegraphics[width=4cm,clip]{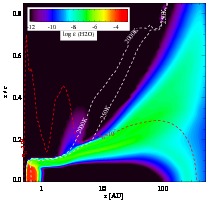}
  \includegraphics[width=4cm,clip]{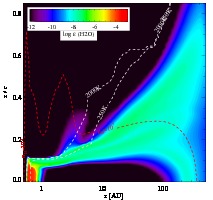}
  \includegraphics[width=4cm,clip]{H2Oabundance_00130.jpg}
  \includegraphics[width=4cm,clip]{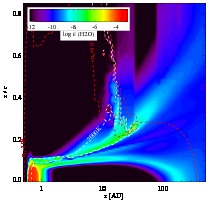}
  \includegraphics[width=4cm,clip]{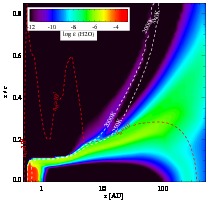}
  \includegraphics[width=4cm,clip]{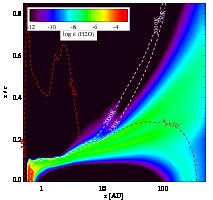}
  \includegraphics[width=4cm,clip]{H2Oabundance_00178.jpg}
  \includegraphics[width=4cm,clip]{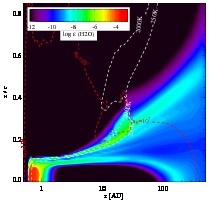}
  \includegraphics[width=4cm,clip]{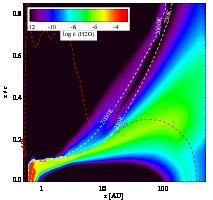}
  \includegraphics[width=4cm,clip]{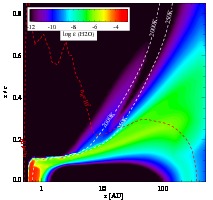}
  \includegraphics[width=4cm,clip]{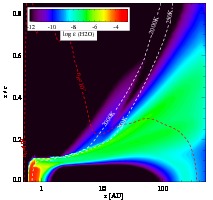}
  \includegraphics[width=4cm,clip]{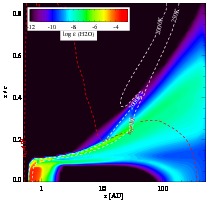}
  \caption{H$_2$O abundances. FUV and X-ray fluxes are the same as Fig. \ref{model_electron_abundance_appendix}.}
  \label{H2Oabundance_struct_appendix}
\end{figure*}

\begin{figure*}
  \centering
  \includegraphics[width=4cm,clip]{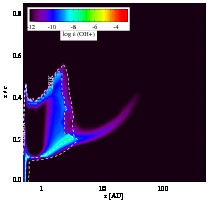}
  \includegraphics[width=4cm,clip]{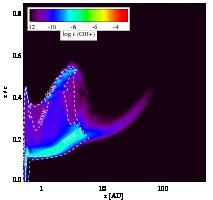}
  \includegraphics[width=4cm,clip]{OH+abundance_00034.jpg}
  \includegraphics[width=4cm,clip]{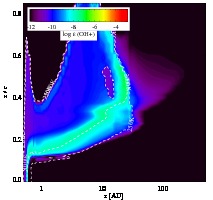}
  \includegraphics[width=4cm,clip]{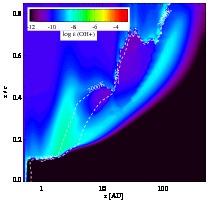}
  \includegraphics[width=4cm,clip]{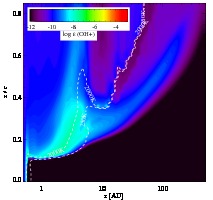}
  \includegraphics[width=4cm,clip]{OH+abundance_00082.jpg}
  \includegraphics[width=4cm,clip]{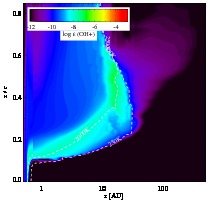}
  \includegraphics[width=4cm,clip]{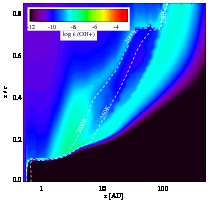}
  \includegraphics[width=4cm,clip]{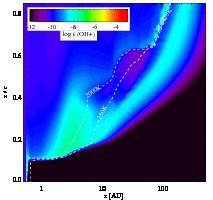}
  \includegraphics[width=4cm,clip]{OH+abundance_00130.jpg}
  \includegraphics[width=4cm,clip]{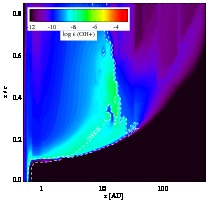}
  \includegraphics[width=4cm,clip]{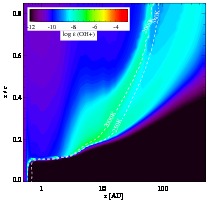}
  \includegraphics[width=4cm,clip]{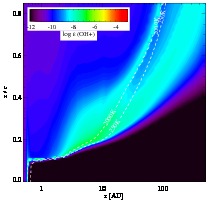}
  \includegraphics[width=4cm,clip]{OH+abundance_00178.jpg}
  \includegraphics[width=4cm,clip]{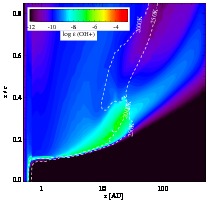}
  \includegraphics[width=4cm,clip]{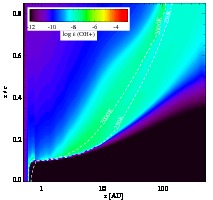}
  \includegraphics[width=4cm,clip]{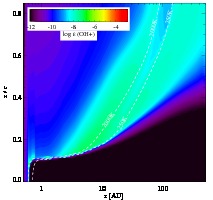}
  \includegraphics[width=4cm,clip]{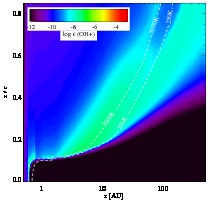}
  \includegraphics[width=4cm,clip]{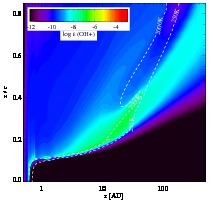}
  \caption{OH$^{+}$ abundances. FUV and X-ray fluxes are the same as Fig. \ref{model_electron_abundance_appendix}.}
  \label{OH+abundance_struct_appendix}
\end{figure*}

\begin{figure*}
  \centering
  \includegraphics[width=4cm,clip]{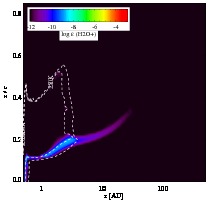}
  \includegraphics[width=4cm,clip]{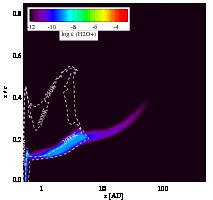}
  \includegraphics[width=4cm,clip]{H2O+abundance_00034.jpg}
  \includegraphics[width=4cm,clip]{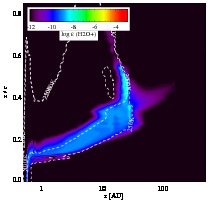}
  \includegraphics[width=4cm,clip]{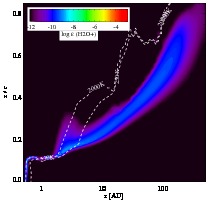}
  \includegraphics[width=4cm,clip]{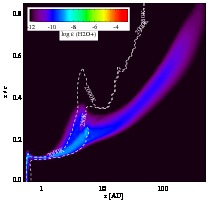}
  \includegraphics[width=4cm,clip]{H2O+abundance_00082.jpg}
  \includegraphics[width=4cm,clip]{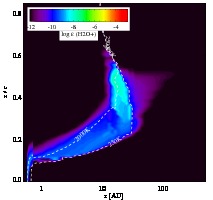}
  \includegraphics[width=4cm,clip]{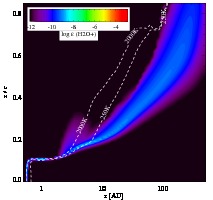}
  \includegraphics[width=4cm,clip]{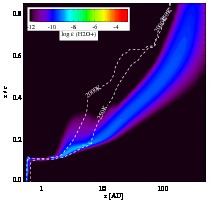}
  \includegraphics[width=4cm,clip]{H2O+abundance_00130.jpg}
  \includegraphics[width=4cm,clip]{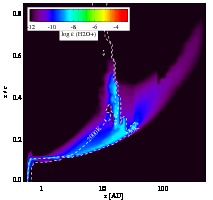}
  \includegraphics[width=4cm,clip]{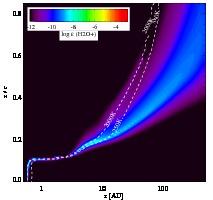}
  \includegraphics[width=4cm,clip]{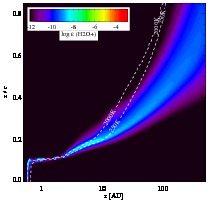}
  \includegraphics[width=4cm,clip]{H2O+abundance_00178.jpg}
  \includegraphics[width=4cm,clip]{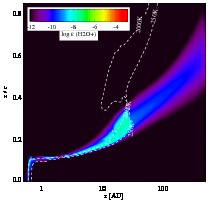}
  \includegraphics[width=4cm,clip]{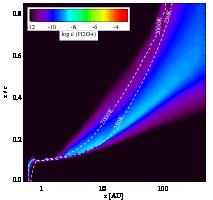}
  \includegraphics[width=4cm,clip]{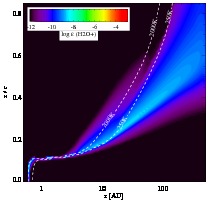}
  \includegraphics[width=4cm,clip]{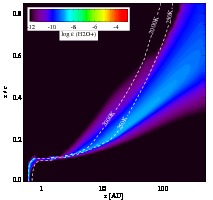}
  \includegraphics[width=4cm,clip]{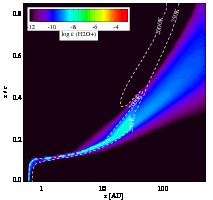}
  \caption{H$_2$O$^+$ abundances. FUV and X-ray fluxes are the same as Fig. \ref{model_electron_abundance_appendix}.}
  \label{H2O+abundance_struct_appendix}
\end{figure*}

\begin{figure*}
  \centering
  \includegraphics[width=4cm,clip]{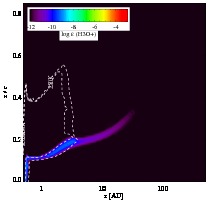}
  \includegraphics[width=4cm,clip]{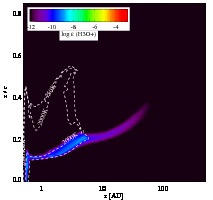}
  \includegraphics[width=4cm,clip]{H3O+abundance_00034.jpg}
  \includegraphics[width=4cm,clip]{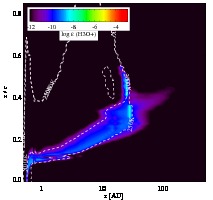}
  \includegraphics[width=4cm,clip]{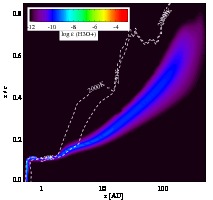}
  \includegraphics[width=4cm,clip]{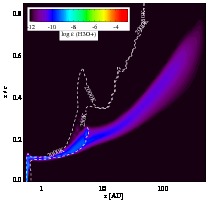}
  \includegraphics[width=4cm,clip]{H3O+abundance_00082.jpg}
  \includegraphics[width=4cm,clip]{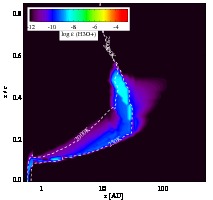}
  \includegraphics[width=4cm,clip]{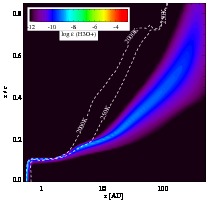}
  \includegraphics[width=4cm,clip]{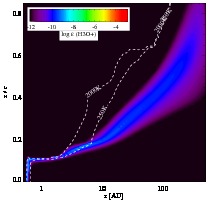}
  \includegraphics[width=4cm,clip]{H3O+abundance_00130.jpg}
  \includegraphics[width=4cm,clip]{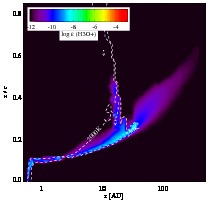}
  \includegraphics[width=4cm,clip]{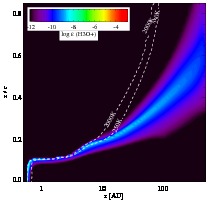}
  \includegraphics[width=4cm,clip]{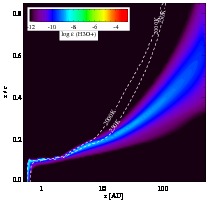}
  \includegraphics[width=4cm,clip]{H3O+abundance_00178.jpg}
  \includegraphics[width=4cm,clip]{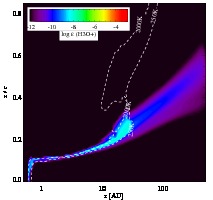}
  \includegraphics[width=4cm,clip]{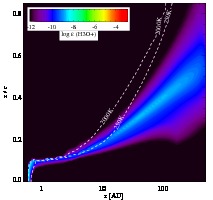}
  \includegraphics[width=4cm,clip]{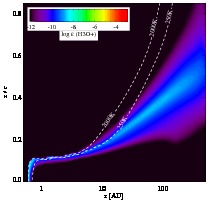}
  \includegraphics[width=4cm,clip]{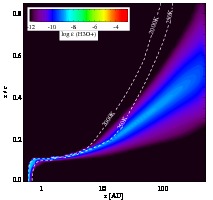}
  \includegraphics[width=4cm,clip]{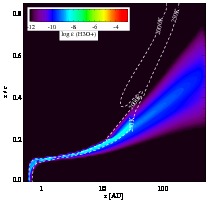}
  \caption{H$_3$O$^+$ abundances. FUV and X-ray fluxes are the same as Fig. \ref{model_electron_abundance_appendix}.}
  \label{H3O+abundance_struct_appendix}
\end{figure*}

\begin{figure*}
  \centering
  \includegraphics[width=4cm,clip]{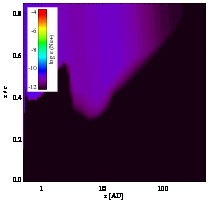}
  \includegraphics[width=4cm,clip]{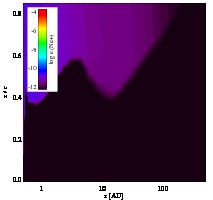}
  \includegraphics[width=4cm,clip]{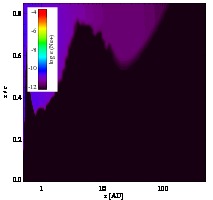}
  \includegraphics[width=4cm,clip]{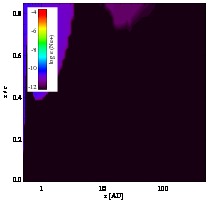}
  \includegraphics[width=4cm,clip]{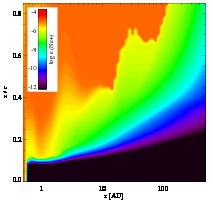}
  \includegraphics[width=4cm,clip]{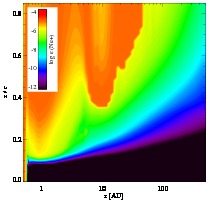}
  \includegraphics[width=4cm,clip]{Ne+abundance_00082.jpg}
  \includegraphics[width=4cm,clip]{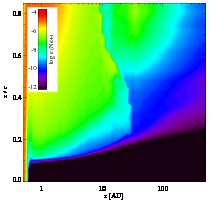}
  \includegraphics[width=4cm,clip]{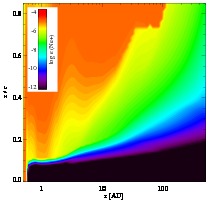}
  \includegraphics[width=4cm,clip]{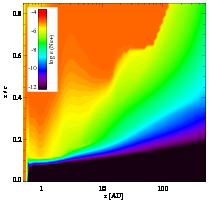}
  \includegraphics[width=4cm,clip]{Ne+abundance_00130.jpg}
  \includegraphics[width=4cm,clip]{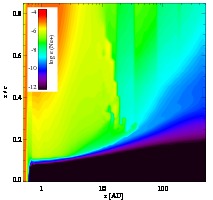}
  \includegraphics[width=4cm,clip]{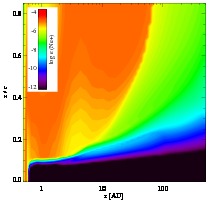}
  \includegraphics[width=4cm,clip]{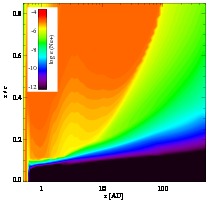}
  \includegraphics[width=4cm,clip]{Ne+abundance_00178.jpg}
  \includegraphics[width=4cm,clip]{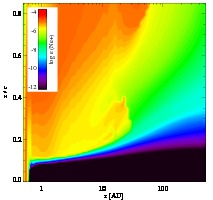}
  \includegraphics[width=4cm,clip]{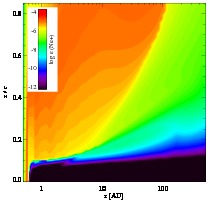}
  \includegraphics[width=4cm,clip]{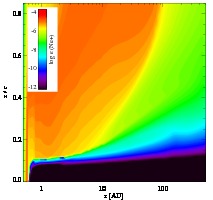}
  \includegraphics[width=4cm,clip]{Ne+abundance_00226.jpg}
  \includegraphics[width=4cm,clip]{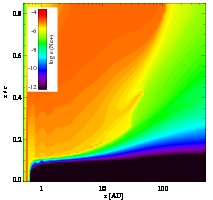}
  \caption{Ne$^+$ abundances. FUV and X-ray fluxes are the same as Fig. \ref{model_electron_abundance_appendix}.}
  \label{Ne+abundance_struct}
\end{figure*}

\begin{figure*}
  \centering
  \includegraphics[width=4cm,clip]{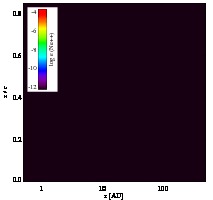}
  \includegraphics[width=4cm,clip]{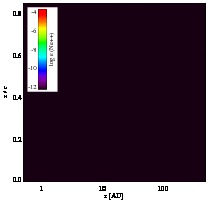}
  \includegraphics[width=4cm,clip]{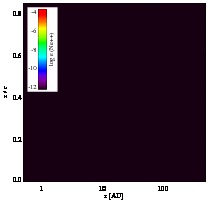}
  \includegraphics[width=4cm,clip]{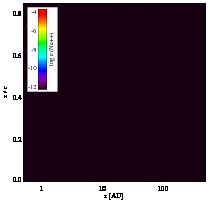}
  \includegraphics[width=4cm,clip]{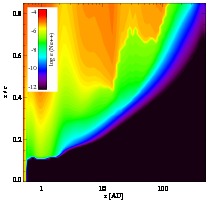}
  \includegraphics[width=4cm,clip]{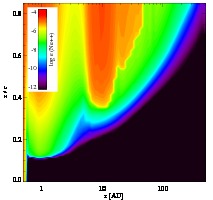}
  \includegraphics[width=4cm,clip]{Ne2+abundance_00082.jpg}
  \includegraphics[width=4cm,clip]{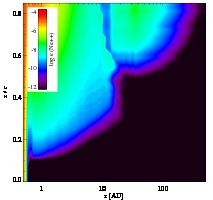}
  \includegraphics[width=4cm,clip]{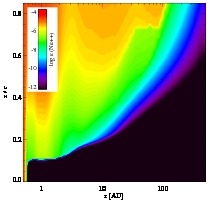}
  \includegraphics[width=4cm,clip]{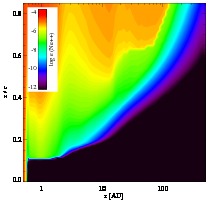}
  \includegraphics[width=4cm,clip]{Ne2+abundance_00130.jpg}
  \includegraphics[width=4cm,clip]{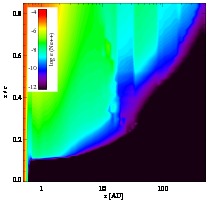}
  \includegraphics[width=4cm,clip]{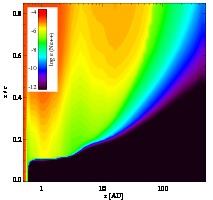}
  \includegraphics[width=4cm,clip]{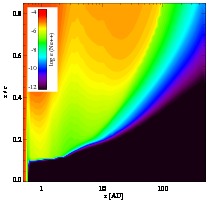}
  \includegraphics[width=4cm,clip]{Ne2+abundance_00178.jpg}
  \includegraphics[width=4cm,clip]{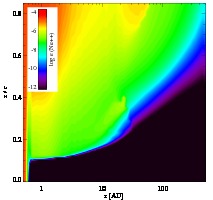}
  \includegraphics[width=4cm,clip]{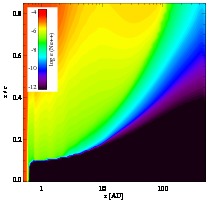}
  \includegraphics[width=4cm,clip]{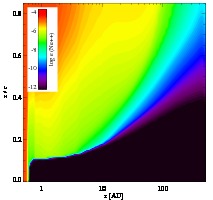}
  \includegraphics[width=4cm,clip]{Ne2+abundance_00226.jpg}
  \includegraphics[width=4cm,clip]{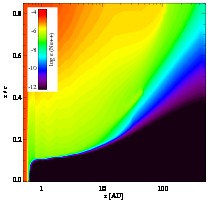}
  \caption{Ne$^{2+}$ abundances. FUV and X-ray fluxes are the same as Fig. \ref{model_electron_abundance_appendix}.}
  \label{Ne2+abundance_struct}
\end{figure*}

\begin{figure*}
  \centering
  \includegraphics[width=4cm,clip]{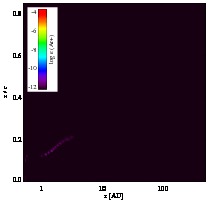}
  \includegraphics[width=4cm,clip]{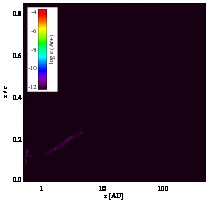}
  \includegraphics[width=4cm,clip]{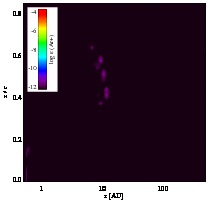}
  \includegraphics[width=4cm,clip]{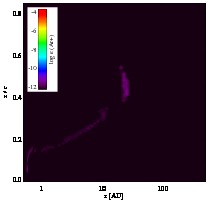}
  \includegraphics[width=4cm,clip]{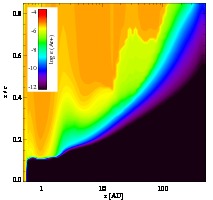}
  \includegraphics[width=4cm,clip]{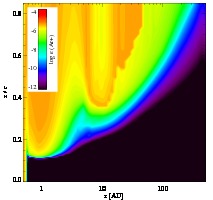}
  \includegraphics[width=4cm,clip]{Ar+abundance_00082.jpg}
  \includegraphics[width=4cm,clip]{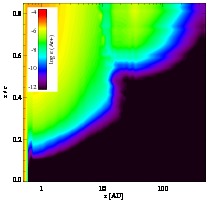}
  \includegraphics[width=4cm,clip]{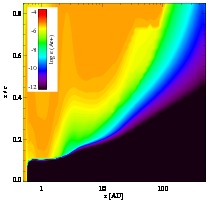}
  \includegraphics[width=4cm,clip]{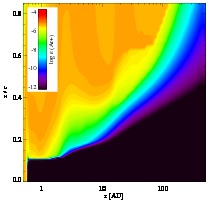}
  \includegraphics[width=4cm,clip]{Ar+abundance_00130.jpg}
  \includegraphics[width=4cm,clip]{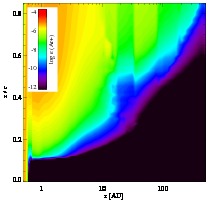}
  \includegraphics[width=4cm,clip]{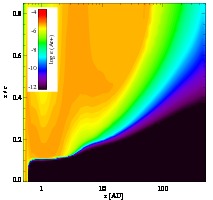}
  \includegraphics[width=4cm,clip]{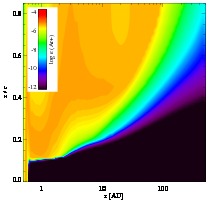}
  \includegraphics[width=4cm,clip]{Ar+abundance_00178.jpg}
  \includegraphics[width=4cm,clip]{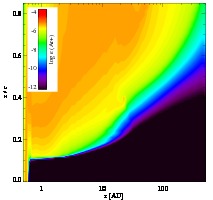}
  \includegraphics[width=4cm,clip]{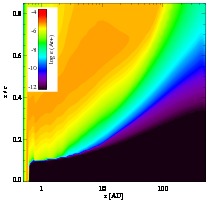}
  \includegraphics[width=4cm,clip]{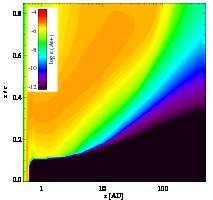}
  \includegraphics[width=4cm,clip]{Ar+abundance_00226.jpg}
  \includegraphics[width=4cm,clip]{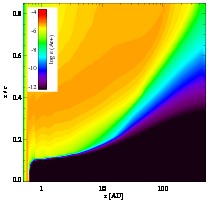}
  \caption{Ar$^+$ abundances. FUV and X-ray fluxes are the same as Fig. \ref{model_electron_abundance_appendix}.}
  \label{Ar+abundance_struct}
\end{figure*}

\begin{figure*}
  \centering
  \includegraphics[width=4cm,clip]{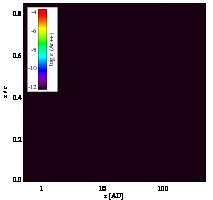}
  \includegraphics[width=4cm,clip]{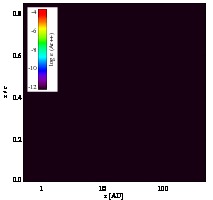}
  \includegraphics[width=4cm,clip]{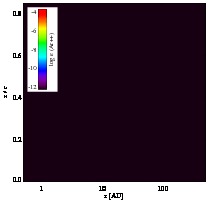}
  \includegraphics[width=4cm,clip]{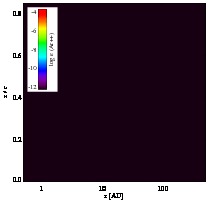}
  \includegraphics[width=4cm,clip]{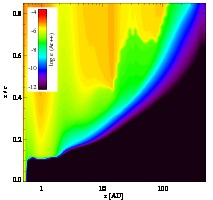}
  \includegraphics[width=4cm,clip]{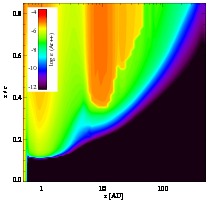}
  \includegraphics[width=4cm,clip]{Ar2+abundance_00082.jpg}
  \includegraphics[width=4cm,clip]{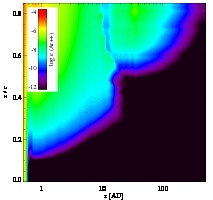}
  \includegraphics[width=4cm,clip]{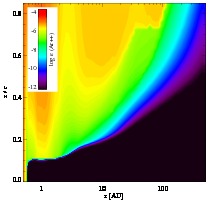}
  \includegraphics[width=4cm,clip]{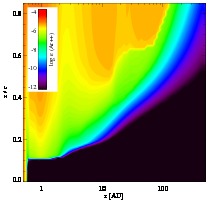}
  \includegraphics[width=4cm,clip]{Ar2+abundance_00130.jpg}
  \includegraphics[width=4cm,clip]{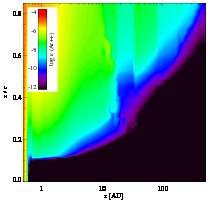}
  \includegraphics[width=4cm,clip]{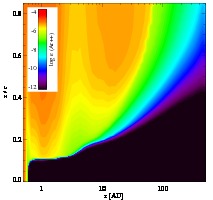}
  \includegraphics[width=4cm,clip]{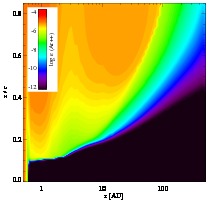}
  \includegraphics[width=4cm,clip]{Ar2+abundance_00178.jpg}
  \includegraphics[width=4cm,clip]{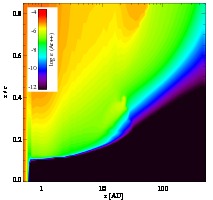}
  \includegraphics[width=4cm,clip]{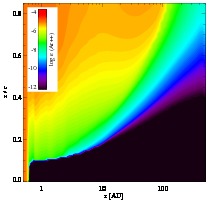}
  \includegraphics[width=4cm,clip]{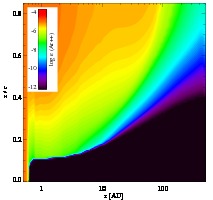}
  \includegraphics[width=4cm,clip]{Ar2+abundance_00226.jpg}
  \includegraphics[width=4cm,clip]{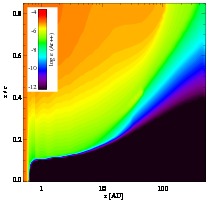}
  \caption{Ar$^{2+}$ abundances. FUV and X-ray fluxes are the same as Fig. \ref{model_electron_abundance_appendix}.}
  \label{Ar2+_abundance_struct_appendix}
\end{figure*}

\begin{figure*}
  \centering
  \includegraphics[width=16cm,clip]{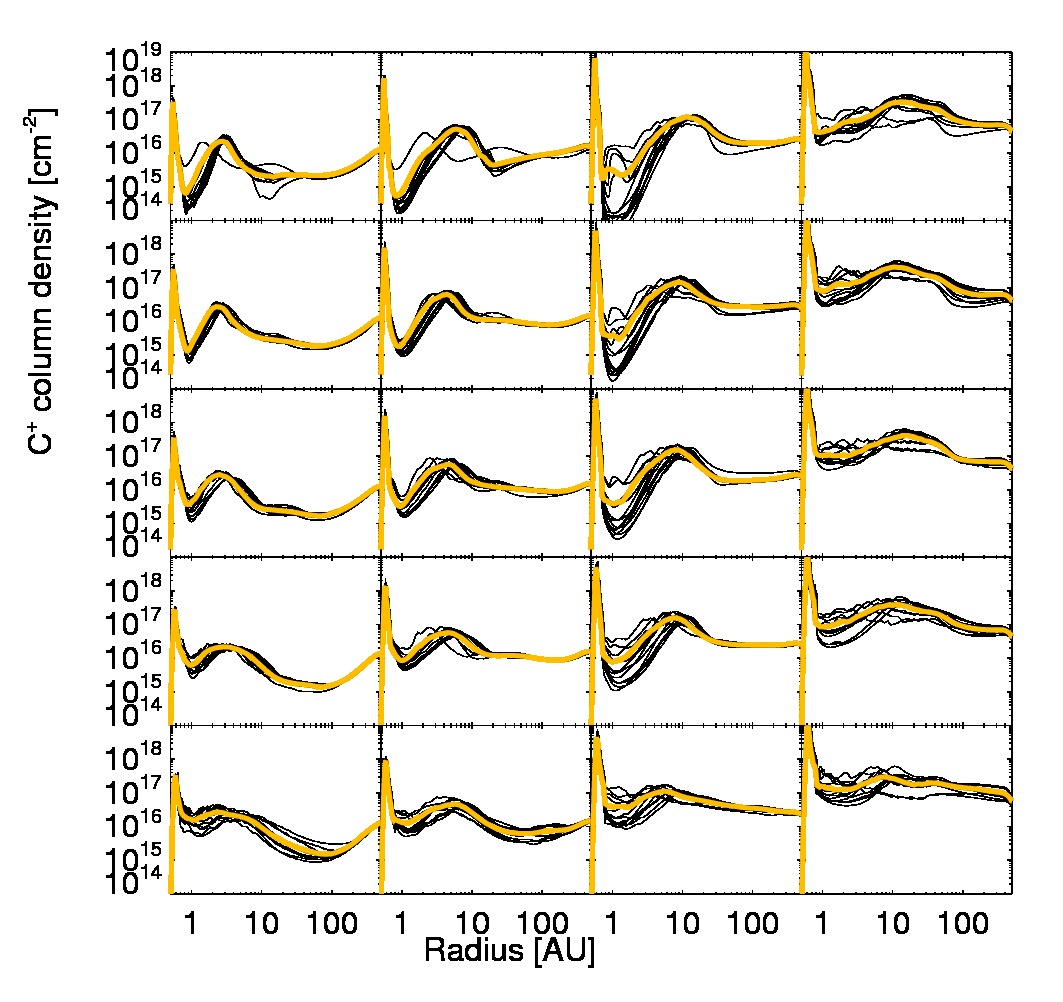}
  \caption{Radial column density distribution of C$^+$. FUV and X-ray fluxes are the same as Fig. \ref{model_electron_abundance_appendix}.}
  \label{C+radial}
\end{figure*}

\begin{figure*}
  \centering
  \includegraphics[width=16cm,clip]{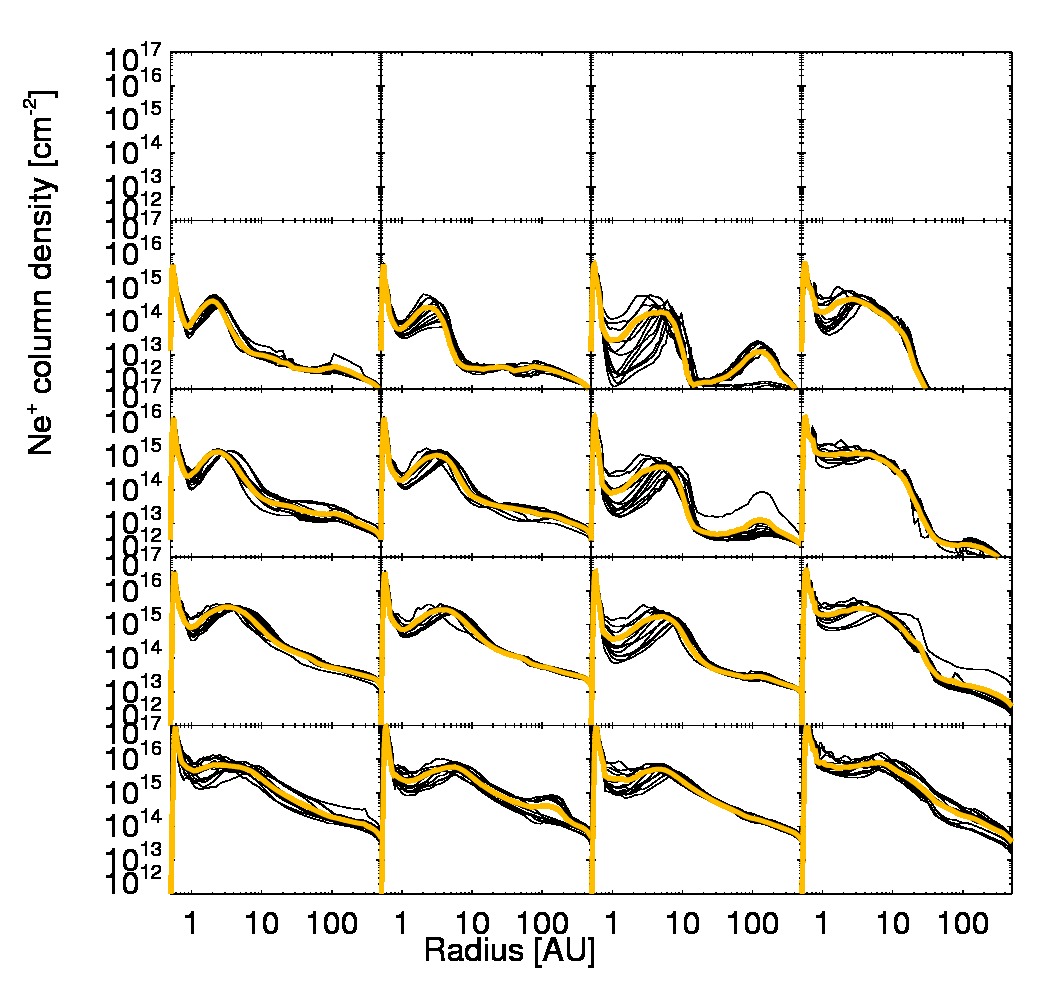}
  \caption{Radial column density distribution of Ne$^{2+}$. FUV and X-ray fluxes are the same as Fig. \ref{model_electron_abundance_appendix}.}
  \label{Ne+radial}
\end{figure*}

\end{appendix}

\end{document}